\documentclass[pra,superscriptaddress,showpacs,preprint,aps,longbibliography]{revtex4-1}
\usepackage{epsfig}             
\usepackage{epstopdf}           
\usepackage{hyperref}           
\usepackage{color}
\usepackage{colortbl}
\usepackage{graphicx}
\usepackage{amssymb,amsmath}
\usepackage{subfigure}
\usepackage{caption}
\usepackage{enumerate}
\usepackage{gensymb}
\usepackage{url}

\begin{document}

\title[ATI]{Above-threshold ionization (ATI) in multicenter molecules: the role of the initial state}
\author{Noslen Su\'arez}
\email[]{noslen.suarez@icfo.es}
\affiliation{ICFO - Institut de Ciencies Fotoniques, The Barcelona Institute of Science and Technology, 08860 Castelldefels (Barcelona), Spain}

\author{Alexis Chac\'on}
\affiliation{Theoretical Division, Los Alamos National Laboratory, Los Alamos, New Mexico 87545, USA}

\author{Emilio Pisanty}
\affiliation{ICFO - Institut de Ciencies Fotoniques, The Barcelona Institute of Science and Technology, 08860 Castelldefels (Barcelona), Spain}

\author{Lisa Ortmann}
\affiliation{Max-Planck Institut f\"ur Physik komplexer Systeme, N\"othnitzer-Strasse 38, D-01187 Dresden, Germany}

\author{Alexandra S. Landsman}
\affiliation{Max-Planck Institut f\"ur Physik komplexer Systeme, N\"othnitzer-Strasse 38, D-01187 Dresden, Germany}
\affiliation{Department of Physics, Max Planck Postech, Pohang, Gyeongbuk 37673, Republic of Korea}

\author{Antonio Pic\'on}
\affiliation{ICFO - Institut de Ciencies Fotoniques, The Barcelona Institute of Science and Technology, 08860 Castelldefels (Barcelona), Spain}

\author{Jens Biegert}
\affiliation{ICFO - Institut de Ciencies Fotoniques, The Barcelona Institute of Science and Technology, 08860 Castelldefels (Barcelona), Spain}
\affiliation{ICREA, Pg. Llu\'{\i}s Companys 23, 08010 Barcelona, Spain}

\author{Maciej Lewenstein}
\affiliation{ICFO - Institut de Ciencies Fotoniques, The Barcelona Institute of Science and Technology, 08860 Castelldefels (Barcelona), Spain}
\affiliation{ICREA, Pg. Llu\'{\i}s Companys 23, 08010 Barcelona, Spain}

\author{Marcelo F. Ciappina}
\affiliation{Institute of Physics of the ASCR, ELI-Beamlines project, Na Slovance 2, 182 21 Prague, Czech Republic}

\date{\today}
\pacs{32.80.Rm,33.20.Xx,42.50.Hz}

\begin{abstract}
A possible route to extract electronic and nuclear dynamics from molecular targets with attosecond temporal and nanometer spatial resolution is to employ recolliding electrons as `probes'.  The recollision process in molecules is, however, very challenging to treat using {\it ab initio} approaches.  Even for the simplest diatomic systems, such as H$_2$, today's computational capabilities are not enough to give a complete description of the electron and nuclear dynamics initiated by a strong laser field.  As a consequence, approximate qualitative descriptions are called to play an important role.  In this contribution we extend the work presented in N. Su\'arez {\it et al.}, Phys.~Rev. A {\bf 95}, 033415 (2017), to three-center molecular targets. Additionally, we incorporate a more accurate description of the molecular ground state, employing information extracted from quantum chemistry software packages. This step forward allows us to include, in a detailed way, both the molecular symmetries and nodes present in the high-occupied molecular orbital. We are able to, on the one hand, keep our formulation as analytical as in the case of diatomics, and, on the other hand, to still give a complete description of the underlying physics behind the above-threshold ionization process. The application of our approach to complex multicenter - with more than 3 centers, targets appears to be straightforward.
\end{abstract}

\maketitle

\section{Introduction}

Strong-field techniques such as high harmonic spectroscopy (HHS) is the workhorse for one of the most stimulating prospects of strong-field and attosecond physics:  the extraction of electronic and nuclear information on the attosecond temporal and sub-{\AA}ngstrom spatial scales using recolliding electrons as `probes'.  HHS employs the quiver motion of an electron, which is liberated by the laser field from the target structure itself, to analyze either the recombination spectrum or the momentum distribution of the rescattering electron.

For instance, the Dyson orbital of an N$_2$ molecule was reconstructed with tomographic techniques, using the information contained in the high-order harmonic generation (HHG) spectrum~\cite{SergeiPRL2006}. The original interpretation of these experiments rely on the strong-field approximation (SFA), which provides a fully quantum description of the well-known "three-step model". Later on, Villeneuve et al.'s~\cite{NatItatani2004} experiment has sparked a true avalanche of experimental and theoretical work on the subject. The use of approximations in the theoretical modelling of HHG in molecules (and in particular the SFA) is, however, not exempt from controversies: the results strongly rely upon the specifics of the model, namely the gauge, the choice of the dipole radiation form, the molecular orbital symmetry and degree of alignment, etc. (cf.~\cite{JensPRA2008, MilosevicPRA2009,Li2013, Qin2012, Senad2012,Nguyen201112,Jens2011, CarlaPRA2010, Jens2010}).  


One step forward in the extraction of molecular structural features is provided by the, so-called, laser-induced electron diffraction (LIED)~\cite{Zuo1996,Lein2002,Lin2010} technique. Here, the recolliding electrons, elastically scattered of the molecular ion, contain information about the multicenter nature of the target that can be easily extracted from the measured photoelectron spectra. Currently, LIED has been used to successfully recover structural information from diatomic and other polyatomic molecules with sub-{\AA}ngstrom spatial resolution. LIED is based on the above-threshold ionization processes. In ATI an electron may directly depart from the molecular target and contribute to the lower-energy region ATI spectrum; this process is termed direct tunneling. On the other hand, the laser-ionized electron might return to the vicinity of the molecular parent ion, driven by the still present laser electric field, and rescatter, thereby gaining much more kinetic energy. This highly energetic electron could excite the remaining ion or even cause the detachment of another electron(s) (for a comprehensive description of these processes within the framework of the SFA and Feynman's path-integral approach, see e.g.~\cite{MaciekScience2001}).

The viability of LIED as a self-imaging technique - the `probe' is a molecular electron that is extracted from the same target it images - has been already established in a series of contributions~\cite{Xu2014,blaga2012, Pullen2015, Wolter308}.
The aim is to gain insight about the electronic structure
of complex molecular targets interpreting the energy spectra and
angular distribution of ATI electrons. Specifically, the high-energy region of the ATI spectra, which is governed by the rescattering process, is particularly sensitive
to the structure and features of the target. In other words, the rescattered electron has acquired information about the target it rescattered off, and hence allows extracting structural information.

The study of the structure
of complex systems, such as molecules, atomic clusters, and
solids, using the ATI spectra is well developed. Investigations of ATI from the simplest molecular systems, i.e.~diatomics, is one the most widely studied processes. Usually, theoretical approaches can be divided in two main groups: those based on a fully quantum-mechanical description, that rely upon on the numerical solution of the time-dependent Schr\"odinger equation (TDSE), and approximate methods based on the SFA and similar quasiclassical approaches. We should mention, however, that the solution of the TDSE within the single active electron approximation (SAE) is only viable for the simplest diatomic molecules, e.g.~, H$_2^+$, H$_2$, D$_2$.  These quantum mechanical results so far largely focused on
the discrepancies between the length and velocity gauge outcomes~\cite{LeinPRA2006}, the influence of the internuclear distance in the interference patterns~\cite{CarlaPRA2007}, or how the molecular alignment ~\cite{LeinJPB2007, Lein2002, LeinPRL2012, MilosevicPRA2006, MilosevicPRL2008} affects the ATI photoelectron spectra. Finally, the importance of the residual Coulomb interaction and 'quality' of the continuum electron wavefunction was also assessed~\cite{Marcelo2007,Marcelo2009}. Even though the TDSE provides the most accurate and complete description of the underlying physics behind ATI and LIED, its numerical integration is very expensive computationally. Furthermore, the TDSE in its full dimensionality can
currently not be solved for complex molecular targets and multielectron
systems. Additionally, it is still not possible to model the time evolution of molecular systems with the required temporal resolution. Thus, approximate descriptions such as the SFA and related methods play a fundamental role in the adequate description of the more complex
instances of LIED and ATI. Similarly, the SFA approach is instrumental in the interpretation of HHG molecular tomography.

We already presented a general theory for symmetric diatomic molecules in the SAE approximation that, amongst other features, allows adjusting both the internuclear separation and molecular potential in a direct and simple way~\cite{PRANoslen2016}. Such approach, relies upon on an analytic approximate solution of the TDSE and is based on a modified version of strong-field approximation (SFA). Using that approach, we were able to find expressions for electron emitted transition amplitudes from two different molecular centers, and accelerated then in the strong laser electric field. In addition, our model directly underpins different underlying physical processes (see e.g.~\cite{PRANoslen2015, PRANoslen2016,PRANoslen2017}). Two important advantages of our
theory are  (i) the dipole matrix elements are free from nonphysical gauge and coordinate system-dependent
terms: this is achieved by adapting the coordinate system, in which SFA is performed, to the center from which
the corresponding part of the time-dependent wave function originates and (ii) we are able to write both the direct and rescattering transition amplitudes in an analytical form, only involving a one-dimensional and two-dimensional time integrals, respectively. 

In the present contribution we build on the theory presented in~\cite{PRANoslen2016}, namely (a) extending the approach to three-center molecular systems and (b) including a more accurate description of the molecular ground state. For (b) a linear combination of atomic orbitals (LCAO), obtained from chemical software suites, is used to model the molecular high-occupied molecular orbital (HOMO). In principle, our approach is capable to manage any basis set, but in order to keep the formulation as analytical as possible, we employ throughout the article a STO-3G basis set.


This paper is organized as follows: in Sec.~II, we present the formulae for the three-centre molecular system ATI transition amplitudes for both direct and rescattered electrons. In Sec.~III we describe how to obtain the bound-free dipole transition matrix element, with the molecular ground state approximated as a LCAO and using a STO-3G basis set. Details of the calculation of the rescattering electron states and the matrix elements that describe the continuum-continuum process are provided in an analytic form. We use the transitions matrix elements obtained in the previous sections (Secs.~II and III) in Sec.~IV to compute photoelectron spectra for diatomic and triatomic molecules. Here, the numerical results retrieved using LCAO approach are compared with those obtained employing the nonlocal short range (SR) potential.
Finally, in Sec.~VI, we summarize the main ideas, present our conclusions and give a brief outlook.

\section{Theory of ATI for a three-center molecular system}\label{cap:Background}

\begin{figure}[htb]
 \centering
            \includegraphics[width=0.6\textwidth] {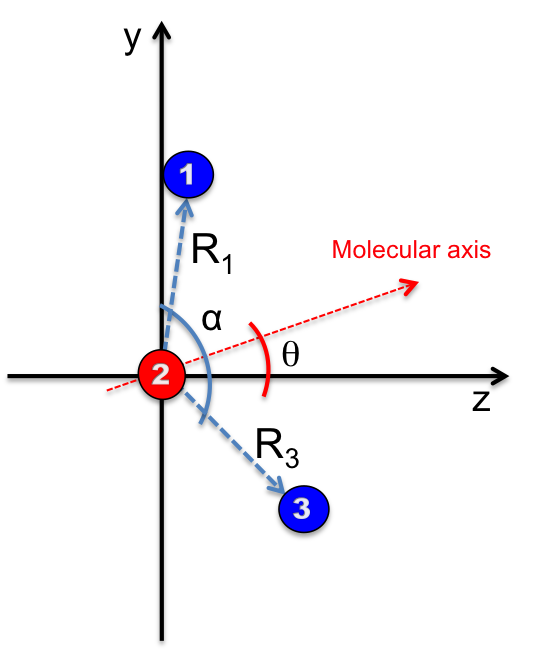}
         \caption{ (color online) A general three-center molecular system aligned a $\theta$ angle with respect to the laser field polarization. The red dashed line defines the molecular axis (see text for details).}
    \label{Fig:1Molecule}
\end{figure}

To obtain the transition probability amplitudes for a triatomic molecule within the generalized strong field approximation (SFA), we extend the analysis presented in our previous works Refs.~\cite{PRANoslen2015,PRANoslen2016,PRANoslen2017}. To this end, we start solving the TDSE for a molecular system of three independent atoms, as is shown in Fig.~\ref{Fig:1Molecule}, under the influence of an intense and short laser pulse ${\bf E}(t)$, linearly polarized along the $z$-axis. 

Our molecular system is defined by a relative position vector ${\bf R}=\textbf{R}_{3}-\textbf{R}_{1}$. We consider the general case of a molecule with three different atoms placed at $\textbf{R}_{1} =[0, \frac{R}{2}\sin(\frac{\alpha}{2} + \theta), \frac{R}{2}\cos(\frac{\alpha}{2}  + \theta)]$,  $\textbf{R}_2 = 0$ and $\textbf{R}_3=[0, -\frac{R}{2}\sin(\frac{\alpha}{2} -\theta), \frac{R}{2}\cos(\frac{\alpha}{2} -\theta)]$, where $\alpha$ and $\theta$ are the angles between the external atoms and the one formed by the molecular axis and the laser electric field polarization, respectively (see Fig.~\ref{Fig:1Molecule}). The molecular axis is defined starting at the origin and bisecting the $\alpha$ angle. A model defined in this way is able to accommodate both linear and angular molecules. For the case of linear configurations $\alpha=180$ degrees. Additionally, our approach allows to study both fixed and randomly oriented molecules (for details see e.g.~\cite{PRANoslen2015,PRANoslen2016,PRANoslen2017}).

In general, as the molecular nuclei are much heavier than the electrons and the laser pulse duration is shorter than the nuclei vibration and rotational dynamics, we fix the nuclei positions and neglect the repulsive interaction between them. Further, throughout the formulation we consider the so-called Single Active Electron (SAE) approximation.

\subsection{Generalized SFA: transition probability amplitudes}
 
The TDSE that governs the whole laser-molecule interactions reads as (atomic units are used throughout this paper unless otherwise stated):

\begin{eqnarray}
\nonumber
i\frac{ \partial}{ \partial t} | \Psi(t) \rangle&=&\Bigg[\frac{-{\nabla}^2}{2} + \hat{V}(\textbf{r}) +\hat{V}_{int}({\bf r},t) \Bigg]| \Psi(t) \rangle,
\label{Eq:SE}
\end{eqnarray}
where $\hat{V}(\textbf{r})$ the potential operator that describes the interaction of the nuclei with the active electron. $\hat{V}_{int}({\bf r},t)=-q_e\hat{{\bf E}}(t)\cdot\hat{{\bf r}}$ represents the interaction of the molecular system with the laser electric field, written in the dipole approximation and length gauge. $q_e$ designates the electron charge (in atomic units $q_e=-1.0$~a.u.).

Our model will be valid in a regime of parameters where the SFA becomes applicable~\cite{Keldysh1965, Faisal1973,Reiss1980,Lewenstein1986, Lewenstein1994,Lewenstein1995}.  As a result, we work in the tunneling regime and we assume that the atomic potential $V(\textbf{r})$ does not play an important role in the electron dynamics once the electron is freed and moving in the laser continuum. Following these observations, we further assume that: (i) Only the ground, $|0 \rangle$, and the continuum  states, $ |\textbf{v}\rangle$, are taken into account in the interaction process. (ii) There is no depletion of the ground state $(U_p < U_{sat})$. (iii) The continuum  states  are approximated by Volkov states; in the continuum the electron is considered as a free particle entirely moving driven by the laser electric field. For a more detailed discussion of the validity of the above statements see e.g.~Refs.~\cite{Lewenstein1994,Lewenstein1995, PRANoslen2015,PRANoslen2016, PRANoslen2017}. 


Based on the assertion (i), we propose a state, $|\Psi(t) \rangle=\sum_{j=1}^{3}| \Psi_j (t)\rangle$, to describe the time-evolution of the three-center system, i.e.~a superposition of three atomic states. In turn, each independent state, $|\Psi_j (t)\rangle$, is a coherent superposition of ground, $|0\rangle= \sum_{j=1}^{3} |0_j\rangle$, and continuum states, $|\textbf{v}\rangle$ ~\cite{Lewenstein1994,Lewenstein1995}:
\begin{equation}
| \Psi_j (t)\rangle= e^{\textit{i}I_p\textit{t}}\bigg[a(t) |0_j\rangle + \: \int{\textit{d}^3 \textbf{v} \: \textit{b}_j( \textbf{v},t) |\textbf{v}\rangle} \bigg],
\label{Eq:PWavef1}
\end{equation}
where the subscript $j=1,2,3$ refers to the positions $\textbf{R}_1$, $\textbf{R}_2$ and $\textbf{R}_3$ of each atom in the three-center molecule, respectively.

The factor $a(t)$ represents  the  amplitude  of the  ground  state,  that is considered constant in time, i.e.~$a(t) \approx 1$, under the assumption  that  there is no depletion (see statement (ii)). The pre-factor $e^{\textit{i}I_p\textit{t}}$, represents the phase oscillations which describe the accumulated  electron energy in the ground state ($I_p=-E_0$ is the ionization potential of the molecular target, with $E_0$ the ground state energy of the three-center molecular system). Furthermore, the continuum states transition amplitudes  are denoted by $\textit{b}_j(\textbf{v},t)$, with $j=1,2$ or $3$ for the atom placed at $\textbf{R}_1$, $\textbf{R}_2$ or $\textbf{R}_3$, respectively. They are obtained following the same procedure as in Ref.~\cite{PRANoslen2016}. Consequently, the time variation of each individual transition amplitude $\dot{b}_j( \textbf{v},t) $ reads:
\begin{eqnarray}
\dot{b}_j( \textbf{v},t) &=& \,\,  -i\bigg(\frac{\textbf{v}^2}{2}+ I_p \bigg) \textit{b}_j( \textbf{v},t) - i \: \textbf{E}(t) \cdot\langle \textbf{v} |  \textbf{r}|0_j \rangle -i \: \textbf{E}(t) \cdot \int{\textit{d}^3 \textbf{v}^{\prime} \: \textit{b}_j( \textbf{v}^{\prime},t)\langle \textbf{v} | \textbf{r} |\textbf{v}^{\prime}\rangle}.
\label{Eq:Newbl}
\end{eqnarray}

The first term on the right-hand of Eq.~(\ref{Eq:Newbl}) represents the  phase evolution  of the  electron in the oscillating laser electric field. In the second term  we have defined the bound-free transition dipole matrix  element as:
\begin{equation}
\textbf{d}_j( \textbf{v})= - \langle \textbf{v}_p |(\textbf{r}-\textbf{R}_j)|0_j \rangle= - \langle \textbf{v}_p |\textbf{r}|0_j \rangle +\textbf{R}_j\langle \textbf{v}_p |0_j \rangle. 
 \label{Eq:DM1}
\end{equation}

The state $ |\textbf{v} \rangle$ represents a scattering state constructed as a plane wave, $| {\bf v}_p \rangle$ plus corrections on each center position $| \delta{\bf v}_j \rangle$. Based on statement (iii) our formulation only considers the continuum state as a plane wave $| {\bf v}_p \rangle$ for the calculation of the bound-free dipole matrix element. Notice that in Eq.~(\ref{Eq:DM1}) we include a correction depending on the relative position of each of the atoms $\textbf{R}_j$. This adjustment allow us to built a bound-free  transition matrix element free of nonphysical gauge and coordinate system-dependent contributions (see~\cite{PRANoslen2016, PRANoslen2017} for more details).

On the third term of Eq.~(\ref{Eq:Newbl}) we define the continuum-continuum transition matrix element $ \textbf{G}( \textbf{v}, \textbf{v}^{\prime} )=\langle \textbf{v} | \textbf{r} |\textbf{v}'\rangle$ that relies upon on the scattering states $|\textbf{v}\rangle$ and $|\textbf{v}'\rangle$ as:
\begin{equation}
\textbf{G}( \textbf{v}, \textbf{v}^{\prime} ) = \textit{i} \:\nabla_\textbf{v}\delta(\textbf{v}-\textbf{v}^{\prime}) - \textbf{R}_j\delta(\textbf{v}-\textbf{v}^{\prime}) + \textbf{g} (\textbf{v},\textbf{v}^{\prime}).
 \label{Eq:CC1m}
\end{equation}

The first two terms on the right-hand side of Eq.~(\ref{Eq:CC1m}) are associated to events where the laser-ionized electron is accelerated by the laser electric field without probability of returning close to any of the ion-cores and rescatters. The last one, the rescattering transition matrix element $\textbf{g}( \textbf{v}, \textbf{v}^{\prime} )$, accounts for all the continuum-continuum processes concerning the entire molecule.  Thus, in $\textbf{g}( \textbf{v}, \textbf{v}^{\prime} )$ the residual Coulomb potential has to be taken into account. In this sense it can be written as a sum of components representing each rescattering channel on the molecule. The second term of Eq.~(\ref{Eq:CC1m}) then reads as:
\begin{eqnarray}
\textbf{g}(\textbf{v},\textbf{v}^{\prime}) &=& \langle \textbf{v}_p | (\textbf{r} -\textbf{R}_1)|\delta\textbf{v}_\mathit{1}'\rangle  + \langle \delta\textbf{v}_\mathit{1} |( \textbf{r} -\textbf{R}_1)|\textbf{v}_p'\rangle + \langle\textbf{v}_p | (\textbf{r} -\textbf{R}_2)|\delta\textbf{v}_\mathit{2}'\rangle \nonumber \\
&&+\: \langle \delta\textbf{v}_\mathit{2} | (\textbf{r} -\textbf{R}_2)|\textbf{v}_p'\rangle+\langle \textbf{v}_p | (\textbf{r} -\textbf{R}_3)|\delta\textbf{v}_\mathit{3}'\rangle +\langle \delta\textbf{v}_\mathit{3} | (\textbf{r} -\textbf{R}_3)|\textbf{v}_p'\rangle.
 \label{Eq:gm}
 \end{eqnarray}
Notice that in Eq.~(\ref{Eq:gm}) we include the corrections related with the relative positions of each of the atoms (see the discussion after Eq.~(\ref{Eq:DM1})). The transition amplitude thus reads as:
\begin{equation}
\begin{split}
 \dot{b}_{j}( \textbf{v},t) =&  -i\left(\frac{\textbf{v}^2}{2}+ I_p -\textbf{R}_j \cdot \textbf{E}(t) \right)\textit{b}_{j}( \textbf{v},t) +i\textbf{E}(t) \cdot {\bf d}_{j}({\bf v}) \\
 & +\,{\bf E}(t)\cdot\nabla_{\bf v} b_{j}({\bf v},t)-i \textbf{E}(t)\cdot \int{\textit{d}^3 \textbf{v}^{\prime}\: \textit{b}_{j}( \textbf{v}^{\prime},t)\:{\bf g}( \textbf{v}, \textbf{v}^{\prime})}.
 \label{Eq:BpuntoL}
\end{split}
\end{equation}

In the following, we use perturbation theory over ${\bf g}( \textbf{v}, \textbf{v}^{\prime})$ in order to solve the partial differential equation Eq.~(\ref{Eq:BpuntoL}). The zeroth-order solution $b_{0,j}({\bf v},t)$ corresponds to the direct transition amplitude describing processes where the laser-ionized electron goes to the continuum and never return to the vicinity of the molecule, i.e.~there is no rescattering with the remaining molecular-ion. On the other hand, the first-order solution $b_{1,jj'}({\bf v},t)$ refers to an electron that, once ionized at a particular center, has a certain probability of rescattering with each of the remaining ions (including the one from which it was laser-ionized). Note that here, as in the case of diatomics~\cite{PRANoslen2016}, for $b_{1,jj'}({\bf v},t)$ we have two indexes: $j$ and $j'$. $j$ denotes the atom from where the electron is released (this index is related to the ionization processes through $\textit{b}_{0,j}( \textbf{v}^{\prime},t)$) and $j'$ represents the position of the rescattering center. The total transition amplitude for the whole system $b({\bf v},t)$ is then the sum over the direct and rescattering processes involving all the molecular centers (see next sections for details).

\subsection{Direct transition amplitude} 

The zeroth-order solution of each center $j$, i.e.~the direct transition amplitude $b_{0,j}({\bf v},t)$,
is obtained solving the equation
\begin{eqnarray}
{\partial }_tb_{0,j}( \textbf{v},t) &=&-\textit{i}  \left(\frac{\textbf{v}^2}{2}+ I_p-\textbf{R}_{j} \cdot \textbf{E}(t) \right) b_{0,j}( \textbf{v},t)+ \textit{i}\: \textbf{E}(t) \cdot  \textbf{d}_j( \textbf{v}) +\textbf{E}(t) \cdot \nabla_{\textbf{v}}b_{0,j}( \textbf{v},t),
\label{Eq:dB1}
\end{eqnarray}
that is obtained considering $\textbf{g} (\textbf{v},\textbf{v}^{\prime})=0$ in Eq.~(\ref{Eq:BpuntoL})~\cite{Lewenstein1995, PRANoslen2015,PRANoslen2016}.
Therefore, the $b_{0,j}({\bf v},t)$ in terms of the canonical momentum~${\bf p}= {\bf v} - {\bf A}(t)$, can be written as:
\begin{eqnarray}
b_{0,j}( \textbf{p},t) &=&\textit{i}   \:\int_0^t{\textit{d} \textit{t}^{\prime}\:\textbf{E}(t^{\prime})}\:\cdot \textbf{d}_{j}\left[ \textbf{p}+\textbf{A}(t^{\prime})\right] \exp\left[-\textit{i} \:{S}_{j}({\bf p},t,t^{\prime})\right]. 
\label{Eq:b_10}
\end{eqnarray}

Equation (\ref{Eq:b_10}) has a direct physical interpretation: it can be understood as the sum of all the ionization events that occur from the time $t'$ to $t$. Then, the instantaneous transition probability amplitude of an electron at a time $t'$, at which it appears into the continuum with momentum ${\bf v}(t')= {\bf v}-{\bf A}(t)+{\bf A}(t')=\textbf{p}+\textbf{A}(t^{\prime})$, is defined by the argument of the time integral in Eq.~(\ref{Eq:b_10}) (note that $\textbf{A}(t) =-\int^{t}{ \textbf{E}(t^{\prime})dt^{\prime}}$ is the associated vector potential). Furthermore, the exponent phase factor in Eq.~(\ref{Eq:b_10}) stands for the ``semi-classical action" and look like,
\begin{eqnarray}
 {S}_{j}({\bf p},t,t^{\prime}) =\int_{t^{\prime}}^{t}{\:d{\tilde t}\left\{[{\bf p}+\textbf{A}({\tilde t})]^2/2 +I_p -\textbf{R}_{j} \cdot  \textbf{E}({\tilde t})  \right \}},
 \label{Eq:Action}
 \end{eqnarray}
 that defines a possible electron trajectory from the birth time $t'$, at position $\textbf{R}_{j}$, until the ``recombination" one $t$. 

Considering we are interested in to obtain the transition amplitude $b_{0,j}({\bf p},t)$ at the end of the laser pulse, the time $t$ is set at $t=t_{\rm F}$. Consequently, the integration time window is defined as $t\in[0,t_{\rm F}]$. Furthermore we set ${\bf E}(0) = {\bf E}(t_{\rm F}) = {\bf 0}$, in such a way to make sure that the laser electric field is a time oscillating wave without static components (the  same arguments apply to the vector potential  ${\bf A}(t)$). Finally, the total transition amplitude for the direct process on our three-center molecular system reads as:
 \begin{eqnarray}
 b_0( \textbf{p},t)= \sum_{j=1}^{3} b_{0,j}( \textbf{p},t).
 \label{Eq:b_0}
\end{eqnarray}
\subsection{Rescattering transition amplitude} 

In order to find the first order correction, i.e.~the transition amplitude for the rescattered photoelectrons $b_{1,jj'}({\bf v},t)$, we set now $\textbf{g} (\textbf{v},\textbf{v}^{\prime}) \not= \textbf{0}$ in Eq.~(\ref{Eq:BpuntoL}). The $b_{1,jj'}({\bf v},t)$ is then obtained by inserting the zeroth-order solution, $b_{0,j}({\bf v},t)$, in the right-hand side of Eq.~(\ref{Eq:BpuntoL}). Thereby, we obtain:
 \begin{eqnarray}
\hspace{-0.55cm}\dot{b}_{1,jj'}( \textbf{v},t) &=&  -i\left(\frac{\textbf{v}^2}{2}+ I_p -\textbf{R}_{j} \cdot \textbf{E}(t) \right)\textit{b}_{1,j'j}( \textbf{v},t)-i \textbf{E}(t)\cdot \int{\textit{d}^3 \textbf{v}^{\prime}\: \textit{b}_{0,j}( \textbf{v}^{\prime},t)\:{\bf g}_{jj'} ( \textbf{v}, \textbf{v}^{\prime})}.
\label{Eq:b_1xy}
 \end{eqnarray}

The solution of Eq.~(\ref{Eq:b_1xy}) in momentum space reads as:
\begin{eqnarray}
\hspace{-0.65cm}b_{1,jj'}( \textbf{p},t) &=& -\int_0^t{\textit{d}t^{\prime}}\int_0^{t^\prime}{\textit{d} \textit{t}^{{\prime}{\prime}}}\int{\textit{d}^3\textbf{p}^{\prime} \:\textbf{E}(t^{\prime})} \cdot \textbf{g}_{jj'} \left[\textbf{p}+\textbf{A}(t^{\prime}),\textbf{p}^{\prime}+\textbf{A}(t^{\prime})\right] \exp{\left[-\textit{i}   S_{j'}({\bf p},t,t')\right]} \nonumber \\
&& \times \:\textbf{E}(t^{{\prime}{\prime}}) \cdot \textbf{d}_{j}\big[ \textbf{p}^{\prime} +\textbf{A}(t^{{\prime}{\prime}})\big]  \exp{\left[-\textit{i}   S_{j}({\bf p}',t',t'')\right]}.
\label{Eq:b_1LR1}
 \end{eqnarray} 

Notice that we have a three-center molecule where ionization and rescattering processes could take place at each of the individual atoms. The total rescattering transition amplitude $b_{1}({\bf p},t)$ is then formed by nine terms and can be written as:
\begin{eqnarray}
b_{1}({\bf p},t)&=& \sum_{j=1}^{3}\sum_{j'=1}^{3} b_{1,jj'}({\bf p},t).
\label{Eq:b_11} 
 \end{eqnarray}
The above equation contains information about all the possible rescattering scenarios that take place in our three-center molecular system. In addition, a direct physical interpretation of each term can be inferred considering: (i) $j=j'$, for the local rescattering processes, and (ii) $j\neq j'$, for the cross and nonlocal ones. 

In general, if  $j=j'$ in Eq.~(\ref{Eq:b_1LR1}), both the electron-tunneling ionization and rescattering take place at the same atom, located at $\textbf{R}_j$. We refer to this process as ``spatially localized'', since the electron performs a local-rescattering with the same atomic core from which it was born at. The total local processes in the case of our three-center molecule then read as:
\begin{eqnarray}
b_{1,jj}({\bf p},t)&=& \sum_{j=1}^{3} b_{1,jj}({\bf p},t) =b_{1,11}({\bf p},t) +b_{1,22}({\bf p},t) +b_{1,33}({\bf p},t) .
\label{Eq:b_loc} 
 \end{eqnarray}

For the case where $j\neq j'$, the processes involve two spatial locations, i.e.~two atomic centers. They represent events where the electron is tunnel-ionized from an atom located at $\textbf{R}_{j}$ and the rescattering process takes place at other atom, located at $\textbf{R}_{j'}$. We call this processes as ``cross processes''. In fact, there exist another processes involving two centers. They occur when the electron is propelled to the continuum from an atom located at $\textbf{R}_{j}$ and rescatters with the same parent ion, but there is certain probability of electron emission from the other ion-core, placed at $\textbf{R}_{j'}$. We label these processes as ``nonlocal processes''. The total  ``nonlocal processes'' can then be calculated as:
\begin{eqnarray}
b_{1,j\neq j'}({\bf p},t)&=& \sum_{j\neq j'}^{3} b_{1,jj'}({\bf p},t), \nonumber\\
&=&b_{1,12}({\bf p},t) +b_{1,21}({\bf p},t) +b_{1,23}({\bf p},t) \nonumber\\
&& +\: b_{1,32}({\bf p},t) +b_{1,13}({\bf p},t) +b_{1,31}({\bf p},t) .
\label{Eq:b_nloc} 
 \end{eqnarray}

Equation~(\ref{Eq:b_1LR1}) has a clear physical interpretation.  As it is expected the rescattering transition amplitude contains two exponential factors, each representing the electron excursions in the laser continuum.  The last factor in Eq.~(\ref{Eq:b_1LR1}), $\exp{\left[-\textit{i} {S}_j({\bf p}^{\prime},t{'},t'')\right]}$, represents the accumulated phase of an electron born at the time $t^{\prime\prime}$ in $\textbf{R}_j$ until it rescatters at time $t^\prime$. In the same way $\exp\left[-\textit{i}{S}_{j'}({\bf p},t,t')\right]$ defines the accumulated phase of the electron after it rescatters at a time $t'$ to the ``final" one $t$,  when the electron is ``measured" at the detector with momentum \textbf{p}. Furthermore, the quantity $\textbf{E}(t^{{\prime}{\prime}}) \cdot \textbf{d}_j\left[\textbf{p}^{\prime} +\textbf{A}(t^{{\prime}{\prime}})\right]$ is the probability amplitude  of an emitted electron  at  the  time $t^{\prime\prime}$ that has a kinetic momentum of ${\bf v}^{\prime}(t'')=\textbf{p}^{\prime}+\textbf{A}(t^{{\prime}{\prime}})$. Finally, the term $\textbf{E}(t^{\prime}) \cdot \textbf{g}_{jj'} \left[\textbf{p}+\textbf{A}(t^{\prime}),\textbf{p}^{\prime}+\textbf{A}(t^{\prime})\right]$ defines the probability amplitude of rescattering at time $t'$.

The rescattering transition amplitude defined in Eq.~(\ref{Eq:b_1LR1}) is a multidimensional highly oscillatory integral: a 2D time-integral embedded in a 3D momentum-integral. The numerical computation of this integral represents a very demanding task from a computational perspective. In order to reduce these complications, and at the same time obtain a physical interpretation of the whole ATI process, we shall employ the stationary phase method to partially evaluate it.
Let us rewrite the quasi-classical action for the three-center molecule as:
 \begin{equation}
 {S}_{j}({\bf p}',t',t'') =  \textbf{R}_{j} \cdot  [ \textbf{A}(t') - \textbf{A}(t'')] + {S}({\bf p}',t',t''),
\end{equation}
where ${S}({\bf p}',t',t'') =   \int_{t''}^{t'}{\:d{\tilde t}\left[({\bf p}' + \textbf{A}({\tilde t}))^2/2 +I_p\right]}$ is the well known semi-classical action that we treat it as in the atomic case (see Ref.~\cite{PRANoslen2015}), i.e.~we only take the contributions to the momentum integral at the saddle or stationary points ${\bf p}'_s$, which are obtained from the equation: $ \nabla_{{\bf p}^{\prime}}
\textit{S}(\textbf{p}^{\prime})|_{{\bf p}'_s}={\bf 0} $. The latter equation implies $\textbf{p}'_s = -\frac{1}{\tau}\int_{t^{{\prime}{\prime}}}^{t^\prime}  { \textbf{A}(\tilde{t}) \textbf{d}\tilde{t}}$, where $\tau=t^{\prime}-t^{\prime \prime}$ defines the excursion time of the electron in the laser continuum. 

Therefore, applying the standard saddle point method to the 3D momentum integral we obtain an expression for rescattering transition amplitude $b_{1,jj'}( \textbf{p},t)$ as:
\begin{eqnarray}
b_{1,jj'}( \textbf{p},t) &=& -\int_0^t{\textit{d}t^{\prime}}\int_0^{t^\prime}{\textit{d} \textit{t}^{{\prime}{\prime}}}\left( \frac{\pi}{\varepsilon +\frac{\textit{i}(t'-t'')}{2}}  \right)^\frac{3}{2} \:\textbf{E}(t^{\prime}) \cdot \textbf{g}_{jj'} \left[\textbf{p}+\textbf{A}(t^{\prime}),\textbf{p}_s^{\prime}+\textbf{A}(t^{\prime})\right] \nonumber \\
&& \times \:\exp{\left[-\textit{i}   S_{j'}({\bf p},t,t')\right]} \: \textbf{E}(t^{{\prime}{\prime}}) \cdot \textbf{d}_{j}\big[ \textbf{p}_s^{\prime} +\textbf{A}(t^{{\prime}{\prime}})\big]  \exp{\left[-\textit{i}   S_{j}(\textbf{p}_s^{\prime},t',t'')\right]}.
\label{Eq:b_1F}
 \end{eqnarray} 

Here, we have introduced an infinitesimal parameter $\varepsilon$, small but non-zero, to avoid the divergence at  $t'=t''$. The character of this not integrable singularity and the simple method to handle it has been pioneered in Ref.~\cite{Lewenstein1994}. For more information see the discussion in Ref.~\cite{PRANoslen2015}.

Using the saddle point approximation in the rescattering transition amplitude $b_{1,jj'}( \textbf{p},t)$, we have substantially reduced the dimensionality of the problem, i.e.~from a 5D integral to a 2D one. This simplification is extremely advantageous from a computational viewpoint.  Moreover,  by invoking  the  saddle  point method,  we obtain a quasi-classical picture for the rescattering transition amplitude of molecular systems.  A similar approach is described in~\cite{Corkum1993, Lewenstein1995, PRANoslen2015, PRANoslen2016}.

As we did for the case of diatomics, Ref.~\cite{PRANoslen2016}, the total rescattering transition amplitude, Eq.~(\ref{Eq:b_11}), is split in two main contributions: the local and the nonlocal + cross processes. In this way we define the $b_{1}( \textbf{p},t)$ as:
 \begin{eqnarray}
b_{1}( \textbf{p},t)&=&b_{\mathrm{Local}}( \textbf{p},t) +b_{\mathrm{NL+Cross}}( \textbf{p},t),\nonumber\\
&=&b_{1,j=j'}({\bf p},t)+b_{1,j\neq j'}({\bf p},t).
\label{Eq:b_1}
 \end{eqnarray} 
The total photoelectron spectra at the end of the laser pulse $t_{\rm F}$, $|b({\bf p},t_{\rm F})|^2$, is then a coherent superposition of both the direct $b_0({\bf p},t_{\rm F})$ and rescattered $b_1({\bf p},t_{\rm F})$ transition amplitudes, i.e.,
 \begin{eqnarray}
|b({\bf p},t_{\rm F})|^2 &=& |b_0({\bf p},t_{\rm F})+b_1({\bf p},t_{\rm F})|^2,\nonumber \\ 
&=& |b_0({\bf p},t_{\rm F})|^2 + |b_1({\bf p},t_{\rm F})|^2 + b_0({\bf p},t_{\rm F}){b_1^*}({\bf p},t_{\rm F}) + c.c.
 \label{Eq:b_T}
\end{eqnarray}

In order to compute the total photoelectron spectra $|b({\bf p},t_{\rm F})|^2$ for the three-center molecular system, we first need to define the ground and the continuum states. After having found them we then could obtain expressions for the bound-free transition dipole matrix elements, ${\bf d}_{j}({\bf v})$, and  the continuum-continuum transition rescattering matrix elements ${\bf g}({\bf v},{\bf v}')$. In the next section, we shall explain how to analytically calculate both the just cited transition matrix elements and the final photoelectron momentum distribution.

All the equations obtained in this subsection are consistent with the atomic and diatomic cases presented in previous publications (see Ref.~\cite{PRANoslen2015, PRANoslen2016}). In fact, all the cases are identical when the internuclear distance goes to zero, ${\textbf R}\to 0$. The verification of this limit for the direct processes is straightforward. Here, the phase factor becomes the well known semi-classical action: $S({\bf p},t,t')$ and the transition amplitude exactly has the same dependency as for an atom, if we replace the atomic matrix elements on it. For the rescattering events, on the other hand, we have to neglect the contribution of the nonlocal and cross terms ($j\neq j'$) in Eq.~(\ref{Eq:b_11}) and follow the same procedure as in the case of the direct processes. In the following sections we obtain the exact dependency of the rescattered matrix elements that also describe the atomic case when ${\textbf R}\to 0$.

\section{Transition matrix amplitudes calculation} 

We aim to employ  two different models to calculate the total photoelectron spectra of molecular systems. 
In the first model we use a nonlocal short-range (SR) potential $V(\textbf{p},\textbf{p}^{\prime})$ to describe the interaction of the electron with the multionic center. Using this potential we are going to compute the bound, $\Psi_{0_{SR}}({\bf p})$, and scattering states, $\Psi_{\textbf{p}_0}(\textbf{p})$, as well as the corresponding dipole $\textbf{d}_{SR}( \textbf{p}_0)$, and rescattering, $\textbf{g}( \textbf{p}_0,\textbf{p}_0')$ transition matrix elements. This model is an extension to three-center molecular systems of the one presented for diatomics~\cite{PRANoslen2016}. 

For the second model we are going to obtain the molecular bound states $\Psi_{0_{LCAO}}({\bf p})$ using a linear combination of atomic orbitals (LCAO). This description appears to be more accurate than the one based on the SR potential. For instance, it is able to model not only $s$-states, as is the case of the SR model, but also $p$-states and more complex orbitals. Additionally it gives a more precise characterization of the bound-free dipole matrix element $\textbf{d}_{LCAO}( \textbf{p}_0)$. In order to maintain the model as analytical as possible, we compute the scattering states and the continuum-continuum transition matrix elements using the nonlocal SR potential.

Considering the model for three-center molecular systems is an extension of the one developed for diatomics, through this sections of the paper we are going to present a brief derivation and we refer the readers to our previous contributions Ref.~\cite{PRANoslen2015, PRANoslen2016, PRANoslen2017} for a more detailed derivation.

\subsection{Bound states and dipole transition matrix element: Nonlocal SR potential}
 
In this section we are going to present the final equations to calculate the bound states and the bound-continuum transition matrix element obtained by using the nonlocal SR potential (for a more detailed description see~\cite{PRANoslen2016, PRANoslen2017}). Let us consider a molecule formed by three fixed nuclei under the SAE. The Hamiltonian describing the molecular system can then be written as:
\begin{equation}
\label{Hm}
H(\textbf{p},\textbf{p}^{\prime}) = \frac{p^2}{2}\delta(\textbf{p}-\textbf{p}^{\prime}) + V(\textbf{p},\textbf{p}^{\prime}).
\end{equation}
The first term on the right-hand side is the kinetic energy of the active electron  and the second one is the interacting nonlocal SR potential defined according to:
\begin{equation} 
V(\textbf{p},\textbf{p}^{\prime}) =  -\gamma' \: \phi (\textbf{p}) \: \phi (\textbf{p}^{\prime}) \:\sum_{j=1}^{3} e^{-\textit{i}\textbf{R}_j \cdot (\textbf{p}-\textbf{p}^{\prime})}.
\label{Eq:Vp}
\end{equation}
This potential describes the interaction between the active electron and each of the nuclei of the molecule, and depends on the their positions $\textbf{R}_{j}$. The function $\phi({\bf p})=\frac{1}{\sqrt{{\bf p}^2 +\Gamma^2}}$ is the same auxiliary function used in previous contributions~\cite{Lewenstein1995, PRANoslen2015, PRANoslen2016,PRANoslen2017}. The parameters $\gamma'$ and $\Gamma $ are constants related with the shape of the ground state (see below for more details). 
 
By using $H(\textbf{p},\textbf{p}^{\prime})$, we write the stationary  Schr\"odinger equation as follows:
\begin{equation}
\label{hstatio}
H(\textbf{p})\Psi(\textbf{p}) = \int{\textit{d}^3 \textbf{p}^{\prime}H(\textbf{p},\textbf{p}^{\prime})\Psi(\textbf{p}^{\prime})} = E_0 \: \Psi(\textbf{p}),
\end{equation}
where $E_0$ denotes the energy of the bound state. Thus, for our three-center system Eq.~(\ref{hstatio}) reads as:
\begin{eqnarray} 
\bigg(\frac{p^2}{2}  + I_p  \bigg) \Psi_{0_{SR}}(\textbf{p}) &= &\: \gamma' \: \phi (\textbf{p})\:\sum_{j=1}^{3}\:e^{- \textit{i} \textbf{R}_j\cdot \textbf{p}} \int{ \textit{d}^3 \textbf{p}^{\prime} \Psi_{0_{SR}}(\textbf{p}^{\prime}) \phi (\textbf{p}^{\prime}) e^{\textit{i}\textbf{R}_j\cdot \textbf{p}^{\prime} } }.
\label{Eq:Sch1}
\end{eqnarray} 
Defining new variables $\check{\varphi} _{j}$ as: 
 \begin{equation}
\check{\varphi} _{j} = \int{ \textit{d}^3 \textbf{p}^{\prime} \Psi_{0_{SR}}(\textbf{p}^{\prime}) \phi (\textbf{p}^{\prime}) e^{\textit{i}\textbf{R}_j \cdot \textbf{p}^{\prime}}} = \int{ \frac{ \textit{d}^3 \textbf{p}^{\prime}  \Psi_{0_{SR}}(\textbf{p}^{\prime}) e^{\textit{i}\textbf{R}_j \cdot \textbf{p}^{\prime} }}{\sqrt{  {p^{\prime}}^2 + \Gamma^2}}},
\label{Eq:gamma1}
\end{equation}
we could analytically obtain the bound states by solving Eq.~(\ref{Eq:Sch1}) in the momentum representation. Explicitly we can thus write:
\begin{equation}
\Psi_{0_{SR}}(\textbf{p}) =  \frac{ \gamma' }{\sqrt{(p^2 + \Gamma^2})(\frac{p^2}{2} + I_p)}\:\sum_{j=1}^{3}\: \check{\varphi}_j e^{- \textit{i}\textbf{R}_j \cdot \textbf{p} } .
\label{Eq:MWF1}
\end{equation}
where $I_p$ denotes the ionization potential that is related to the ground state potential energy by $E_{0} = -I_p$.  The exact values of the variables $\check{\varphi} _{j} $ are determined by solving an eigenvalues problem. Finally, the explicit expression for the bound state is obtained using the normalization condition (see~\cite{PRANoslen2016, PRANoslen2017} for more details).

Once we obtain the bound states, the dipole transition matrix element $\textbf{d}_{SR}( \textbf{p}_0)$ can then be computed using Eq.~(\ref{Eq:DM1}), with $| 0_j \rangle=\Psi_{0_{SR_j}}$, as:

\begin{eqnarray}
\textbf{d}_{SR}( \textbf{p}_0)&=&\sum_{j=1}^3 \Big[ - \textit{i}\nabla_{\textbf{p}}\Psi_{0_{SR_j}}(\textbf{p}) \Bigg\rvert_{{\bf p}_0} +\textbf{R}_j  \Psi_{0_{SR_j}}(\textbf{p}_0) \Big].
\end{eqnarray} 
The explicit expression is obtained in~\cite{PRANoslen2017} and reads as:
\begin{eqnarray}
\textbf{d}_{SR}( \textbf{p}_0)&=&\sum_{j=1}^3 {\bf d}_{SR_j}({\bf p}_0)=  -2i\:\mathcal{M} \mathcal{A} ( \textbf{p}_0)
\Bigg[ \frac{I_2}{I_1-I_3} \: \Big(e^{-i\textbf{R}_1 \cdot  \textbf{p}_0} + e^{-i\textbf{R}_3 \cdot  \textbf{p}_0}\Big) + 1 \Bigg],\nonumber\\
 \label{Eq:d3N-SR}
\end{eqnarray} 
where $\mathcal{M}$ is a normalization constant and $\mathcal{A} ( \textbf{p}_0)$, $I_1$, $I_2$ and $I_3$ are the same as those defined in~\cite{PRANoslen2017}. Similarly, we can write the dipole transition matrix for the two-center system (see Appendix~\ref{App:01} for more details), defined as a sum over each of the atoms placed at $\textbf{R}_j$, with $j=1,2$~\cite{PRANoslen2016} .

\subsection{Scattering waves and the continuum-continuum transition matrix element}
To obtain the scattering states we are going to consider the same Hamiltonian as before, i.e.~the one defined in Eq.~(\ref{Hm}). Let us first consider a scattering wave $\Psi_{{\bf p}_0}({\bf p})$, with asymptotic momentum ${\bf p}_0$, as a coherent superposition of a plane wave and an extra correction:
\begin{eqnarray}
\Psi_{\textbf{p}_0}(\textbf{p})  = \delta(\textbf{p}-\textbf{p}_0) +  \delta\Psi_{{\bf p}_0}(\textbf{p}).
\label{Eq.ScatState}
 \end{eqnarray}
This state has an energy $E={{\bf p}_0^2}/2$. Then, the Schr\"odinger equation in momentum representation reads as:
 \begin{equation}
\bigg(\frac{p^2}{2} - \frac{p_0^2}{2}  \bigg) \delta\Psi_{ \textbf{p}_0}(\textbf{p}) = -V(\textbf{p},\textbf{p}_0)  
- \int{\textit{d}^3 \textbf{p}^{\prime} V(\textbf{p},\textbf{p}^{\prime})\delta\Psi_{\textbf{p}_0}(\textbf{p}^{\prime})}.
\label{Eq:ScSt1}
\end{equation}

This procedure is a natural extension of the one presented in Ref. \cite{PRANoslen2016} for the diatomic system. Moreover, after some algebra and substituting the nonlocal SR potential, the correction then results:
\begin{eqnarray}
 \delta\Psi_{ \textbf{p}_0}(\textbf{p}) = & \frac{2  \gamma' \: \phi (\textbf{p}) \phi (\textbf{p}_0) }{\big(p^2 - p_0^2 \big)}\:\sum_{j=1}^{3} e^{-\textit{i}\textbf{R}_j \cdot (\textbf{p}-\textbf{p}_0)}  + \frac{2 \gamma' \: \phi (\textbf{p})}{\big(p^2 - p_0^2 \big)} \:\sum_{j=1}^{3} \:  \check{\varphi}'_j\:  e^{-i\textbf{R}_{j} \cdot  \textbf{p}},
\end{eqnarray}  
where  the variables $\check{\varphi}'_j$ are defined by,
\begin{eqnarray}
\check{\varphi}'_j &=&    \int{ \textit{d}^3 \textbf{p}^{\prime}  \delta\Psi_{ \textbf{p}_0} (\textbf{p}^{\prime}) \phi (\textbf{p}^{\prime}) e^{\textit{i}\textbf{R}_{j} \cdot \textbf{p}^{\prime} }}.
\end{eqnarray}

The correction $\delta\Psi_{ \textbf{p}_0}(\textbf{p}) $ depends on the position of each atom and can thus be written as:
\begin{eqnarray}
\delta\Psi_{ \textbf{p}_0}(\textbf{p}) = \sum_{j=1}^{3} \: \delta\Psi_{\textbf{R}_j \textbf{p}_0}(\textbf{p}),
\end{eqnarray}  
where each of the individual terms $\delta\Psi_{\textbf{R}_j \textbf{p}_0}(\textbf{p})$ is defined by:
\begin{eqnarray}
\delta\Psi_{\textbf{R}_j \textbf{p}_0}(\textbf{p})= \frac{2  \gamma' \: \phi (\textbf{p}) }{\big(p^2 - p_0^2 +\textit{i}\epsilon \big)} \Big\{ \phi (\textbf{p}_0) \: e^{-\textit{i}\textbf{R}_j \cdot (\textbf{p}-\textbf{p}_0)}  +  \check{\varphi}'_j\:  e^{-i\textbf{R}_{j} \cdot  \textbf{p}} \Big\},
\end{eqnarray}  

As in the case of the bound states we need to obtain the explicit form of $\check{\varphi}'_j$. In the case of diatomics it can be obtained from $\Psi_{2-{\bf p}_0}= \delta(\textbf{p}-\textbf{p}_0)+ \sum_{j=1}^{2}\: \delta\Psi_{\textbf{R}_j \textbf{p}_0}(\textbf{p})$ (see~Appendix~\ref{App:01} and \cite{PRANoslen2016} for more details).

For our three-center molecular system the scattering wavefunction can also be written as a composition of three functions, each centered at $\textbf{R}_1$, $\textbf{R}_2$ and $\textbf{R}_3$. Explicitly, we then have:
\begin{equation}
\Psi_{\textbf{p}_0}(\textbf{p})=  \delta(\textbf{p}-\textbf{p}_0) +\sum_{j=1}^{3}\: \delta \Psi_{\textbf{R}_j \textbf{p}_0}(\textbf{p}),
\label{Eq:RescM}
\end{equation}
where
\begin{equation} 
\delta\Psi_{\textbf{R}_1\textbf{p}_0}(\textbf{p})=  \frac{- \mathcal{D}_1(\textbf{p}_0) \: e^{- i \textbf{R}_1 \cdot (\textbf{p} - \textbf{p}_0) } - \mathcal{D}_3(\textbf{p}_0) \:e^{- i \textbf{R}_1 \cdot \textbf{p} \: + \: i \textbf{R}_2 \cdot  \textbf{p}_0 } -\mathcal{D}_4(\textbf{p}_0) \:e^{- i \textbf{R}_1 \cdot \textbf{p} \: + \: i \textbf{R}_3 \cdot  \textbf{p}_0 } }{ \sqrt{p^2 + \Gamma^2} \:(p_0^2 -p^2 + i \epsilon) },
\label{Eq:dphiMR1}
\end{equation}
\begin{equation} 
\delta\Psi_{\textbf{R}_2\textbf{p}_0}(\textbf{p})=\frac{- \mathcal{D}_2 (\textbf{p}_0) \: e^{-i \textbf{R}_2 \cdot (\textbf{p} - \textbf{p}_0 )} - \mathcal{D}_5 (\textbf{p}_0) \: e^{-i \textbf{R}_2 \cdot \textbf{p} \: + \:i \textbf{R}_1 \cdot\textbf{p}_0 } - \mathcal{D}_5 (\textbf{p}_0) \: e^{-i \textbf{R}_2 \cdot \textbf{p} \: + \: i \textbf{R}_3 \cdot \textbf{p}_0 } }{ \sqrt{p^2 + \Gamma^2} \:(p_0^2 -p^2 +\textit{i}\epsilon) },
\label{Eq:dphiMR2}
\end{equation}
and 
\begin{equation} 
\delta\Psi_{\textbf{R}_3\textbf{p}_0}(\textbf{p})=\frac{- \mathcal{D}_1 (\textbf{p}_0) \: e^{-i \textbf{R}_3 \cdot (\textbf{p} - \textbf{p}_0 )} - \mathcal{D}_3 (\textbf{p}_0) \: e^{-i \textbf{R}_3 \cdot \textbf{p} \: + \: i \textbf{R}_2 \cdot \textbf{p}_0 } - \mathcal{D}_4 (\textbf{p}_0) \: e^{-i \textbf{R}_3 \cdot \textbf{p} \: + \: i \textbf{R}_1 \cdot \textbf{p}_0 } }{ \sqrt{p^2 + \Gamma^2} \:(p_0^2 -p^2 +\textit{i}\epsilon) }.
\label{Eq:dphiMR3}
\end{equation}
$\epsilon$ is another infinitesimal parameter to avoid the divergence at the ``energy shell", $p^2=p^2_0$. This singularity  is avoided due to the finite spread of the involved wavepackets. In the numerical calculations we smooth this singularity, allowing $\epsilon$ to be of the order of 1 (see~\cite{PRANoslen2015,PRANoslen2016} for more details).

The integration ``constants" for the scattering states in Eq.~(\ref{Eq:RescM}) have the following functional form:
\begin{equation}
\mathcal{D}_1(\textbf{p}_0)=\frac{2\gamma'}{\sqrt{p_0^2 + \Gamma^2}} \bigg(  \frac{1+ 2I'_{11} +{I'_{11}}^2 -{I'_{12}}^2 }{ [(1+I'_{11}) - I'_{13}]\times[(1+I'_{11})^2 + (1+I'_{11}) I'_{13} -2 {I'_{12}}^2] }  \bigg); 
\end{equation} 
\begin{equation}
\mathcal{D}_2(\textbf{p}_0)=\frac{2\gamma'}{\sqrt{p_0^2 + \Gamma^2}} \bigg( \frac{1+ I'_{11} + I'_{13}}{1+ 2I'_{11} + {I'_{11}}^2 + I'_{13} + I'_{11}I'_{13} -2{I'_{12}}^2}  \bigg);
\end{equation} 
\begin{equation}
\mathcal{D}_3(\textbf{p}_0)=\frac{2\gamma'}{\sqrt{p_0^2 + \Gamma^2}} \bigg( \frac{-I'_{12} -I'_{12} I'_{11} + I'_{12} I'_{13}}{[(1+I'_{11}) - I'_{13}]\times[(1+I'_{11})^2 + (1+I'_{11}) I'_{13} -2 {I'_{12}}^2]}  \bigg); 
\end{equation} 
\begin{equation}
\mathcal{D}_4(\textbf{p}_0)=\frac{2\gamma'}{\sqrt{p_0^2 + \Gamma^2}} \bigg( \frac{{I'_{12}}^2 - I'_{13} -I'_{11} I'_{13}}{[(1+I'_{11}) - I'_{13}]\times[(1+I'_{11})^2 + (1+I'_{11}) I'_{13} -2 {I'_{12}}^2]}  \bigg);
\end{equation} 
and 
\begin{equation}
\mathcal{D}_5(\textbf{p}_0)=\frac{2\gamma'}{\sqrt{p_0^2 + \Gamma^2}} \bigg( \frac{- I'_{12} }{1+ 2I'_{11} + {I'_{11}}^2 + I'_{13} + I'_{11}I'_{13} -2 {I'_{12}}^2}  \bigg),
\end{equation} 
where
\begin{equation} 
I'_{11} =2\gamma'  \int {\frac{ \textit{d}^3  \textbf{p} }{( p^2 +\Gamma^2)(p_0^2 -p^2 +\textit{i}\epsilon)} }=\frac{-4\gamma' \:\pi^2}{( \Gamma - i\sqrt{|p_0^2 + \textit{i}\:\epsilon|})},
\end{equation}   

\begin{equation}
I'_{12} = 2\gamma' \int {\frac{ \textit{d}^3  \textbf{p}\: e^{\pm \textit{i} (\textbf{R}_1 - \textbf{R}_2) \cdot \textbf{p}}}{( p^2 +\Gamma^2)(p_0^2 -p^2 +\textit{i}\epsilon)} }=\frac{- 4 \pi^2 \:\gamma'} {\frac{R}{2}\:(p_0^2 + \Gamma^2  + i\epsilon)} \bigg[ e^{i\frac{R}{2}\:\sqrt{p_0^2 +i\epsilon}} - e^{-\frac{R}{2}\: \Gamma} \bigg ],
\end{equation} 
and
\begin{eqnarray}
I'_{13} &=& 2 \gamma' \int {\frac{ \textit{d}^3  \textbf{p} \: e^{\pm \textit{i}(\textbf{R}_1 - \textbf{R}_3) \cdot \textbf{p}} }{(p^2 +\Gamma^2)(p_0^2 -p^2 +\textit{i}\epsilon) } },\nonumber\\
&&=\frac{- 4 \pi^2 \:\gamma'} {R\sin(\alpha/2)\:(p_0^2 + \Gamma^2  + i\epsilon)} \bigg[ e^{i R\sin(\alpha/2)\:\sqrt{p_0^2 +i\epsilon}} - e^{- R\sin(\alpha/2)\: \Gamma} \bigg ].
\end{eqnarray} 

Once the rescattering states are completely defined we proceed to the evaluation of the continuum-continuum transition matrix element using Eq.~(\ref{Eq:gm}). We thus have:
\begin{eqnarray}
\textbf{g}(\textbf{p}_1,\textbf{p}_2) &=&  \sum_{j=1}^{3} \Bigg[   \textit{i}\nabla_{\textbf{p}} \delta\Psi_ 
{\textbf{R}_j \textbf{p}_2}(\textbf{p}) \rvert_{{\bf p}_1}   -\textbf{R}_j \delta\Psi_{\textbf{R}_j \textbf{p}_2}(\textbf{p}) \Bigg] \Bigg\rvert_{{\bf p}_1}\nonumber\\
&&+\sum_{j=1}^{3} \Bigg[   \textit{i}\nabla_{\textbf{p}} \delta\Psi_{\textbf{R}_j \textbf{p}_1}(\textbf{p})
-\textbf{R}_j \delta\Psi_{\textbf{R}_j \textbf{p}_1}(\textbf{p}) \Bigg]^*\Bigg \rvert_{{\bf p}_2}.
\end{eqnarray}

For our three-center molecular system we get an independent transition matrix element for each of the possible rescattering scenarios, i.e.~ 
 \begin{eqnarray}
\textbf{g}_{11}(\textbf{p}_1, \textbf{p}_2) &=&  \mathcal{Q}_{1}(\textbf{p}_1, \textbf{p}_2)\: e^{-i \textbf{R}_1 \cdot (\textbf{p}_1 - \textbf{p}_2) } \nonumber\\
 \textbf{g}_{22}(\textbf{p}_1, \textbf{p}_2)&=& \mathcal{Q}_{2} (\textbf{p}_1, \textbf{p}_2) \:e^{- i \textbf{R}_2 \cdot (\textbf{p}_1-\textbf{p}_2) } \nonumber\\
\textbf{g}_{33}(\textbf{p}_1, \textbf{p}_2)&=&\mathcal{Q}_{1}(\textbf{p}_1, \textbf{p}_2)\: e^{-i \textbf{R}_3 \cdot (\textbf{p}_1 - \textbf{p}_2) } \nonumber\\
\textbf{g}_{12}(\textbf{p}_1, \textbf{p}_2) &=&  \mathcal{Q}_{5}(\textbf{p}_1, \textbf{p}_2) \:e^{- i \textbf{R}_2 \cdot\textbf{p}_1 +i \textbf{R}_1 \cdot\textbf{p}_2 }\nonumber\\
\textbf{g}_{21}(\textbf{p}_1, \textbf{p}_2) &=&\mathcal{Q}_{3}(\textbf{p}_1, \textbf{p}_2) \: e^{-i \textbf{R}_1 \cdot \textbf{p}_1 + i \textbf{R}_2 \cdot  \textbf{p}_2 } \nonumber\\
\textbf{g}_{13}(\textbf{p}_1, \textbf{p}_2) &=&  \mathcal{Q}_{4}(\textbf{p}_1, \textbf{p}_2) \:e^{- i \textbf{R}_3 \cdot\textbf{p}_1 + i \textbf{R}_1 \cdot\textbf{p}_2 },\nonumber\\
\textbf{g}_{31}(\textbf{p}_1, \textbf{p}_2) &=& \mathcal{Q}_{4} (\textbf{p}_1, \textbf{p}_2) \:e^{- i \textbf{R}_1 \cdot\textbf{p}_1 +i \textbf{R}_3 \cdot\textbf{p}_2 }\nonumber\\
\textbf{g}_{23}(\textbf{p}_1, \textbf{p}_2) &=&\mathcal{Q}_{3} (\textbf{p}_1, \textbf{p}_2) \:e^{- i \textbf{R}_3 \cdot\textbf{p}_1 +i \textbf{R}_2 \cdot\textbf{p}_2 },\nonumber\\
\textbf{g}_{32}(\textbf{p}_1, \textbf{p}_2) &=& \mathcal{Q}_{5} (\textbf{p}_1, \textbf{p}_2) \:e^{- i \textbf{R}_2 \cdot\textbf{p}_1 + i \textbf{R}_3 \cdot\textbf{p}_2 },
\label{Eq:g3N}
\end{eqnarray}
where the $\mathcal{Q}_{j}$, with $j=1-5$, are defined by:
\begin{eqnarray}
\mathcal{Q}_{1} (\textbf{p}_1, \textbf{p}_2) &=& -i \Big[ \mathcal{D}_1 (\textbf{p}_2) \mathcal{C}_1(\textbf{p}_1, \textbf{p}_2) - \mathcal{D}^*_1(\textbf{p}_1) \mathcal{C}_2(\textbf{p}_1, \textbf{p}_2) \Big],\\
\mathcal{Q}_{2} (\textbf{p}_1, \textbf{p}_2) &=& -i \Big[ \mathcal{D}_2 (\textbf{p}_2) \mathcal{C}_1(\textbf{p}_1, \textbf{p}_2) -\mathcal{D}^*_2(\textbf{p}_1) \mathcal{C}_2(\textbf{p}_1, \textbf{p}_2) \Big],\\
\mathcal{Q}_{3} (\textbf{p}_1, \textbf{p}_2) &=& -i \Big[ \mathcal{D}_3 (\textbf{p}_2) \mathcal{C}_1(\textbf{p}_1, \textbf{p}_2) -\mathcal{D}^*_5(\textbf{p}_1) \mathcal{C}_2(\textbf{p}_1, \textbf{p}_2) \Big],\\\mathcal{Q}_{4} (\textbf{p}_1, \textbf{p}_2) &=& -i \Big[ \mathcal{D}_4 (\textbf{p}_2) \mathcal{C}_1(\textbf{p}_1, \textbf{p}_2) -\mathcal{D}^*_4(\textbf{p}_1) \mathcal{C}_2(\textbf{p}_1, \textbf{p}_2) \Big],\\
\mathcal{Q}_{5} (\textbf{p}_1, \textbf{p}_2) &=& -i \Big[ \mathcal{D}_5 (\textbf{p}_2) \mathcal{C}_1(\textbf{p}_1, \textbf{p}_2) -\mathcal{D}^*_3(\textbf{p}_1) \mathcal{C}_2(\textbf{p}_1, \textbf{p}_2) \Big],
\end{eqnarray}
and
\begin{equation}
\mathcal{C}_1 (\textbf{p}_1, \textbf{p}_2) =  \Bigg[  \frac{ \textbf{p}_1(3p^2_1 -p^2_2 + 2\Gamma^2 ) }{  (p_1^2 +\Gamma^2) ^{\frac{3}{2}}(p_2^2 - p_1^2 +i\epsilon)^2 }\Bigg],
\mathcal{C}_2(\textbf{p}_1, \textbf{p}_2) =  \Bigg[  \frac{\textbf{p}_2(3 p^2_2- p^2_1 + 2\Gamma^2 )}{  (p_2^2 +\Gamma^2)^{\frac{3}{2}} (p_1^2 - p_2^2 -\textit{i}\epsilon)^2 } \Bigg].
\end{equation}

\subsection{Bound states and dipole transition matrix element: Molecular Orbital as a LCAO}

In this section we are going to calculate the molecular bound states as a linear combination of atomic orbitals (LCAO) of Gaussian-like functions. Our formulation takes full advantage of the GAMESS package~\cite{GAMESS1,GAMESS2}. For simplicity we use a STO-3G basis set, but note that our approach is quite general and other basis sets could be employed. 

Let us define the bound state of the molecular system as:
\begin{equation}
\Psi_{0_{LCAO}}({\bf p}) =\sum_{j=1}^{3} \sum_{i=1}^{5} G_{j(i)} \Phi_{j(i)}(\textbf{p}),
\end{equation}
where the index $j$ represents the number of the atoms in the molecule. Furthermore, the index $i$ accounts for the different atomic orbitals (throughout our contribution we model molecular systems using only $1s$, $2s$ and $2p$ states, but states with other quantum numbers could be implemented), i.e.~
\begin{eqnarray}
&&i=1\rightarrow 1s,\nonumber\\
&&i=2\rightarrow 2s,\nonumber\\
&&i=3\rightarrow 2p_x,\nonumber\\
&&i=4\rightarrow 2p_y,\nonumber\\
&&i=5\rightarrow 2p_z.
\end{eqnarray}
Furthermore, $G_{j(i)}$ is a constant defining the {\it weight} of each atom orbital. In our case we consider the Highest Occupied Molecular Orbital (HOMO) and the particular values are obtained using GAMESS. Finally, the functions $\Phi_{j (i)}(\textbf{p})$ define the atomic orbitals. For instance, an atomic orbital based on $s$ states can be written as:

\begin{eqnarray}
\Phi_{j(s)}(\textbf{p}) &=& e^{-i \textbf{R}_j\cdot \textbf{p}} \:\frac{1}{2^{3/2}} \sum_{n=1}^{3} \: \frac{C_{n; j (s)}}{\zeta^{3/2}_{n; j (s)}} \: e^{\frac{ - \textbf{p}^2}{ 4 \zeta_{n; j (s)}}},
\end{eqnarray}
meanwhile that for $2p-3p$ states are:
\begin{equation}
\Phi_{j (2p_r)}(\textbf{p}) =-i \: p_r \: e^{-i \textbf{R}_j\cdot \textbf{p}} \:\frac{1}{2^{5/2}} \sum_{n=1}^{3} \: \frac{C_{n; j(p_r)}}{\zeta^{5/2}_{n; j (p_r)}} \: e^{\frac{ - \textbf{p}^2}{ 4 \zeta_{n; j(p_r)}}}.
\end{equation}
Here the index $r$ can take the values $x$, $y$, or $z$.
The coefficients $C_{n; j(s,p_r)}$ and $\zeta_{n; j (s,p_r)}$ are obtained using, for example, a Roothaan-Hartree-Fock optimization scheme (see~\cite{GAMESS1,GAMESS2} for more details).

The dipole transition matrix element within this model, that describes the transition of the electron from the bound to the continuum state, then reads as:
\begin{equation}
\textbf{d}_{LCAO}( \textbf{v}) = - \langle \textbf{v} |(\textbf{r} - \textbf{R}_j)| \Psi_{0_{LCAO}} \rangle = -\langle \textbf{v} |\textbf{r} |\Psi_{0_{LCAO}} \rangle + \textbf{R}_j \langle \textbf{v} |\Psi_{0_{LCAO}} \rangle.
\end{equation}
More explicitly,
\begin{eqnarray}
\textbf{d}_{LCAO}( \textbf{p}_0) =-\sum_{j=1}^{3} \sum_{i=1}^{5} G_{j(i)}  \Big\{ \textit{i}\nabla_{\textbf{p}}\:\Phi_{j(i)}(\textbf{p}) \Big\rvert_{{\bf p}_0} -  \textbf{R}_j\:\Phi_{j(i)}(\textbf{p}_0) \Big\},
\end{eqnarray}
where depending on the states character the gradient results
\begin{equation}
\textit{i}\nabla_{\textbf{p}}\:\Phi_{j(1s)}(\textbf{p}) \Big\rvert_{{\bf p}_0}=\textbf{R}_j\:\Phi_{j(1s)}(\textbf{p}_0) -\frac{i\:{\bf p}_0}{2\:\zeta_{n; j (s)}}\:\Phi_{j(1s)}(\textbf{p}_0),
 \end{equation}
for the $s$ states and
 \begin{equation}
\textit{i}\nabla_{\textbf{p}}\:\Phi_{j (2p_r)}(\textbf{p})\Big\rvert_{{\bf p}_0} =\textbf{R}_j\:\Phi_{j (2p_r)}(\textbf{p}_0)
-\frac{i \:{\bf p}_0\Phi_{j (2p_r)}(\textbf{p}_0)}{2\:\zeta_{n; j (2p_r)}} +  \delta \Phi_{j (2p_r)}({\bf p}_0) \: \hat{\textbf{r}},
\end{equation} 
where
\begin{equation}
\delta \Phi_{j (2p_r)}({\bf p}_0)= e^{-i \textbf{R}_j\cdot \textbf{p}_0}\: \frac{1}{2^{5/2}} \sum_{n=1}^{3} \frac{C_{n; j (2p_z)}} {\zeta^{5/2}_{n; j (2p_z)}}  \:\: e^{\frac{ - \textbf{p}_0^2}{ 4 \zeta_{n; j (2p_z)}}},
\end{equation}
for the $p$ states.

Using the above equations we are able to obtain analytical expressions for the molecular dipole transition matrix elements.  As was noted in this Section, we introduced the formulation particularized for three-center molecular systems. Nevertheless, and for completeness, we present in the next subsections expressions for two prototypical two-center molecules, O$_2$ and CO, as well.  Additionally, our three-center examples will be based on the CO$_2$ and CS$_2$ molecules.
 
\subsubsection{O$_2$}

The bound state (HOMO) for the O$_2$ molecule oriented on the  $y$-axis, written as a LCAO, reads as:
\begin{eqnarray}
\Psi_{0-\textrm{O$_2$}}(\textbf{p})&=& G_{1 (2p_z)} \Phi_{1 (2p_z)}(\textbf{p}) + G_{2 (2p_z)} \Phi_{2 (2p_z)}(\textbf{p}).
\end{eqnarray}
Furthermore,  the dipole transition matrix element can be computed from
\begin{eqnarray}
\textbf{d}_{\textrm{O$_2$}}( \textbf{p}_0) &=-&\sum_{j=1}^{2} G_{j(2p_z)} \:  \Big\{ \textit{i}\nabla_{\textbf{p}}\:\Phi_{j(2p_z)}(\textbf{p}) \Big\rvert_{{\bf p}_0} -  \textbf{R}_j\:\Phi_{j(2p_z)}(\textbf{p}_0)  \Big\},
\end{eqnarray}
where explicitly we then have,
 \begin{eqnarray}
\textbf{d}_{\textrm{O$_2$}}( \textbf{p}_0) &=& \:G_{1(2p_z)} \: \Big\{ \textit{i} \:\textbf{p}_0 \frac{\Phi_{1(2p_z)}(\textbf{p}_0) }{2\: \zeta_{n; 1(2p_z)} }  -   \delta \Phi_{1 (2p_z)}(\textbf{p}_0) \: \hat{\textbf{k}}\Big\} \nonumber\\
&& +\:G_{2(2p_z)} \:  \Big\{ \textit{i} \:\textbf{p}_0 \frac{\Phi_{2(2p_z)}(\textbf{p}_0) }{2\: \zeta_{n; 2(2p_z)} }  -   \delta \Phi_{2 (2p_z)}(\textbf{p}_0) \: \hat{\textbf{k}}\Big\}.
\label{Eq:dO2}
\end{eqnarray}
The parameters $G_{(1,2)(2p_z)}$ and $\zeta_{n;1/2(2p_z)}$ are obtained setting the molecule in its equilibrium position via an optimization procedure using GAMESS~\cite{GAMESS1,GAMESS2}.

\subsubsection{CO}

For the case of CO the bound state is a composition of $1s$, $2s$ and $2p$ states. It can then be written as:
\begin{eqnarray}
\Psi_{0-\textrm{CO}}(\textbf{p})&=&\sum_{j=1}^{2} \Big[G_{j (1s)} \Phi_{j (1s)}(\textbf{p})+ G_{j (2s)} \Phi_{j (2s)}(\textbf{p}) + G_{j (2p_z)} \Phi_{j (2p_z)}(\textbf{p})\Big].
\end{eqnarray}

The dipole transition matrix element reads as:
\begin{eqnarray}
\textbf{d}_{\textrm{CO}}( \textbf{p}_0) &=&-\sum_{j=1}^{2} \Bigg[G_{j(1s)} \:  \Big\{ \textit{i}\nabla_{\textbf{p}}\:\Phi_{j(1s)}(\textbf{p}) \Big\rvert_{{\bf p}_0} -  \textbf{R}_j\:\Phi_{j(1s)}(\textbf{p}_0)  \Big\} +G_{j(2s)} \:  \Big\{ \textit{i}\nabla_{\textbf{p}}\:\Phi_{j(2s)}(\textbf{p}) \Big\rvert_{{\bf p}_0}\nonumber\\
&& -  \textbf{R}_j\:\Phi_{j(2s)}(\textbf{p}_0)  \Big\} 
+G_{j(2p_z)} \:  \Big\{ \textit{i}\nabla_{\textbf{p}}\:\Phi_{j(2p_z)}(\textbf{p}) \Big\rvert_{{\bf p}_0} -  \textbf{R}_j\:\Phi_{j(2p_z)}(\textbf{p}_0)  \Big\} \Bigg],
\end{eqnarray}
where, more explicitly, we then have:
 \begin{eqnarray}
\textbf{d}_{\textrm{CO}}( \textbf{p}_0) &=&\sum_{j=1}^{2}\Bigg[G_{j(1s)} \:  \Big\{\textit{i} \:\textbf{p}_0  \frac{\Phi_{j(1s)}(\textbf{p}_0) }{2\: \zeta_{n; j(1s)} }   \Big\} +G_{j(2s)} \:  \Big\{ \textit{i} \:\textbf{p}_0\frac{\Phi_{j(2s)}(\textbf{p}_0) }{2\: \zeta_{n; j(2s)} }    \Big\} \nonumber\\
&& +\:G_{j(2p_z)} \:  \Big\{ \textit{i} \:\textbf{p}_0 \frac{ \Phi_{j(2p_z)}(\textbf{p}_0)}{2\: \zeta_{n; j(2p_z)} }  -  \delta \Phi_{j (2p_z)}(\textbf{p}_0) \: \hat{\textbf{k}}\Big\} \Bigg].
\label{Eq:dCO}
\end{eqnarray}

\subsubsection{CO$_2$}

For the case of CO$_2$ the bound state is:
\begin{eqnarray}
\Psi_{0-\textrm{CO$_2$}}(\textbf{p})&=&\sum_{j=1}^{3} \Bigg[G_{j (2p_x)} \Phi_{j (2p_x)}(\textbf{p}) +G_{j (2p_z)} \Phi_{j (2p_z)}(\textbf{p})\Bigg].
\end{eqnarray}

\begin{figure}[htb]
\includegraphics[width=0.5\textwidth]{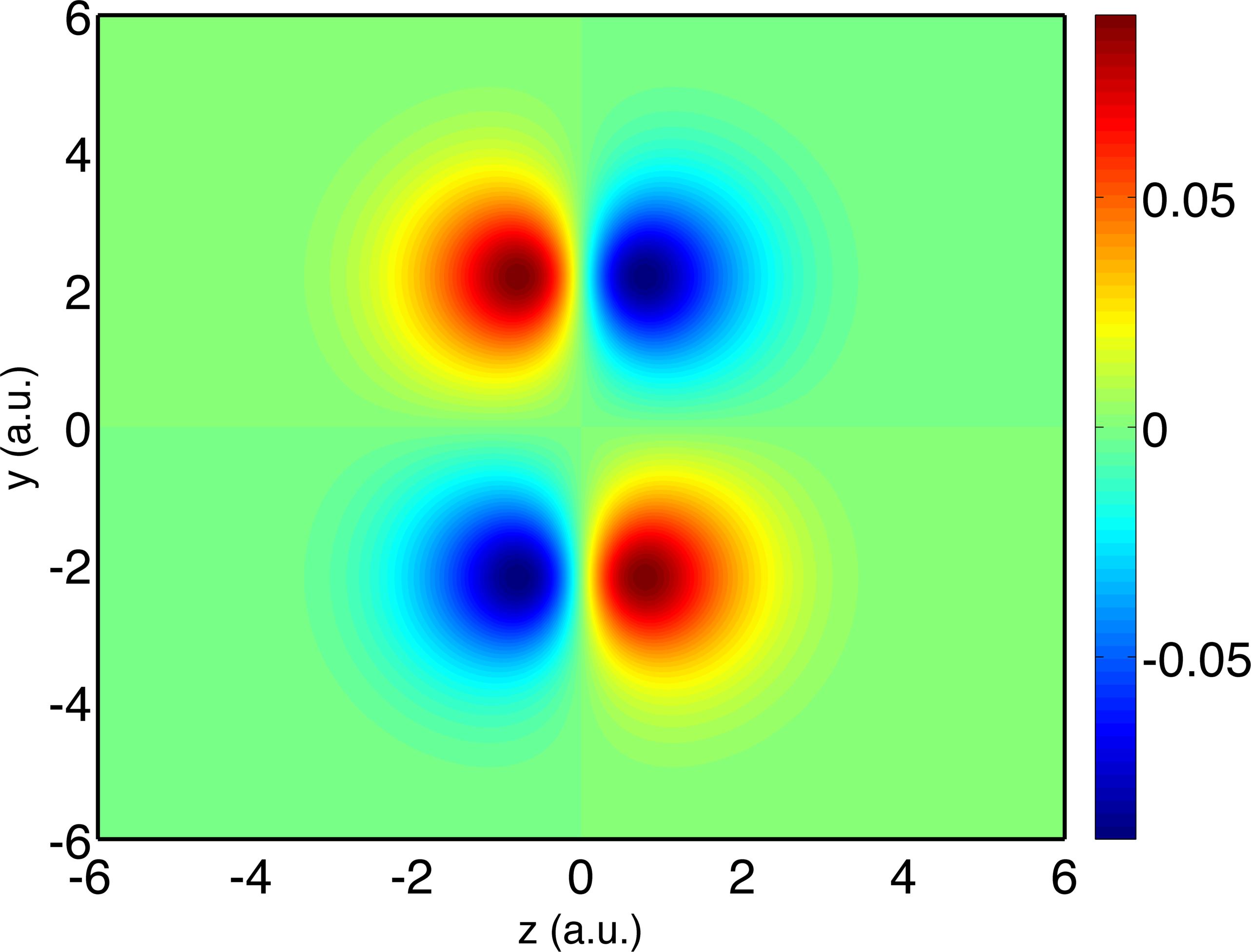}
		    \caption{(color online) CO$_2$ HOMO presented in the $z-y$ plane, calculated using the LCAO method (see the text for details).}
		  	\label{Fig:CO2_MO}
		\end{figure}

A plot of the HOMO for this molecule is shown in Fig.~\ref{Fig:CO2_MO}, where we consider the molecule in equilibrium - each of the C$-$O bonds length is set to $2.2$ a.u. (1.164 \AA), and oriented parallel to the laser field (linearly polarized along the $z$-axis). 

The dipole transition matrix element can be explicitly written as:
\begin{eqnarray}
\textbf{d}_{\textrm{CO$_2$}}( \textbf{p}_0) &=& \sum_{j=1}^{3} \Bigg[G_{j(2p_x)} \:  \Big\{ \textit{i} \:\textbf{p}_0 \frac{\Phi_{j(2p_x)}(\textbf{p}_0) }{2\: \zeta_{n; j(2p_x)} }  -   \delta \Phi_{j (2p_x)}(\textbf{p}_0) \: \hat{\textbf{i}}\Big\}\nonumber\\
&& +\:G_{j(2p_z)} \:  \Big\{ \textit{i} \:\textbf{p}_0 \frac{\Phi_{j(2p_z)}(\textbf{p}_0) }{2\: \zeta_{n ;j(2p_z)} }  -   \delta \Phi_{j (2p_z)}(\textbf{p}_0) \: \hat{\textbf{k}}\Big\} \Bigg].
\label{Eq:dCO2}
\end{eqnarray}

\subsubsection{CS$_2$}

The CS$_2$ the bound state within the LCAO approach reads as:
\begin{eqnarray}
\Psi_{0-\textrm{CS$_2$}}(\textbf{p})&=& \sum_{j=1}^{3} \Bigg[ G_{j (2p_x)} \Phi_{j (2p_x)}(\textbf{p}) +G_{j (2p_z)} \Phi_{j (2p_z)}(\textbf{p})\nonumber\\
&&+ G_{j (3p_x)} \Phi_{j (3p_x)}(\textbf{p}) +G_{j (3p_z)} \Phi_{j (3p_z)}(\textbf{p})\Bigg].
\end{eqnarray}

\begin{figure}[htb]
 \includegraphics[width=0.5\textwidth]{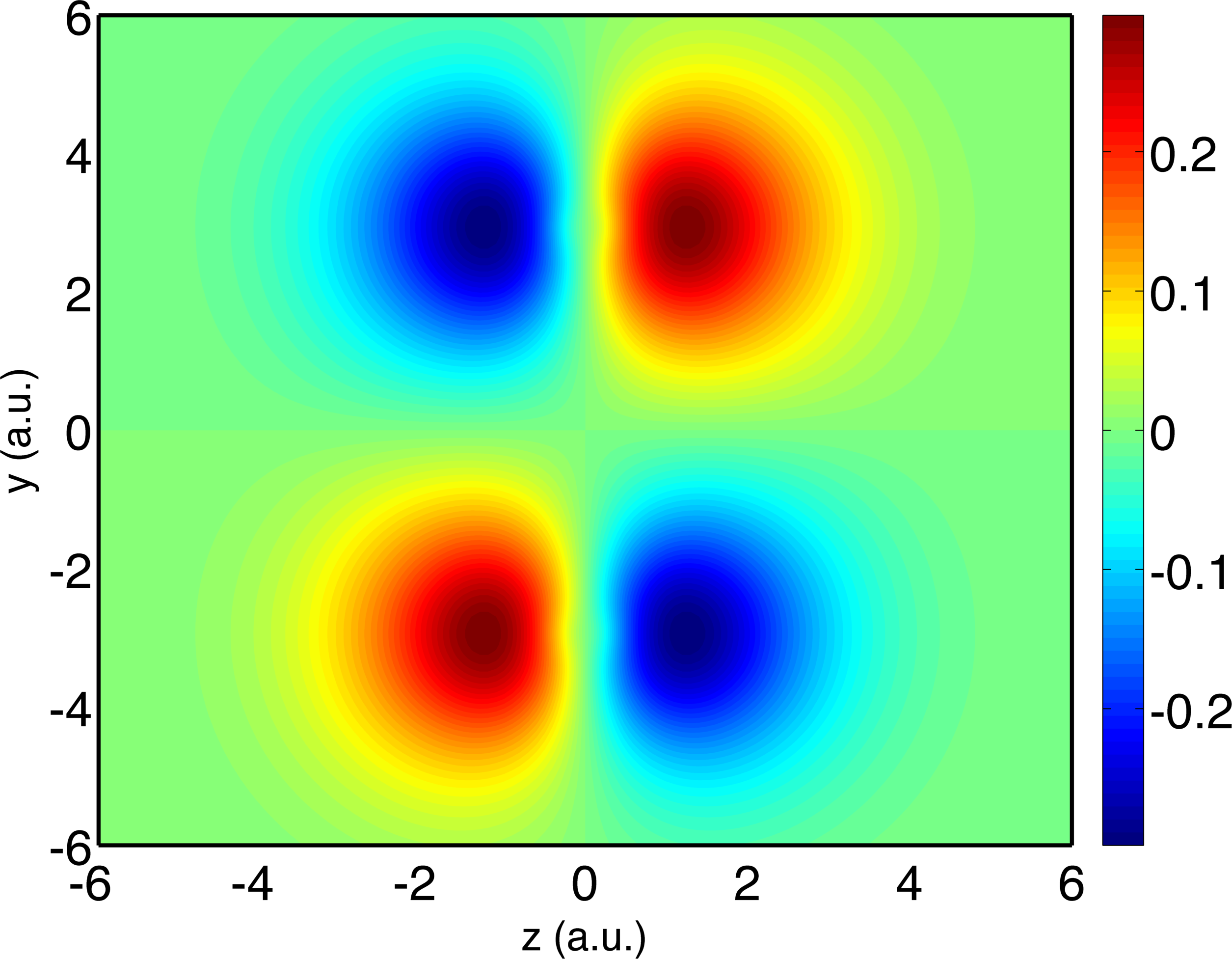}
		    \caption{(color online) CS$_2$ HOMO presented in the $z-y$ plane, calculated using the LCAO method (see the text for details).}
		  	\label{Fig:CS2_MO}
		\end{figure}

As in the previous case, we consider the CS$_2$ molecule in equilibrium - each of the C$-$S bonds length is set to $2.92$ a.u. (1.545 \AA), and oriented parallel to the laser field (polarized along the $z$-axis). A plot of the CS$_2$ HOMO is depicted in Fig.~\ref{Fig:CS2_MO}.

Finally, the dipole transition matrix element for the CS$_2$ molecule reads as:
 \begin{eqnarray}
\textbf{d}_{\textrm{CS$_2$}}( \textbf{p}_0) &=& \sum_{j=1}^{3} \Bigg[G_{j(2p_x)} \:  \Big\{ \textit{i} \:\textbf{p}_0 \frac{\Phi_{j(2p_x)}(\textbf{p}_0) }{2\: \zeta_{j;n(2p_x)} }  -   \delta \Phi_{j (2p_x)}(\textbf{p}_0) \: \hat{\textbf{i}}\Big\} +\:G_{j(2p_z)} \:  \Big\{ \textit{i} \:\textbf{p}_0 \frac{\Phi_{j(2p_z)}(\textbf{p}_0) }{2\: \zeta_{j; n(2p_z)} }\nonumber\\
&&  -   \delta \Phi_{j (2p_z)}(\textbf{p}_0) \: \hat{\textbf{k}}\Big\}
 +\:G_{j(3p_x)} \: \Big\{ \textit{i} \:\textbf{p}_0 \frac{\Phi_{j(3p_x)}(\textbf{p}_0) }{2\: \zeta_{j; n(3p_x)} }  -   \delta \Phi_{j (3p_x)}(\textbf{p}_0) \: \hat{\textbf{i}}\Big\} \nonumber\\
 &&+\:G_{j(3p_z)} \: \Big\{ \textit{i} \:\textbf{p}_0 \frac{\Phi_{j(3p_z)}(\textbf{p}_0) }{2\: \zeta_{j;n(3p_z)} }  -   \delta \Phi_{j (3p_z)}(\textbf{p}_0) \: \hat{\textbf{k}}\Big\} \Bigg].
\end{eqnarray}

After obtaining both the dipole and the continuum-continuum transition matrix elements it is then possible to compute the Eqs.~(\ref{Eq:b_10}) and (\ref{Eq:b_1F}) to obtain the direct, the rescattering and the total photoelectron transition amplitudes. We stress out that our model only involves, in the worst case scenario, the numerical calculation of a 2D integral, being the rest of the expressions written in terms of fully analytical functions. 


\section{Results and Discussion}

Along this section, we compute the ATI spectra generated from different molecular systems using two different approaches, namely: (i) \textit{Model A}: we employ a nonlocal SR potential to calculate both the bound and scattering states, as well the bound-free and continuum-continuum matrix element and (ii) \textit{Model B}: the initial ground state is modeled as an LCAO and a nonlocal SR potential is used to compute the scattering states and the continuum-continuum matrix element. The bound-free transition matrix element is then obtained employing an LCAO for the bound part and a nonlocal SR potential scattering state for the continuum part.

We compare the ATI spectra computed using models (i) and (ii) for four different molecular systems in order to establish similarities and differences. We present calculations of ATI for two different diatomic and two triatomic molecular systems. Furthermore, the splitting of the contributions to the photoelectron spectra helps us to distinguish which of the direct and rescattering scenarios is relevant in the different energy/momentum regions. 

In the simulations we use an ultrashort laser pulse with a central frequency $\omega_0=0.057$~a.u. (wavelength $\lambda=800$ nm), a
$\sin^2$ envelope shape and $N_c=4$ total cycles (this
corresponds to a full-width at half-maximum FWHM~$=5.2$ fs). The
CEP is set to $\phi_{0}=0$~rad and the time step to $\delta t=
0.02$~a.u. The numerical integration time window is then
$t$:~$[0,t_{\rm F}]$, where $t_{\rm F}= N_cT_0\approx 11$ fs and
$T_0=2\pi/\omega_0$ denote the final ``detection" time and the
cycle period of the laser field, respectively.

\subsection{Results on diatomic molecules: O$_2$ and CO}

In this Section, we apply our analytical model using the equations presented in the Appendix~\ref{App:01} to calculate the photoelectron spectra of two prototypical diatomic systems: O$_2$ and CO. The numerical integration of the photoelectron spectra by means of  Eqs.~(\ref{Eq:b_10}) and (\ref{Eq:b_1F}) has been performed via a rectangular rule with particular emphasis on the convergence of the results. As the final momentum
distribution, Eq.~(\ref{Eq:b_T}), is ``locally" independent of the
momentum ${\bf p}$, i.e.~$|b({\bf p},t)|^2$ can be computed
concurrently for a given set of ${\bf p}$ values. We have
optimized the calculation of the whole transition amplitude,
$|b({\bf p},t)|^2$, by using the OpenMP parallel
package~\cite{OpenMPBook}. The final momentum photoelectron distribution, $|b({\bf p},t)|^2$, is computed in a 1D-momentum line along $p_z$ and a 2D-momentum $p_z-p_y$ plane.

\subsubsection{O$_2$ molecule}

The computation of the photoelectron spectra was performed by using, Eqs.~(\ref{Eq:b_T}), (\ref{Eq:d2N-SR}) and (\ref{Eq:g2N}) for the case of \textit{Model A}. Here we set $\Gamma=1$ and $\gamma=0.08$~a.u. in our nonlocal SR potential in order to match the dioxygen ionization potential obtained with GAMESS $I_p=0.334$~a.u for the singlet state. Next, in the case of the calculation using the \textit{Model B}, we use Eqs.~(\ref{Eq:b_T}), (\ref{Eq:dO2}) and (\ref{Eq:g2N}).  

\begin{figure}[htb]
            \subfigure[]{ \includegraphics [width=0.45\textwidth] {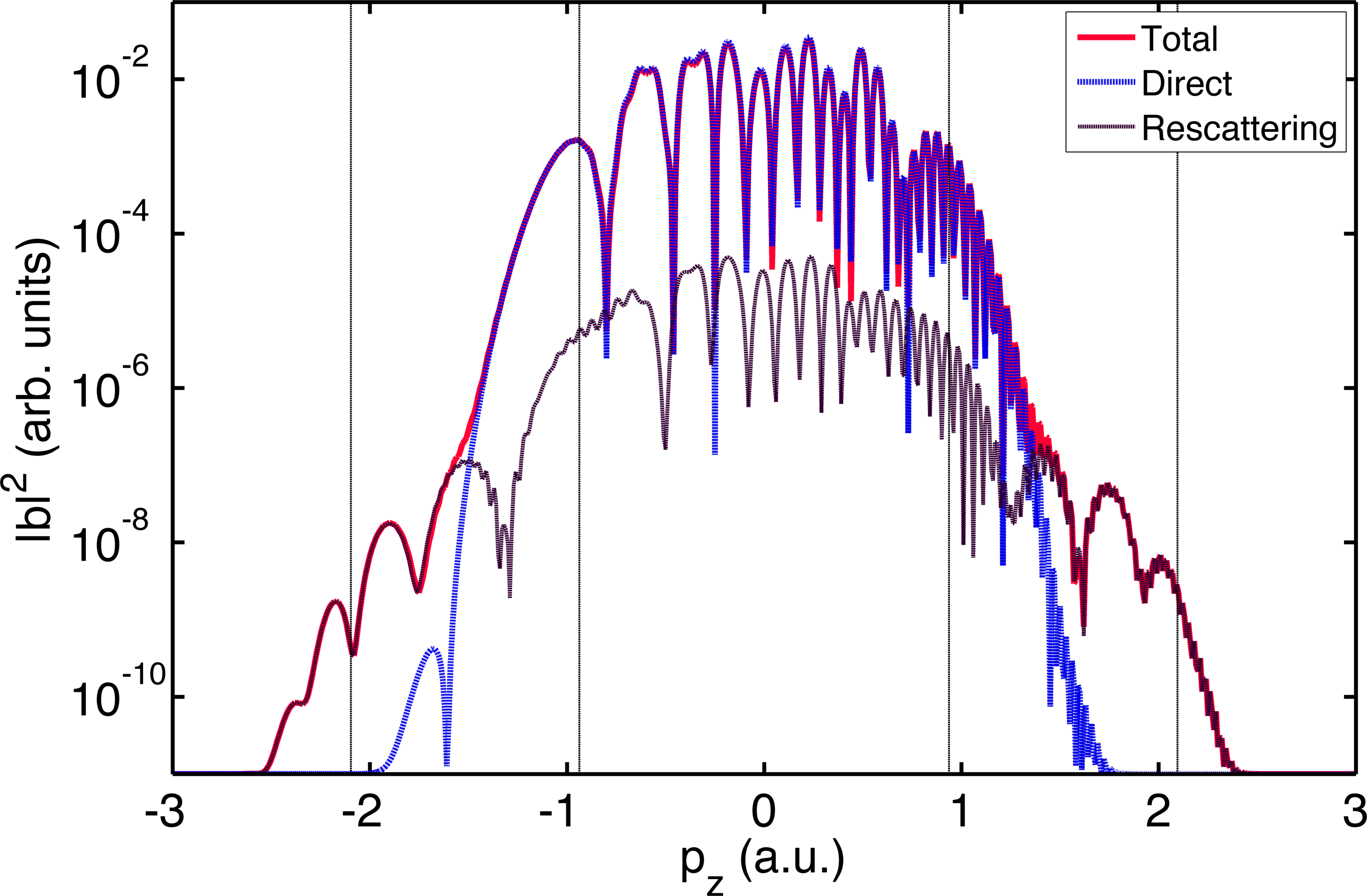}
          }
             \subfigure[]{ \includegraphics [width=0.39\textwidth] {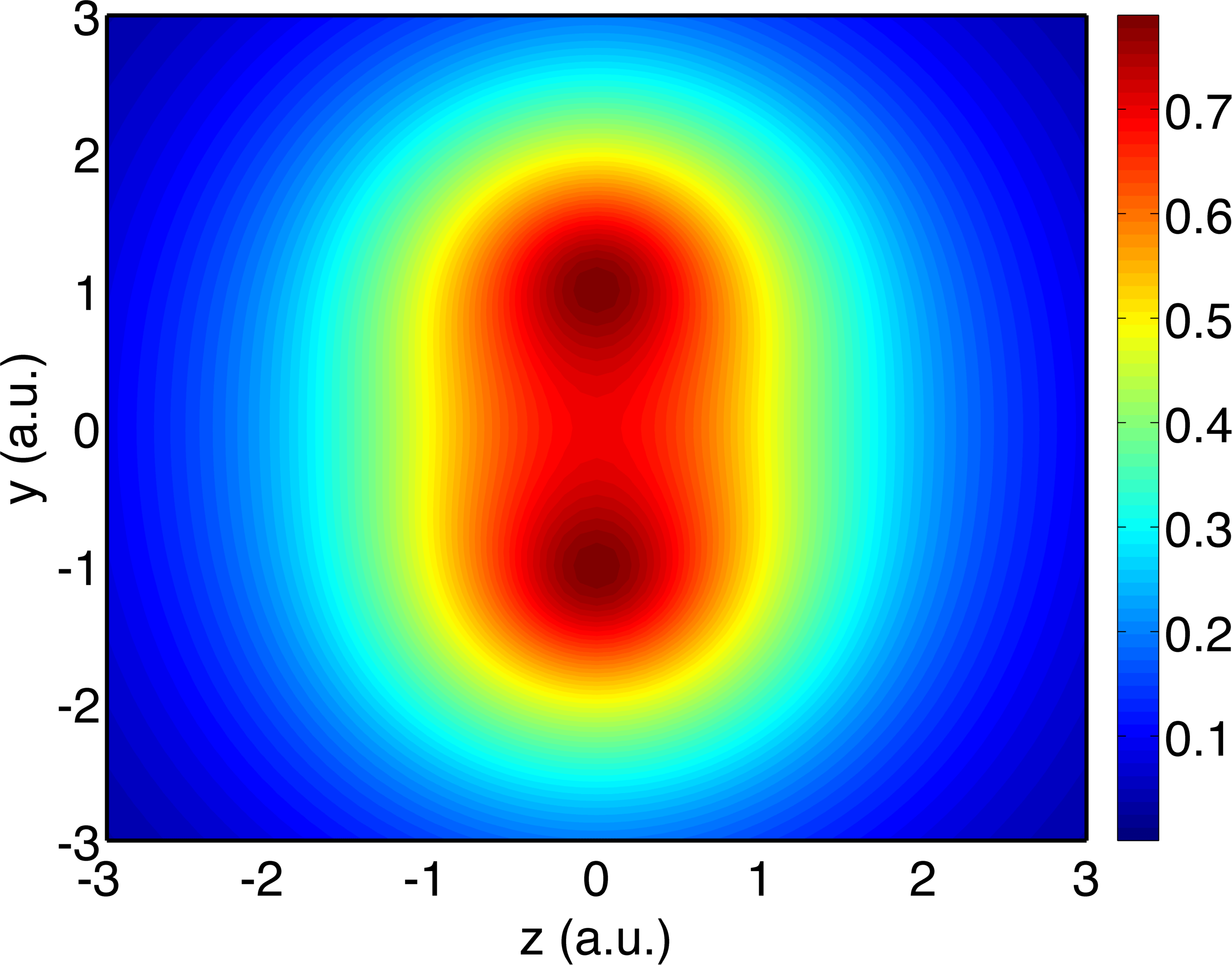}
           }
             \subfigure[]{ \includegraphics [width=0.45\textwidth] {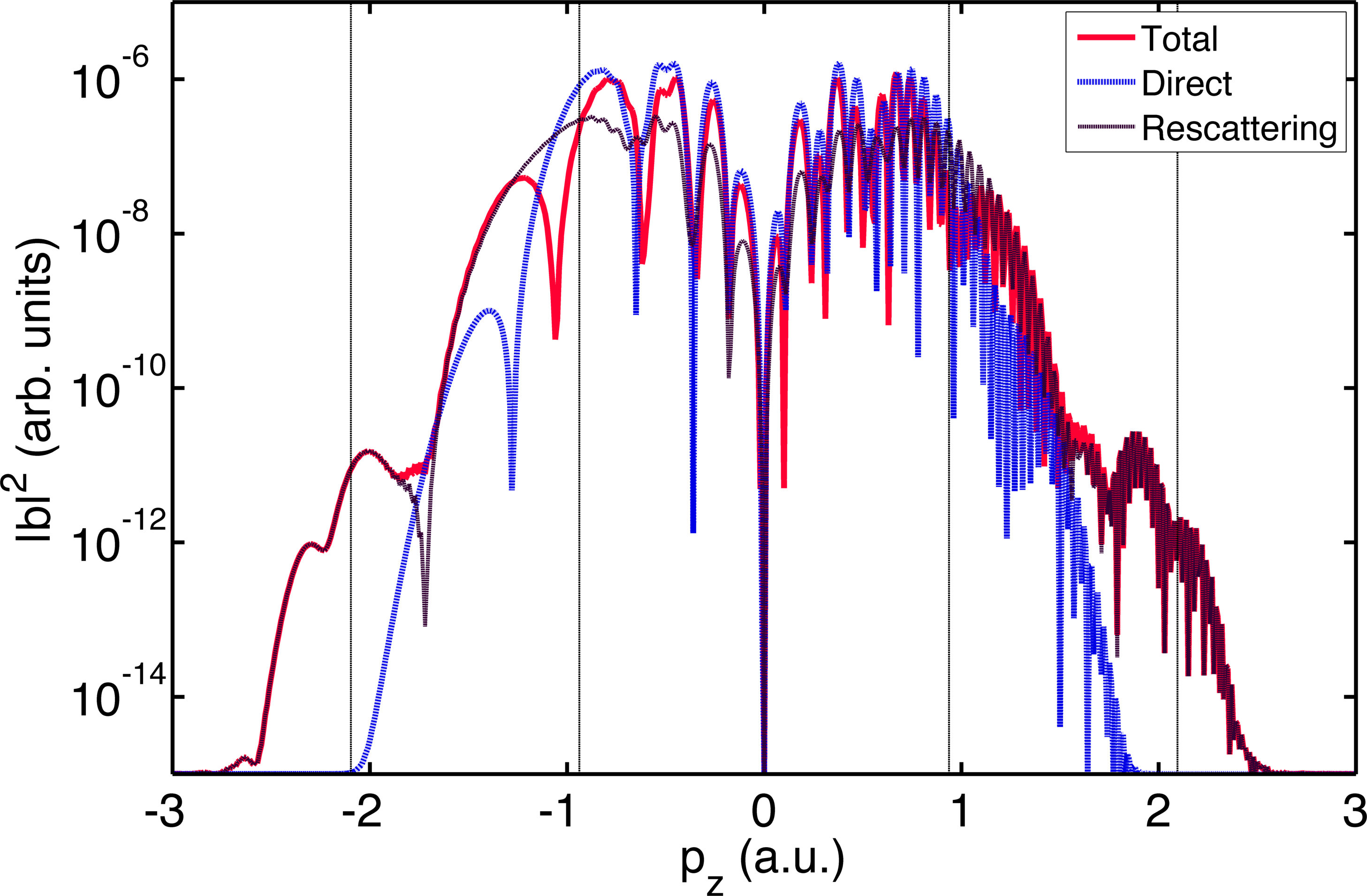}
            }
                         \subfigure[]{\includegraphics[width=0.39\textwidth]{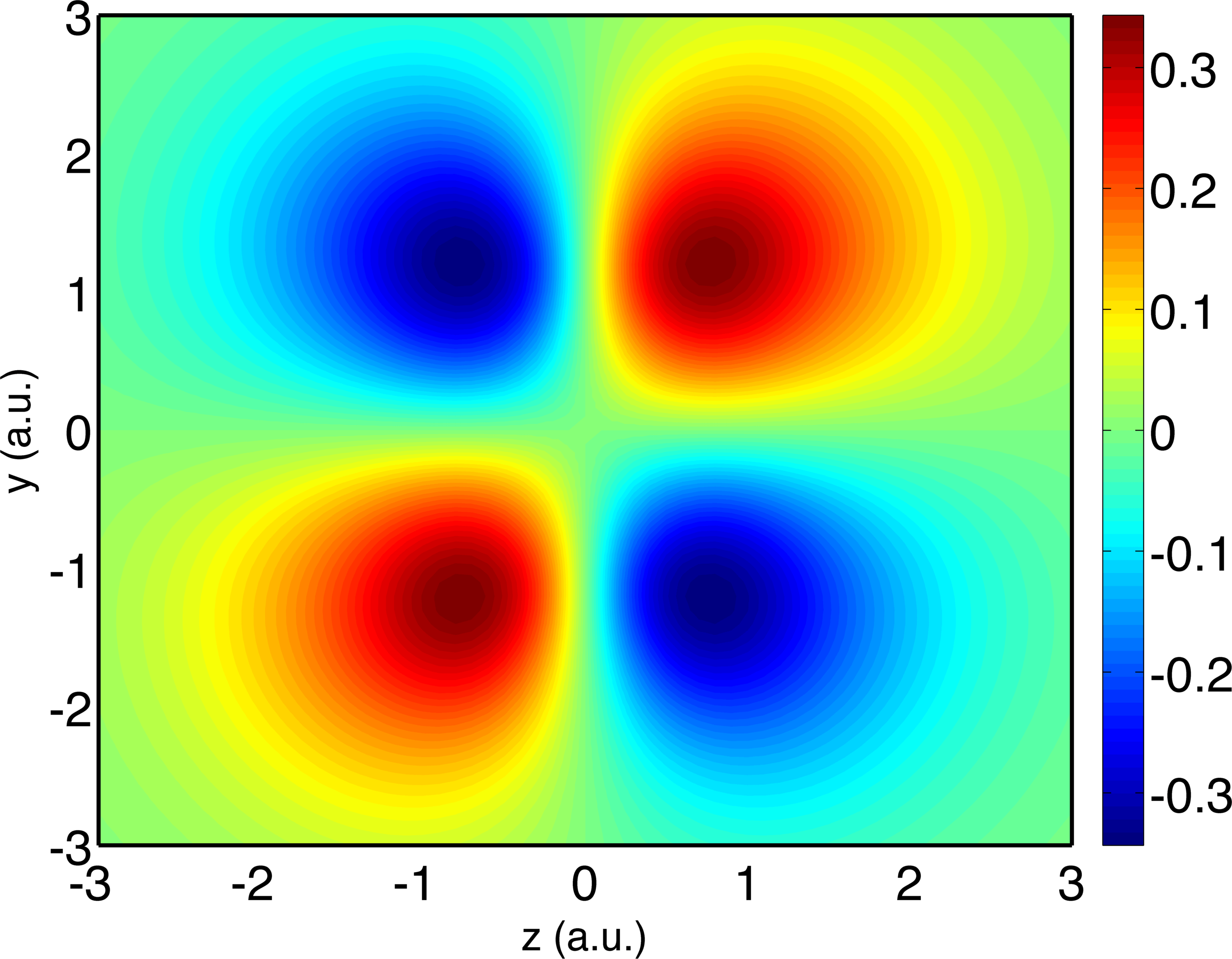}}
		    \caption{(color online) (a)-(b) Total, direct and rescattering contributions to the photoelectron spectra as a function of the final momentum and HOMO calculated using the nonlocal SR potential, respectively; (c)-(d) the same as in (a)-(b) but now the HOMO is modeled using a LCAO. In all the cases, we used, an O$_2$ molecule oriented parallel to the laser field polarization (see text for more details).}
		  	\label{Fig:O2_BT}
		\end{figure}

In Fig.~\ref{Fig:O2_BT}, we display the respective molecular orbitals (right panels) and the results for the 1D photoelectron spectra (left panels) using each of the models. In this case the molecule is oriented parallel to the laser field polarization. 

In the upper panels of Fig.~\ref{Fig:O2_BT} we present the results using the \textit{Model A}. Here, we use the nonlocal SR potential to obtain the ground state and the bound-free dipole matrix element. This kind of potential only support $s$ states as we can see in Fig.~\ref{Fig:O2_BT}(b). On the contrary, the \textit{Model B} gives a more accurate description of the O$_2$ molecular orbital (MO) (see Fig.~\ref{Fig:O2_BT}(d)).  The shape of the MO introduces noticeable differences in the final total photoelectron spectra (red line) as well in the different, direct (blue line) and rescattering (black line), contributions.

Figures~\ref{Fig:O2_BT}(a) and ~\ref{Fig:O2_BT}(c) show the main contributions to the full final photoelectron spectra, namely: the total $|b(\textbf{p},t)|^2$, Eq.~(\ref{Eq:b_T}), the direct $|b_0(\textbf{p},t)|^2$, Eq.~(\ref{Eq:b_0}) and the rescattering $|b_1(\textbf{p},t)|^2$, Eq.~(\ref{Eq:b_1}) terms, respectively. The black solid lines define the two cutoffs defined by $2U_p$ and $10U_p$. As we can infer from the latter figures the two models show slightly different behaviors. In the case of \textit{Model A}, that describes the HOMO as a superposition of two one-electron
$1s$ atomic orbitals (AOs), Fig.~\ref{Fig:O2_BT}(a), we see an overestimation of the direct processes. This fact could be caused by the kind of SR potential used to get the molecular ground state. This SR potential does not properly describe the attraction force felt by the electron both when it is bound and in the continuum. In this way this electron could `escape' more easily from the ionic core and becomes a `direct electron'. 

Results from the two models also show some similarities: stronger oscillations for small values of the electron momentum followed by a rapid decrease of the ATI yield (at $|p_z|\lesssim1.0$ a.u.), a plateau, where the amplitude remains almost constant, and both approaches end up with an abrupt cutoff around the same value of $|p_z|\lesssim 2.1$~a.u. (black solid  lines)~\cite{Lewenstein1995,Milosevic2006}.
 
We can also observe from Figs.~\ref{Fig:O2_BT}(a) and~\ref{Fig:O2_BT}(c) that the differences start to disappear for high electron energies, where the spectra are dominated by the rescattered electrons. This is so because the core potential plays a minor role in this energy region. Furthermore, our model captures the CEP asymmetry as well: electrons with positive final momentum are more influenced by the laser field polarization and this creates a stronger interference pattern.

\subsubsection{CO molecule}

\begin{figure}[htb]
        
                         \subfigure[]{\includegraphics[width=0.5\textwidth]{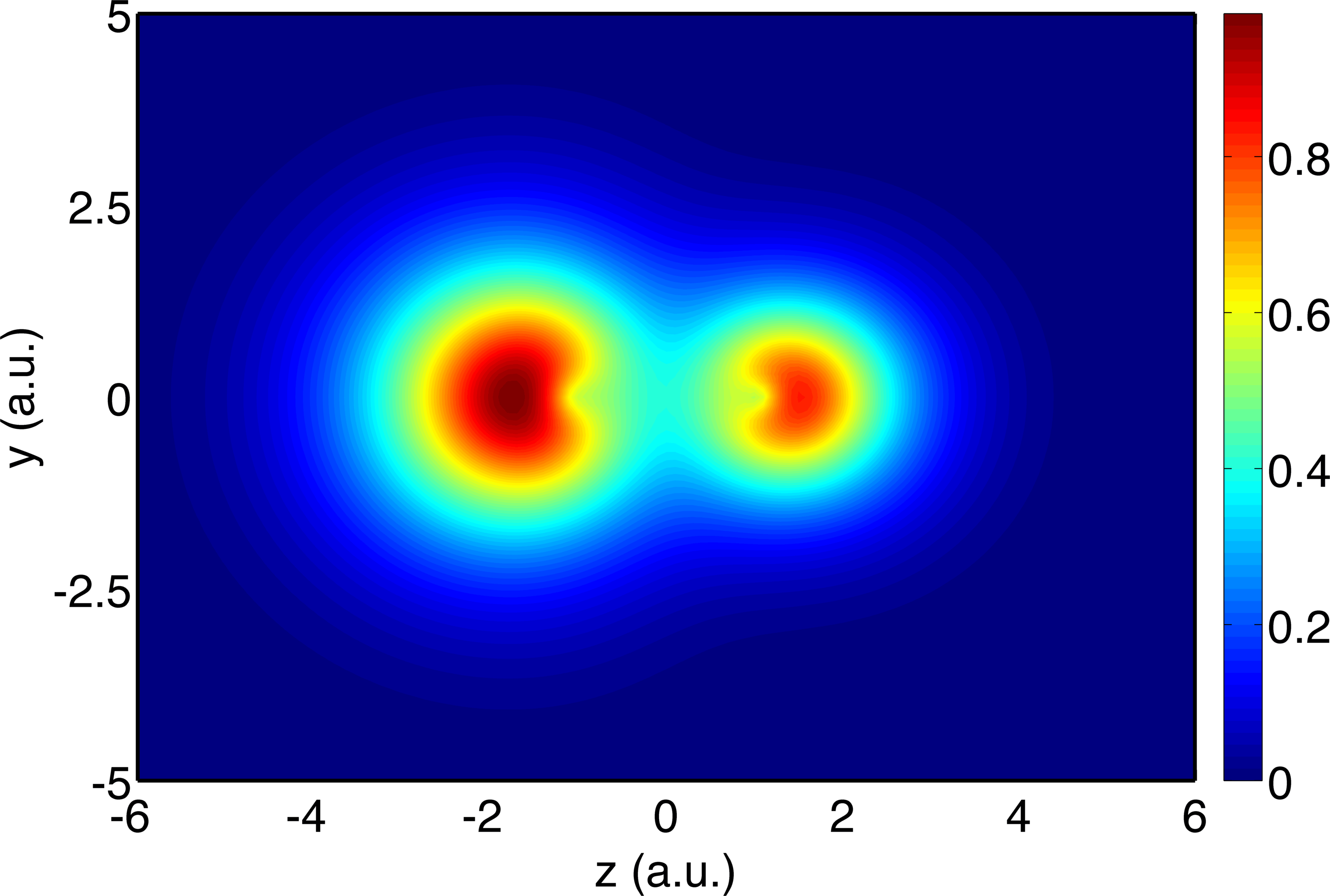}}
             
		    \caption{(color online) CO HOMO presented in the $z-y$ plane calculated using GAMESS (see the text for more details).}
		  	\label{Fig:CO_MO}
		\end{figure}

In the CO calculation we set the parameters of our nonlocal SR potential to $\Gamma=1$ and $\gamma=0.09$~a.u., in order to match the ionization potential obtained with GAMESS, $I_p=0.44$~a.u. As we already mentioned, this nonlocal SR potential only describe MOs as a composition of $s$ states, see Fig.~\ref{Fig:O2_BT}(b). On the other hand, the main advantage to use GAMESS is that it describes the MO much more accurately. Additionally, with GAMESS we have the possibility to easily model more complex molecules. The MO of the CO molecule obtained from GAMESS is a superposition of $s$ and $p$ states and it is shown in Fig.~\ref{Fig:CO_MO}. We consider the CO molecule is in equilibrium, the internuclear distance is set to $R=2.13 $ a.u. (1.127 \AA), and oriented parallel to the laser field polarization. 

\begin{figure}[htb]
            \subfigure[]{ \includegraphics [width=0.45\textwidth] {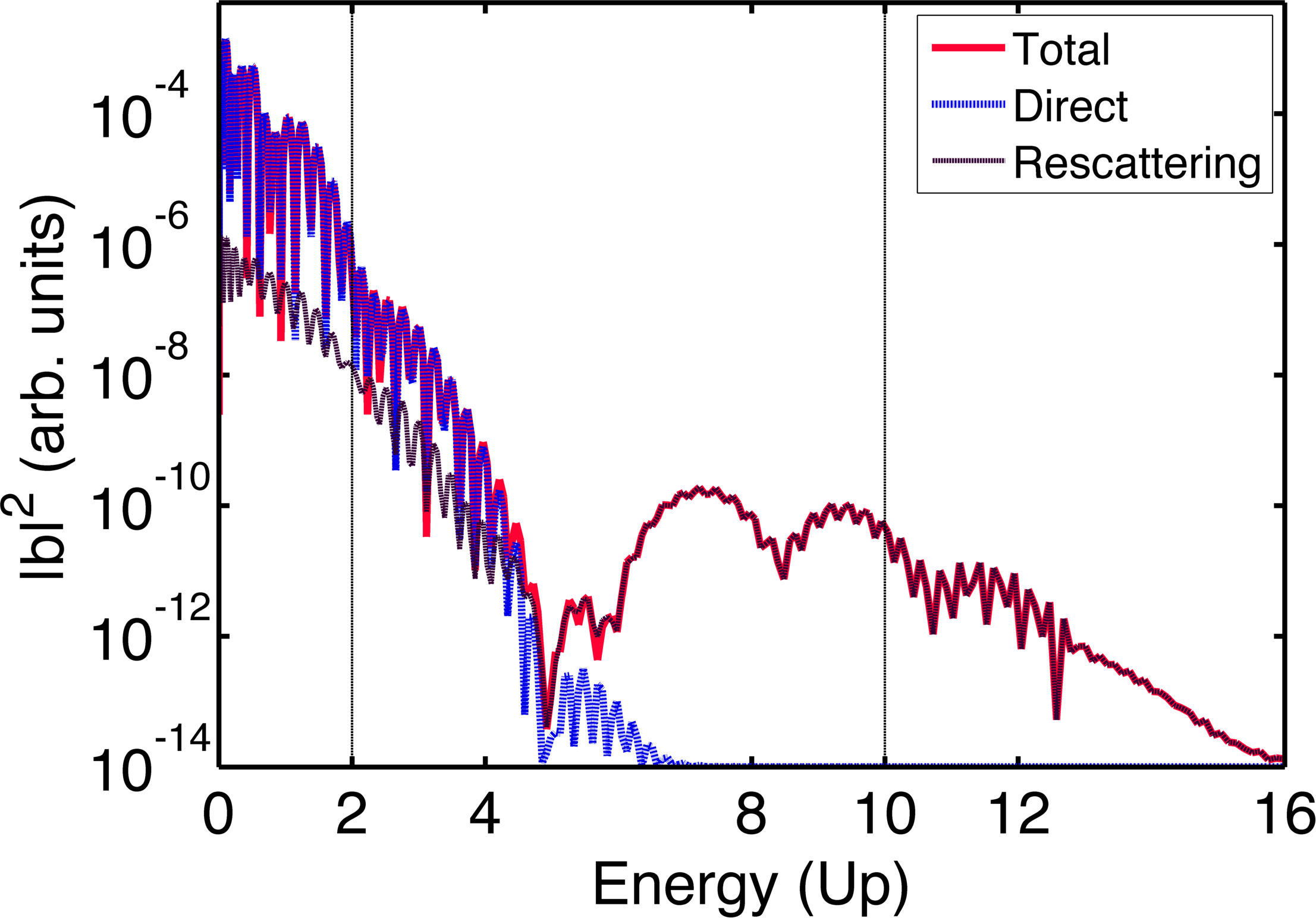}
          }
             \subfigure[]{ \includegraphics [width=0.45\textwidth] {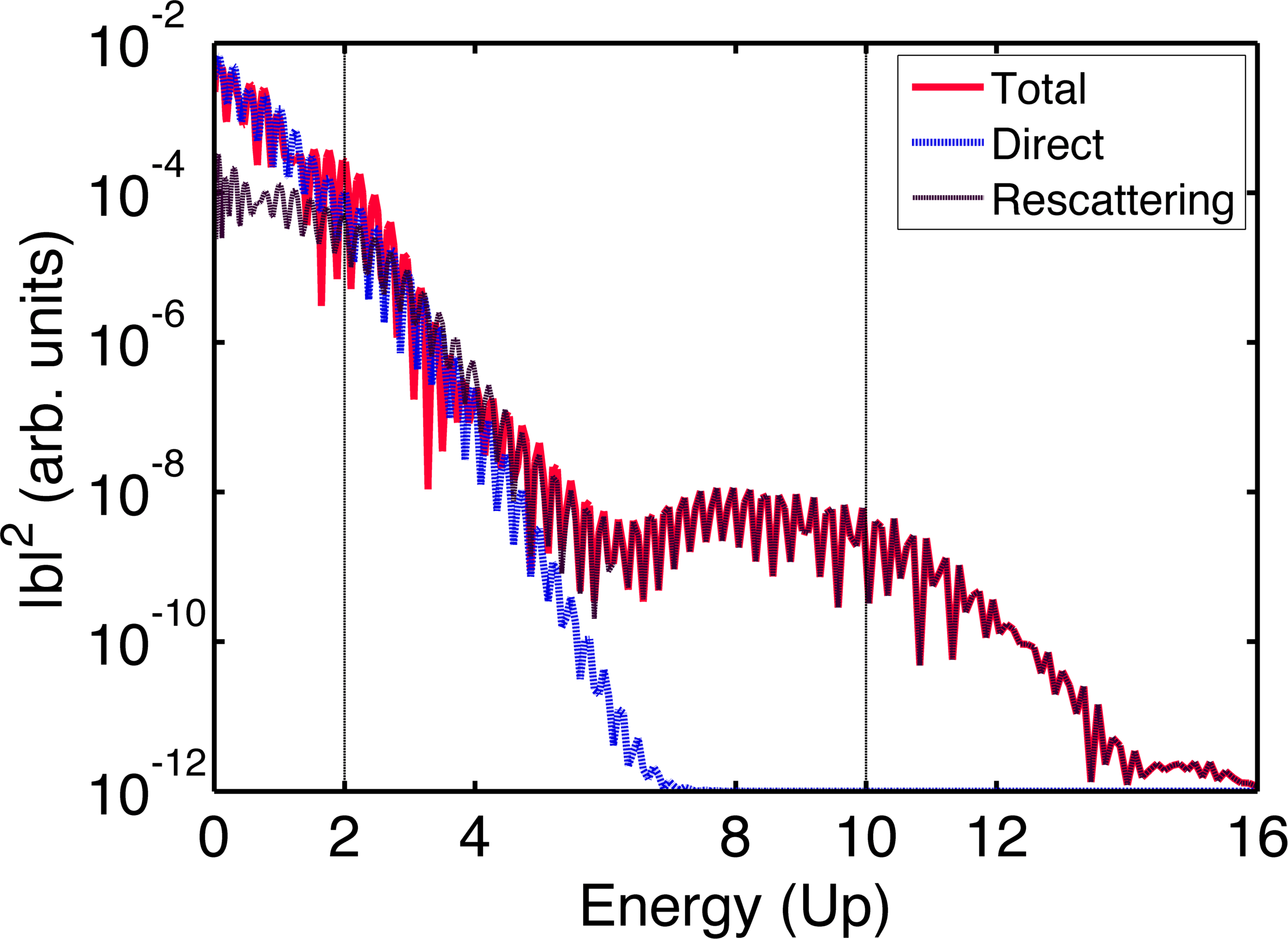}
           }
		    \caption{(color online) Total, direct and rescattering contributions to the photoelectron spectra, as a function of the electron energy in $U_p$ units, for the CO molecule. (a) \textit{Model A} and (b) \textit{Model B} (see text for more details).}
		  	\label{Fig:CO_BT}
		\end{figure}

Figure~\ref{Fig:CO_BT} shows the main contributions to the final photoelectron spectra for the CO molecule: the total $|b(\textbf{p},t)|^2$, Eq.~(\ref{Eq:b_T}), the direct $|b_0(\textbf{p},t)|^2$, Eq.~(\ref{Eq:b_0}) and the rescattering $|b_1(\textbf{p},t)|^2$, Eq.~(\ref{Eq:b_1}) terms, respectively. In the Fig.~\ref{Fig:CO_BT}(a) we display the results using the \textit{Model A}, meanwhile the ones from the \textit{Model B} are shown in Fig.~\ref{Fig:CO_BT}(b).

A clear observation from these plots is that each term contributes to different regions of the photoelectron spectra, i.e.~for electron energies $E_{p}\lesssim2 U_p$ the direct term $|b_0({\textbf p},t)|^2$ dominates the spectrum and, on the contrary, it is the rescattering term, $|b_1({\textbf p},t)|^2$ the one that wins in the high-energy electron region. Both photoelectron spectra shows the expected two cutoffs defined by $2U_p$ and $10U_p$ (black solid lines) which are ubiquitously present in both atomic and diatomic molecular ATI~\cite{Lewenstein1995, Milosevic2006}.

One of the main differences between the two models is that the total maximum yield amplitude is two orders of magnitude higher in the case of the \textit{Model B}, than in the \textit{Model A}. Besides this contrast the dynamic range of both spectra is quite similar: about ten orders of magnitude until the end of the signal. In here we only show the electrons moving to the `right', i.e.~with positive momentum; but as in the above case of O$_2$ the total spectra show CEP asymmetries.

The two spectra show some remarkable similarities; both have a deep minimum around $5U_p$, more pronounced for the \textit{Model A} case [Fig.~\ref{Fig:CO_BT}(a)], from where the yield of the direct processes starts to decrease. The contribution of the direct processes is negligible for energies $\gtrsim 7 U_p$, from where the spectra is dominated by the scattering processes. The two CO spectra show, in general, more similarities than in the O$_2$ case; this is due to the nature of the CO HOMO: in the CO molecule the MO is a composition of not only $2p$ also $1s$ AOs and our SR potential is able to partially include the contribution of the latter.

 \begin{figure}[htb]
 
            \subfigure[]{ \includegraphics [width=0.45\textwidth] {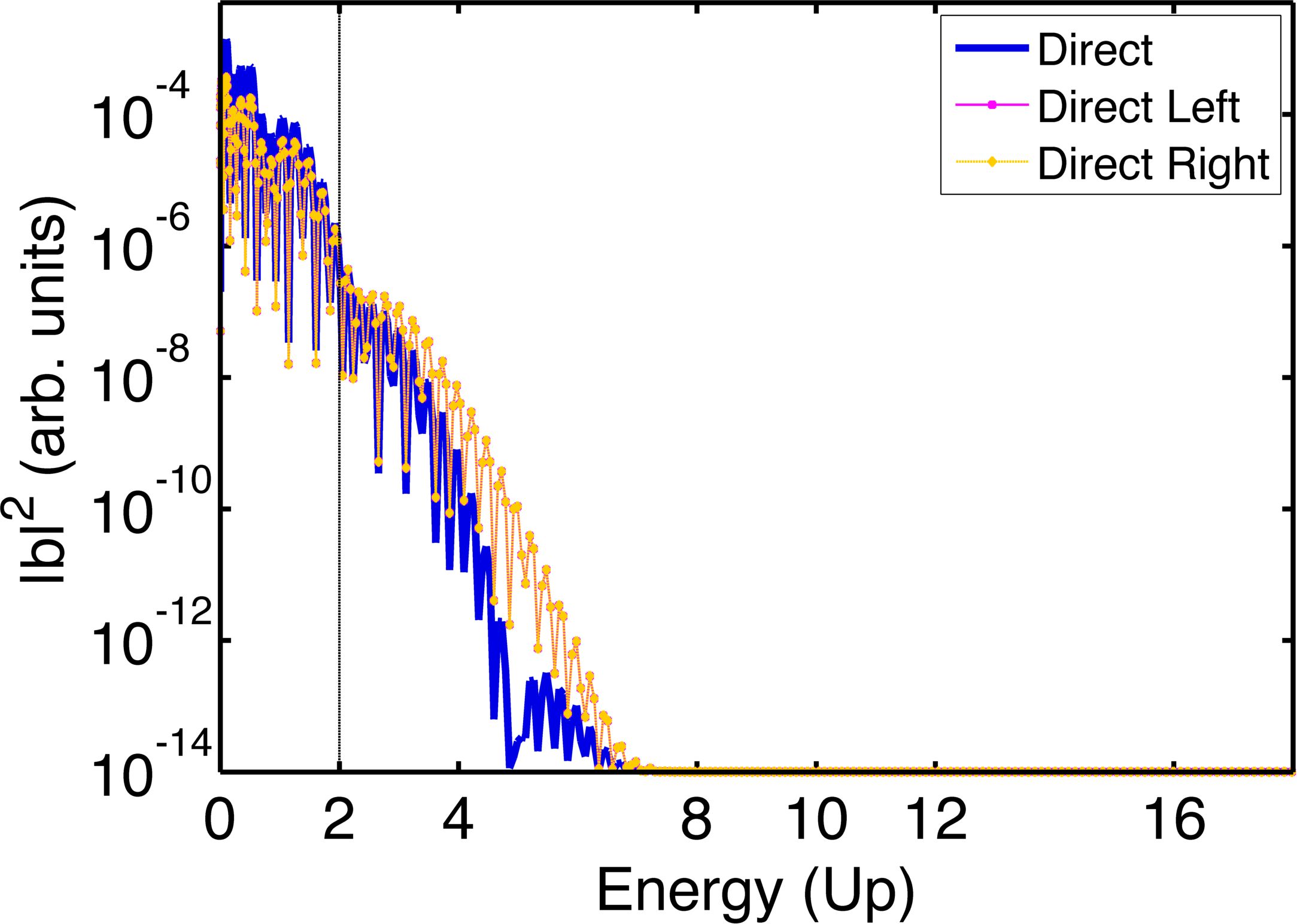}
            \label{fig:subfiga}}
		 \subfigure[]{\includegraphics[width=0.45\textwidth]{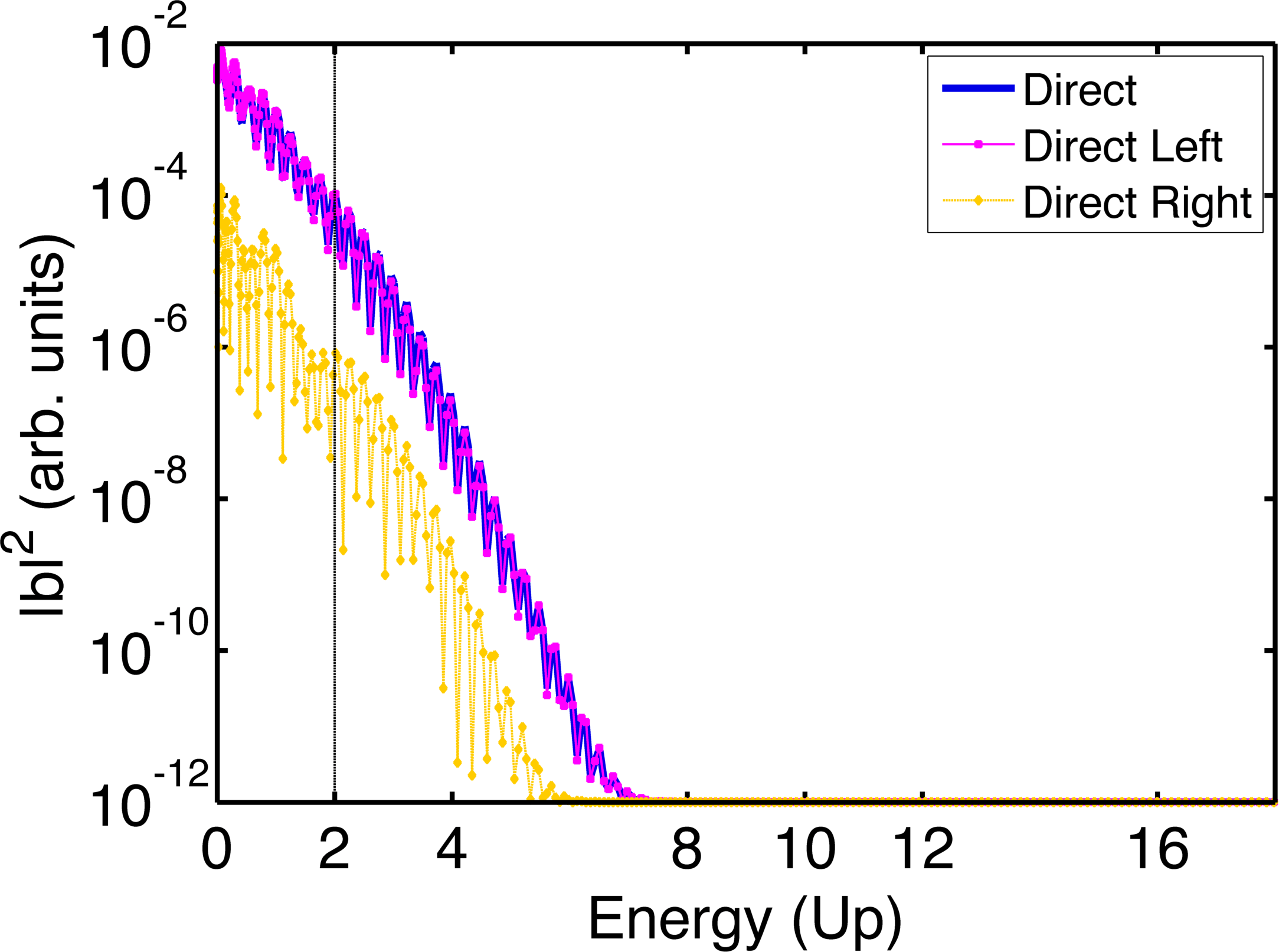}}
%
             \caption{(color online) Direct contributions to the CO photoelectron spectra (in logarithmic scale) as a function of the electron energy, in $U_p$ units, calculated by using \textit{Model A} (a) and \textit{Model B} (b) (see text for more details).
                                    }                  \label{Fig:CO_Contributions}                
                    \end{figure}


In order to have a more complete picture of the underlying mechanisms we present in Fig.~\ref{Fig:CO_Contributions} the different direct processes contributions to the total ATI spectra. In Fig.~\ref{Fig:CO_Contributions}(a) we show the split of the direct processes obtained using the \textit{Model A}, whereas in Fig.~\ref{Fig:CO_Contributions}(b) we depict the results using the \textit{Model B}. The first observation in this comparison arises from the fact that the contributions from the atom on the \textit{left} ($|b_{0,1}(\textbf{p},t)|^2$), i.e.~carbon, and the atom on the \textit{right} ($|b_{0,2}(\textbf{p},t)|^2$), i.e.~oxygen, are different in the case of the \textit{Model B} [Fig.~\ref{Fig:CO_Contributions}(b)]. The amount of photoelectrons ionized from the carbon atom (pink dotted line) is much larger than the one from the oxygen (yellow dotted line). This is in agreement with the shape of the CO HOMO, see Fig.~\ref{Fig:CO_MO}, where the electronic cloud around the carbon atom is much bigger. The same effect is observed for the recattering terms (not shown), where the total local term is dominated by the local processes coming from the carbon atom.

 \begin{figure}[htb]
		 \subfigure[]{\includegraphics[width=0.5\textwidth]{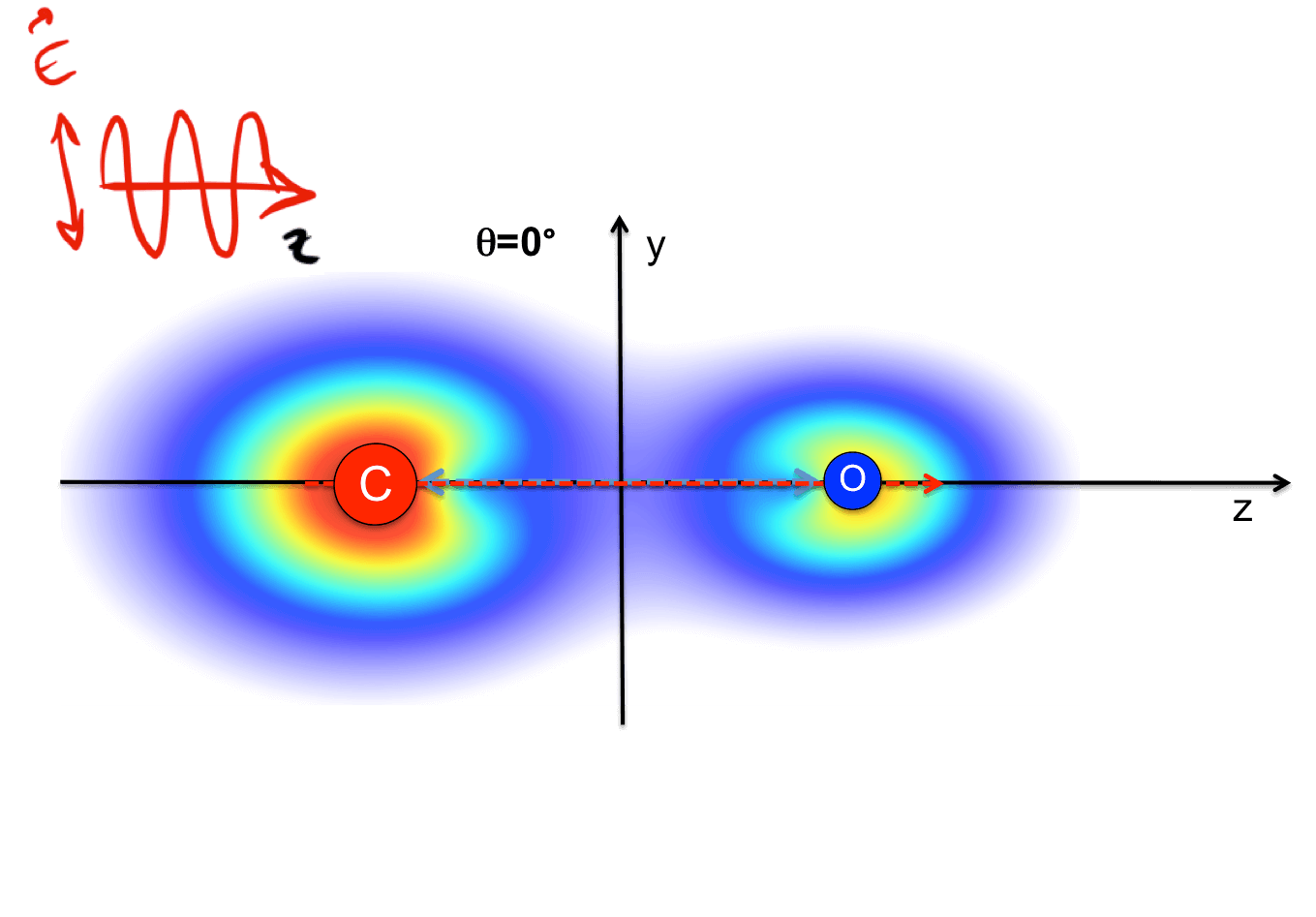}}
		 \subfigure[]{\includegraphics[width=0.45\textwidth]{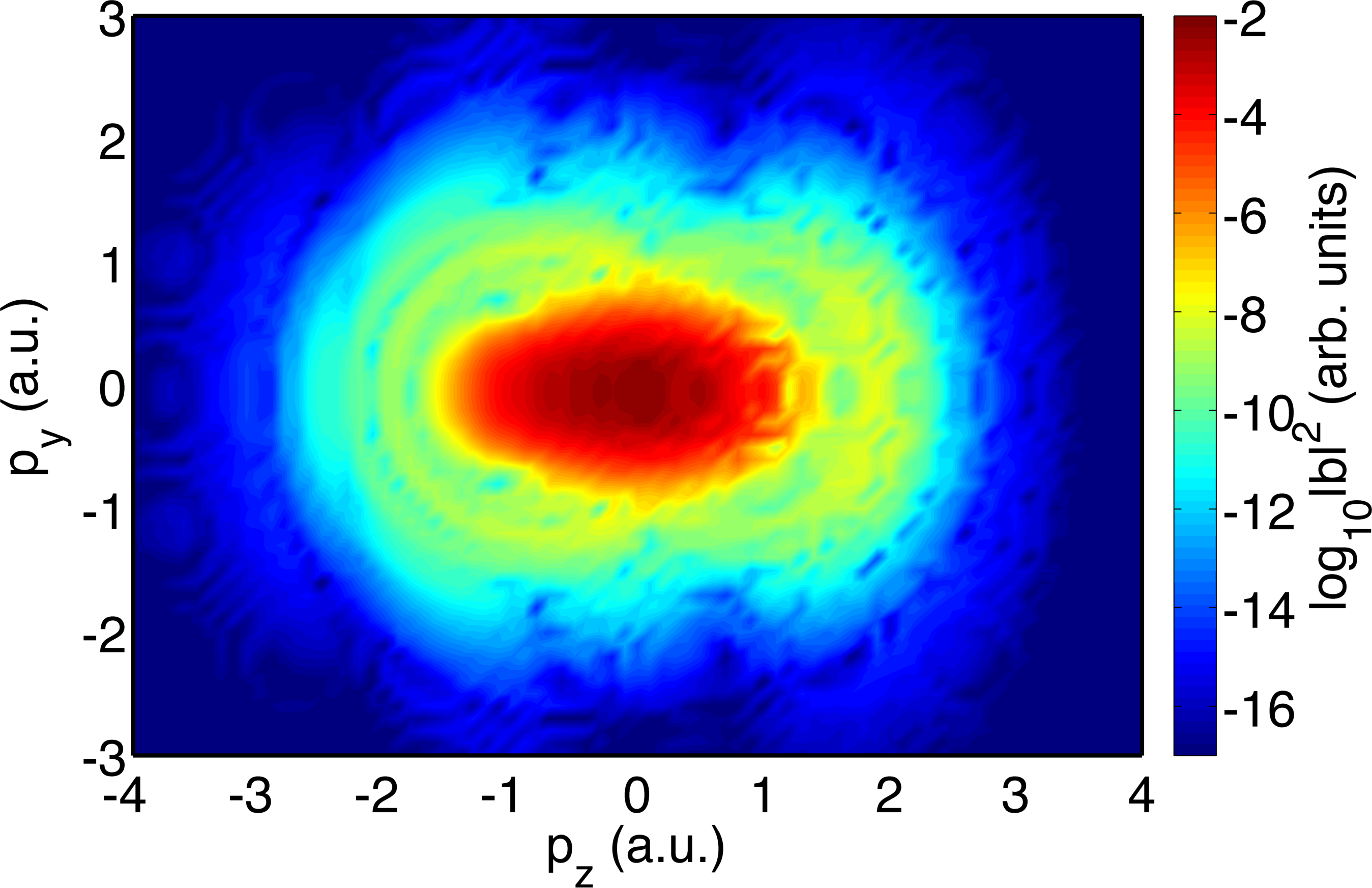}}
		  \subfigure[]{\includegraphics[width=0.5\textwidth]{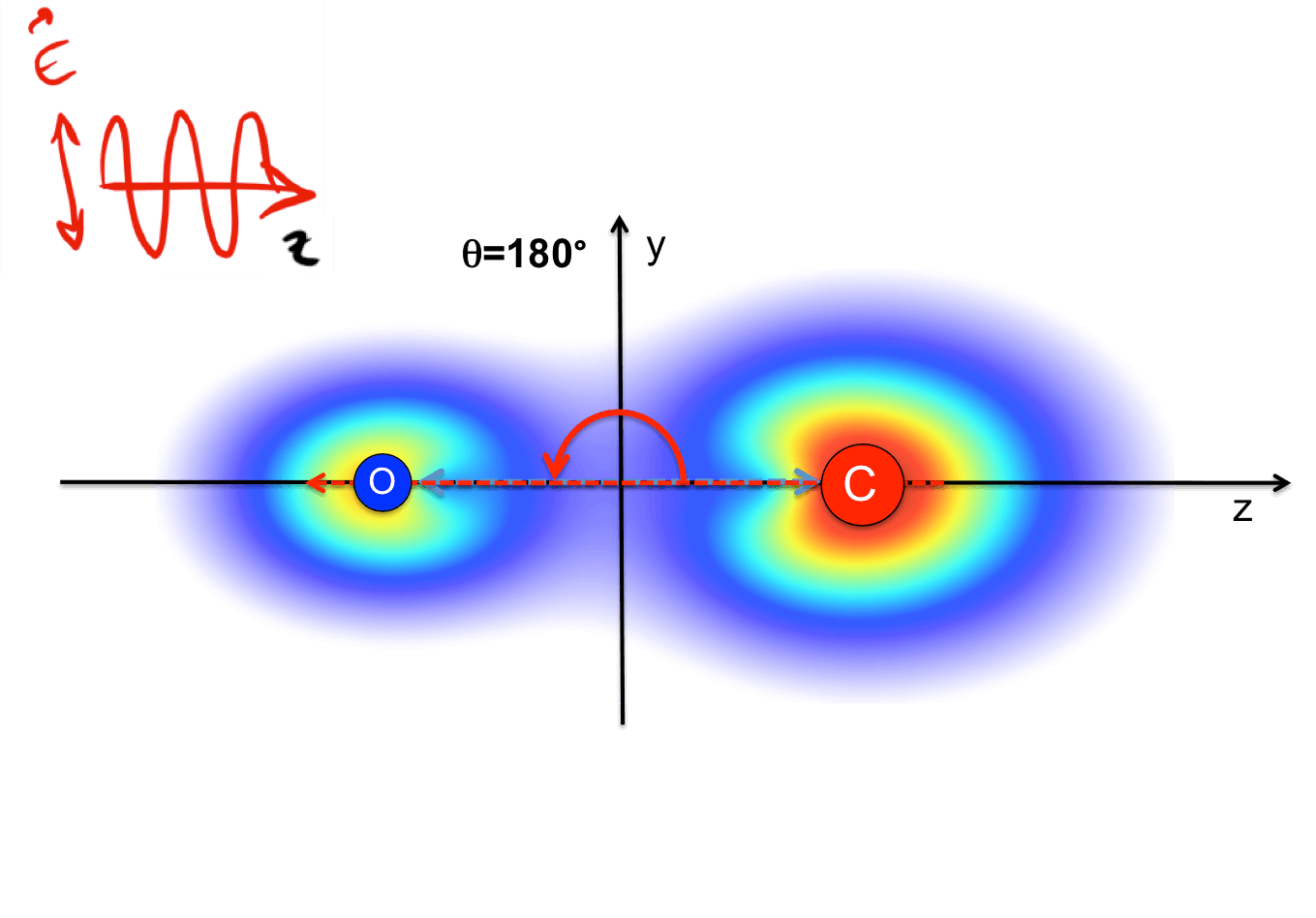}}
		  \subfigure[]{\includegraphics[width=0.45\textwidth]{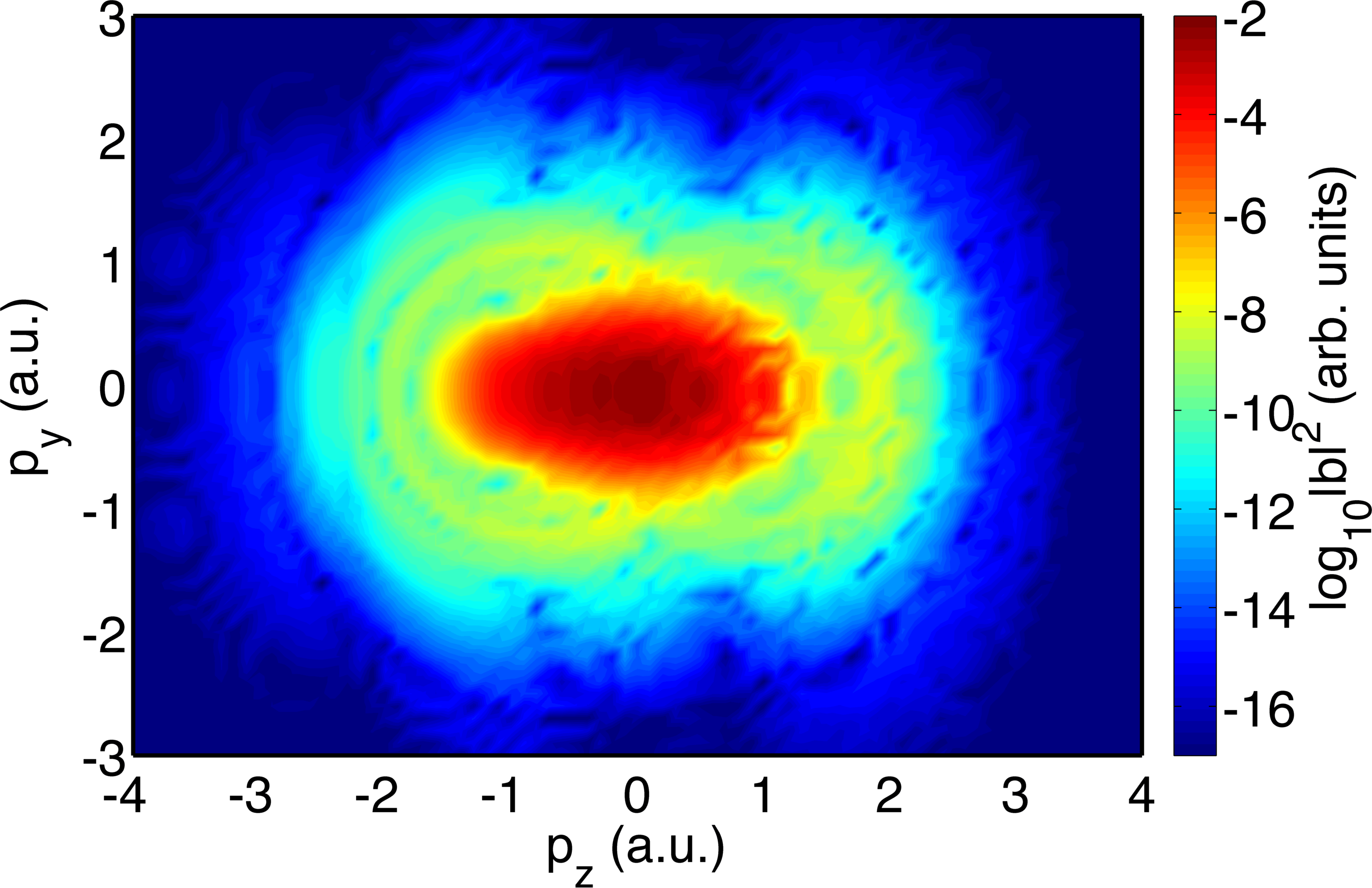}}
  \caption{(Color online) Total photoelectron spectra for a 2D-momentum plane $(p_z,p_y)$. ATI photoelectron spectra (in logarithmic scale) as a function of the momentum $(p_z, p_y)$ computed by our quasiclassical model. (a) Representation of the CO molecule aligned at $0^{\circ}$ with respect to the laser field polarization; (b) total ATI photoelectron spectra; (c)-(d) the same as in (a)-(b) but for the molecule aligned at $180^{\circ}$ with the laser field polarization.
                                    }                  \label{Fig:CO_2D180}                
                    \end{figure}
                    
In the case of the calculations using \textit{Model A} those differences are not so pronounced, see Fig.~\ref{Fig:CO_Contributions}(a): we can observe that the contributions of both atoms are equal in amplitude and shape. This is so because the bound state obtained from the SR potential does not properly describe the CO HOMO: this potential is unable to take into account the heteronuclear character of the CO molecule and describes its HOMO similar to the one shown in Fig.~\ref{Fig:O2_BT}(b).


Considering the nuclear asymmetry features discussed before we next study the differences in the ATI spectra for the molecule aligned parallel ($0^{\circ}$) or antiparallel ($180^{\circ}$) with respect to the laser field polarization. The results of a 2D calculation, for both orientations and using the LCAO within the \textit{Model B} is presented in Fig.~\ref{Fig:CO_2D180}. Figures~\ref{Fig:CO_2D180}(a) and~\ref{Fig:CO_2D180}(c) show a sketch of the molecular orientation, superimposed over the MO. Here we can see that for the case of the CO molecule aligned parallel (antiparallel) the carbon atom is on the 'left' ('right'), meanwhile the oxygen atom is on the 'right' ('left'). Furthermore, Figs.~\ref{Fig:CO_2D180}(b) and~\ref{Fig:CO_2D180}(d) depict the total ATI spectra for both the parallel and antiparallel cases, respectively.

The total ATI spectra presented in Figs~\ref{Fig:CO_2D180}(b)-(d) show the typical CEP asymmetry, but surprisingly any features related to the heteronuclear character of the CO molecule appear to be missing: the two ATI spectra, the one obtained for the molecule at $0^{\circ}$ (Fig.~\ref{Fig:CO_2D180}(b)) and the one for $180^{\circ}$ (Fig.~\ref{Fig:CO_2D180}(d)), look almost identical.

 \begin{figure}[htb]
 
            \subfigure[]{ \includegraphics [width=0.4\textwidth] {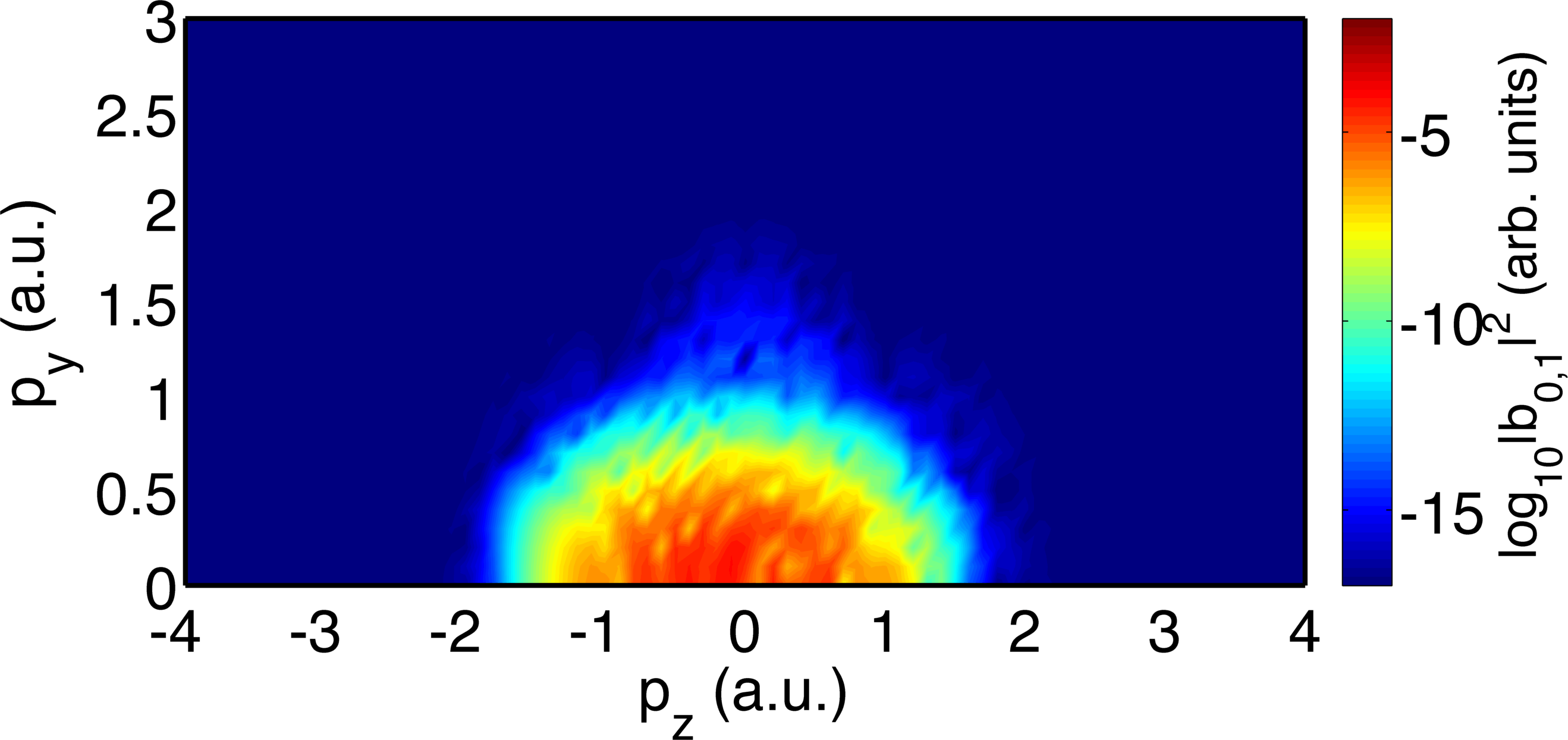}
            \label{fig:subfiga}}
		 \subfigure[]{\includegraphics[width=0.4\textwidth]{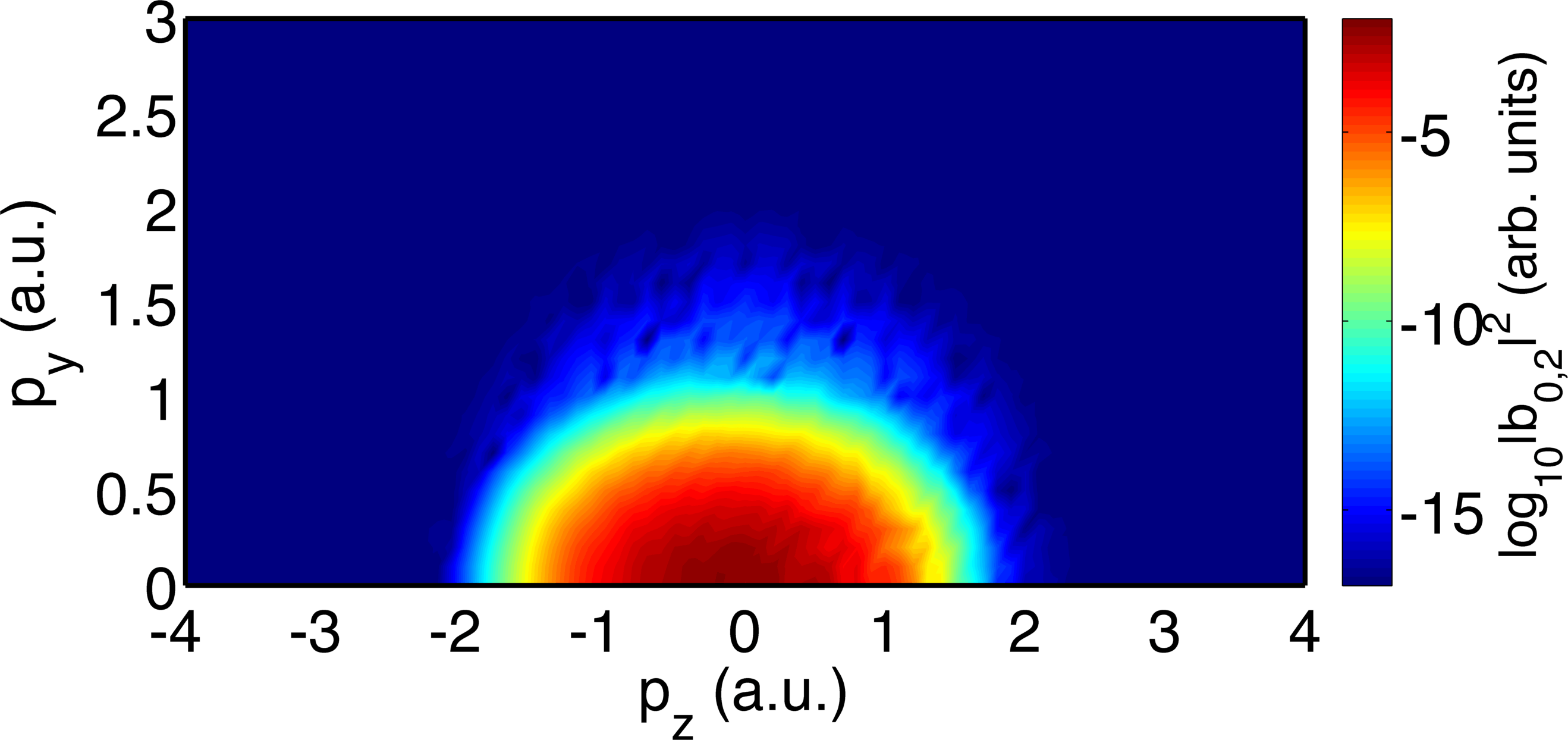}}
     		 \subfigure[]{\includegraphics[width=0.4\textwidth]{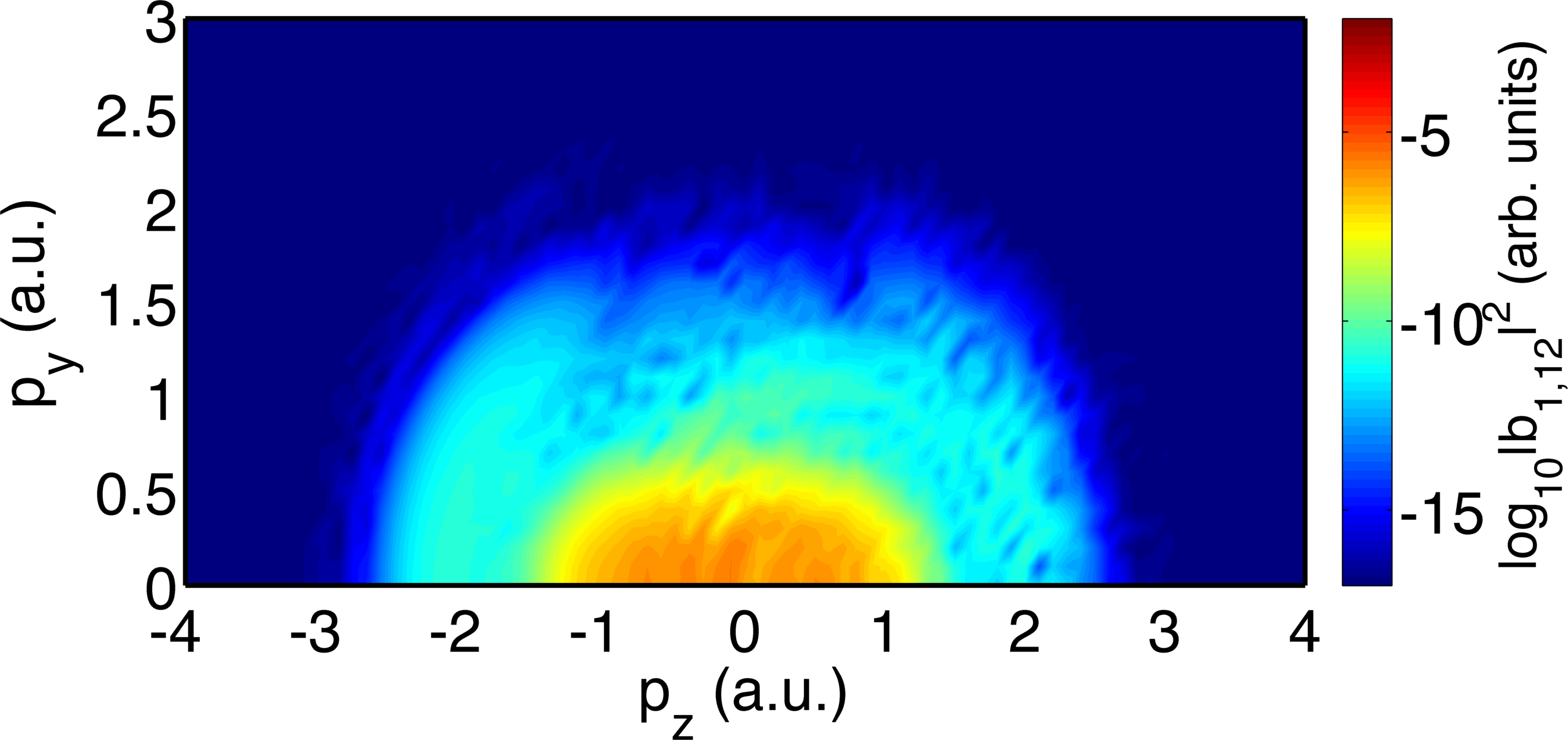}}
		 \subfigure[]{\includegraphics[width=0.4\textwidth]{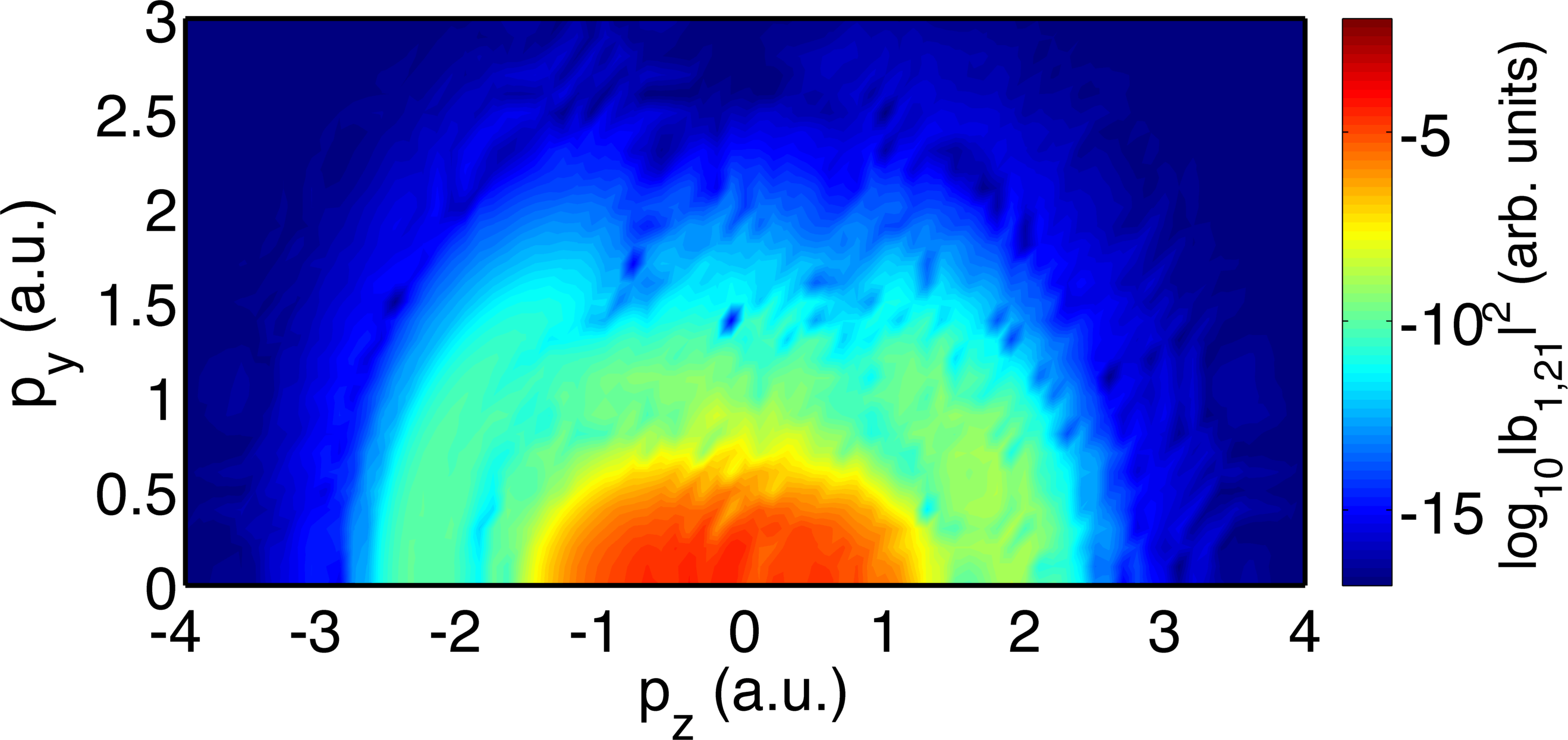}}

             \caption{(color online) Rescattering contributions to the photoelectron spectra for a 2D-momentum plane $(p_z, p_y)$. ATI photoelectron spectra (in
logarithmic scale) calculated by using the \textit{Model B} for a CO molecule at equilibrium $R=2.12$ a.u. and oriented antiparallel ($\theta=180^{\circ}$) to the laser field polarization. (a) local term of the atom on the left; (b) local term of the atom on the right; (c) nonlocal and cross term with ionization from the left; (d) nonlocal and cross term with ionization from the right.
                                    }                  \label{Fig:CO_180}                
                    \end{figure}

In order to get a more detailed description of the CO ATI spectra presented in Figs~\ref{Fig:CO_2D180}(b)-(d) in Fig.~9 we plot the contribution of the rescattering processes to the total ATI. On the other hand, the direct contributions show the expected behavior: a symmetry inversion. In this case the major contribution also comes from the carbon atom on the \textit{right} ($|b_{0,2}(\textbf{p},t)|^2$), whereas the direct ionization from the oxygen atom, on the \textit{left} ($|b_{0,1}(\textbf{p},t)|^2$), is much more smaller.


In Figs.~\ref{Fig:CO_180}(a) and ~\ref{Fig:CO_180}(b) we present the local processes contributions $|b_{1,11}(\textbf{p},t)|^2$ and $|b_{1,22}(\textbf{p},t)|^2$, respectively. On the other hand, Figs.~\ref{Fig:CO_180}(c) and~\ref{Fig:CO_180}(d) depict the cross and nonlocal contributions, namely $|b_{1,12}(\textbf{p},t)|^2$ and $|b_{1,21}(\textbf{p},t)|^2$. In all the cases the molecule is aligned $180^{\circ}$ with the laser field polarization, i.e.~the oxygen atom is on the left, meanwhile the carbon atom is on the right. Interestingly, for the case of $0^{\circ}$ we obtain the same plots, but with the terms interchanged, i.e.~now the higher contribution comes from the carbon atom now located in the \textit{left} at the position $R_1$. This is the same asymmetry feature observed in the direct terms, see Fig.~\ref{Fig:CO_Contributions}(b). The heteronuclear character of the molecule can now be seen in the local and rescattering components but, as we observed, not in the total photoelectron spectra. This fact could be related to the compensation of the MO differences when the direct and rescattering terms are coherently added.  
	
\subsection{Results on triatomic molecules: CO$_{2}$ and CS$_{2}$}

In this section we are going to extend our analysis to more complicated molecular systems, formed now by three atomic centers.  We start our analysis computing the ATI for the CO$_{2}$ molecule. We present the different contributions, direct and rescattering, to the total photoelectron spectra and discuss their differences and similarities. We use next the CS$_{2}$ molecule as another three-center prototypical system. For this case we also calculate the different processes contributing to the total spectra and make a similar study to the one done for CO$_{2}$. In this way we are able to highlight both the discrepancies and coincidences between these two comparable molecular systems.

\subsubsection{CO$_{2}$ molecule}

We consider a CO$_{2}$ molecule in equilibrium, i.e.~the two oxygen atoms are separated a distance $R=4.4$~a.u. (2.327 \AA) with the carbon atom located in the mid point. The ionization potential of the outer electron predicted by GAMESS is $I_p=0.39$~a.u. The corresponding parameters of our nonlocal SR potential to obtain this $I_p$ are $\Gamma=0.8$ and $\gamma=0.1$~a.u.

\begin{figure}[htb]
          
          \subfigure[]{ \includegraphics [width=0.4\textwidth] {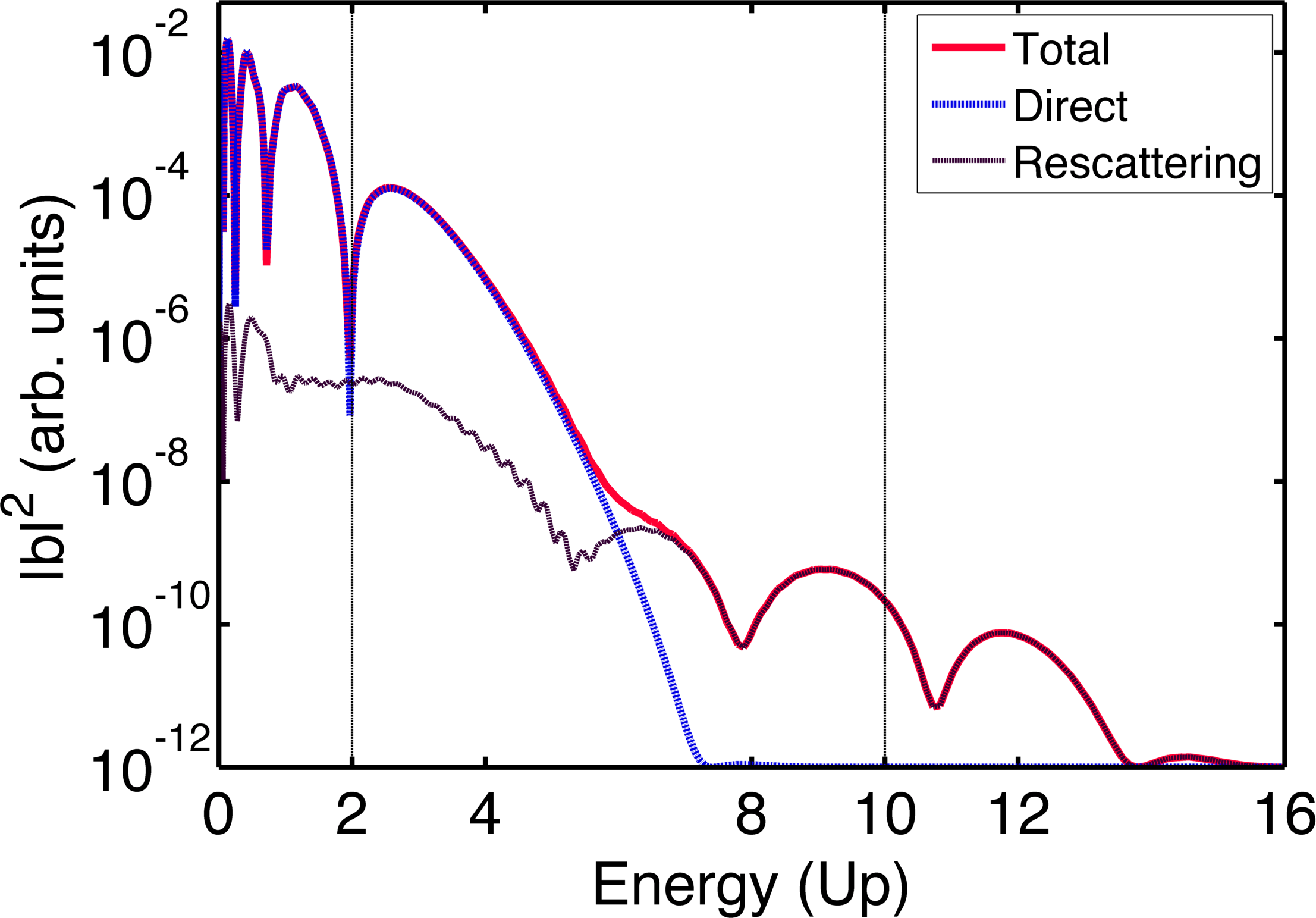}
            \label{fig:subfiga}}
             \subfigure[]{ \includegraphics [width=0.4\textwidth] {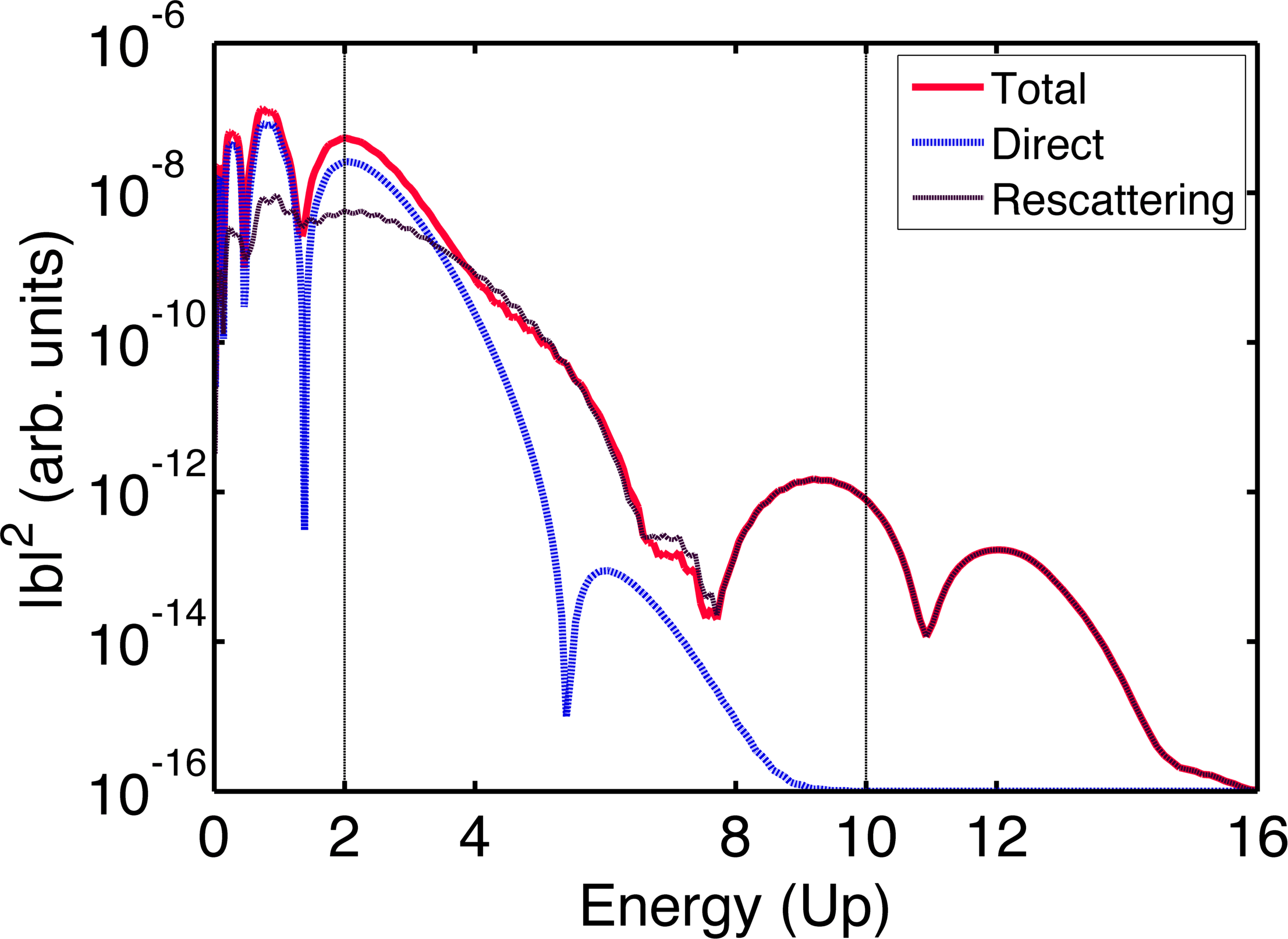}}
            
		    \caption{(color online) CO$_2$ molecular ATI spectra (in logarithmic scale) as a function of the electron energy in $U_p$ units. (a) spectra calculated using \textit{Model A}; (b) spectra computed using \textit{Model B}. In both calculations the CO$_2$ molecule is oriented parallel to the laser polarization (see text for more details).}
		  	\label{Fig:CO2_BT}
		\end{figure}
		
\begin{figure}[htb]

 \subfigure[]{ \includegraphics [width=0.4\textwidth] {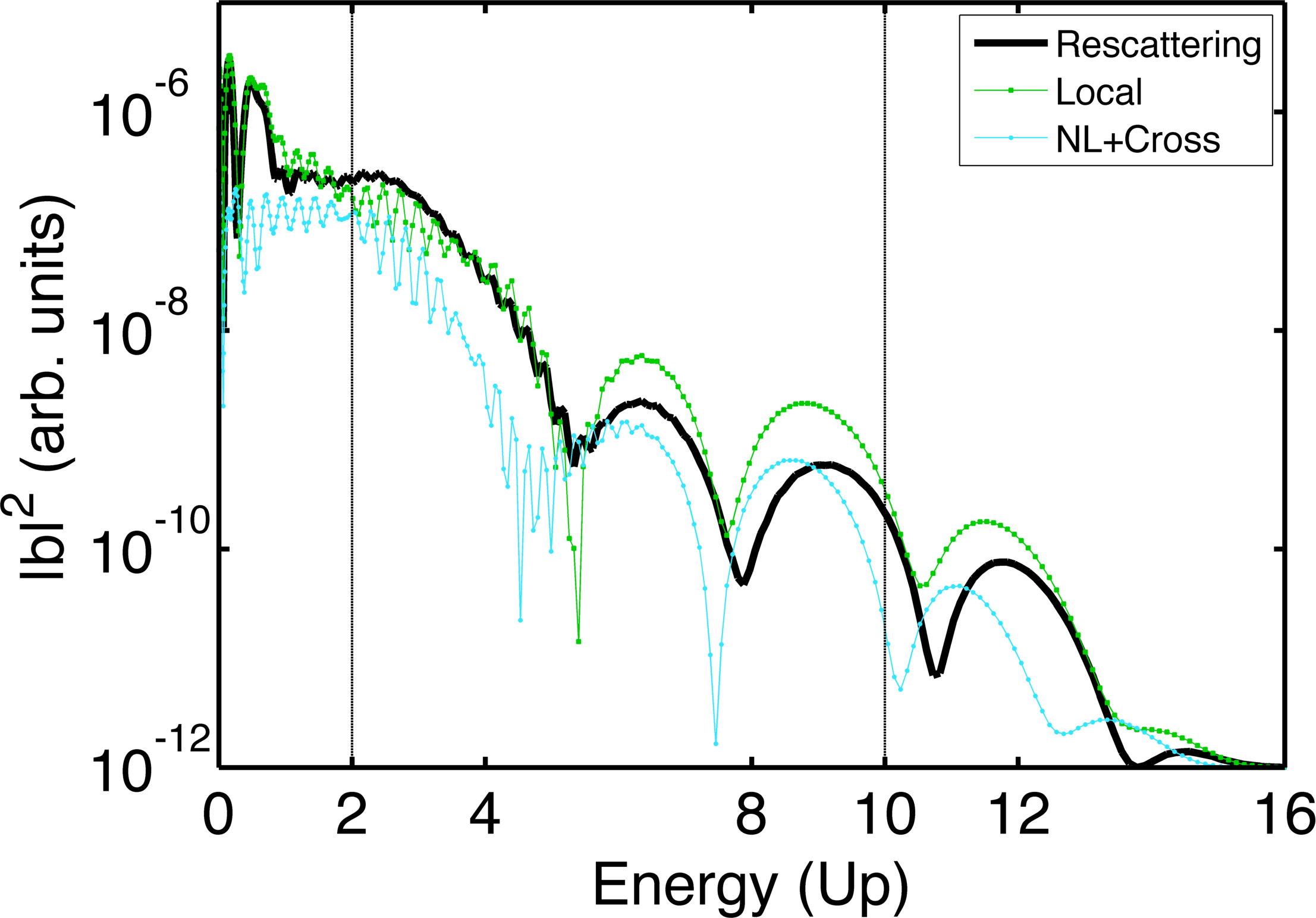}}
               \subfigure[]{ \includegraphics [width=0.4\textwidth] {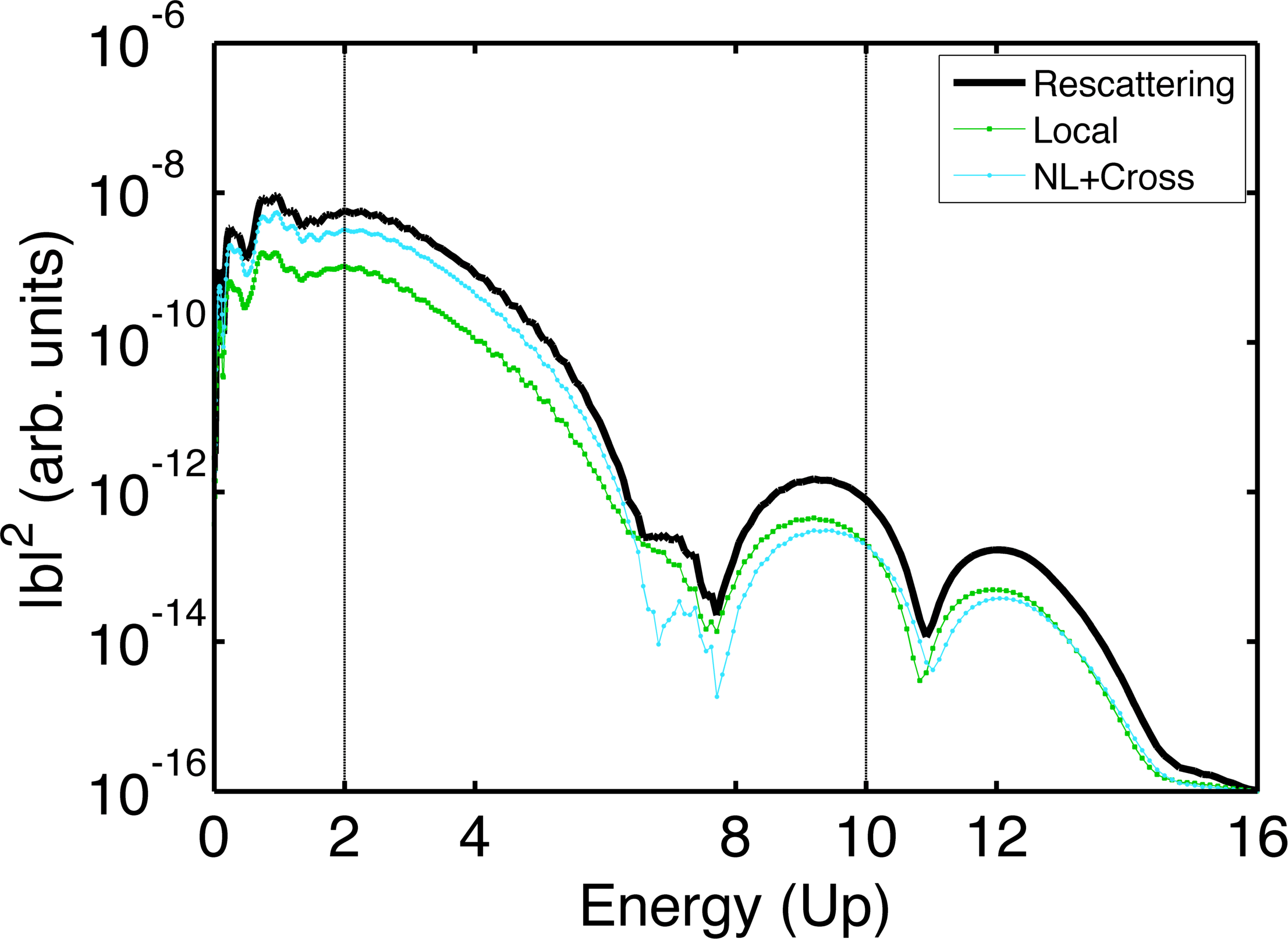}}
               \subfigure[]{ \includegraphics [width=0.4\textwidth] {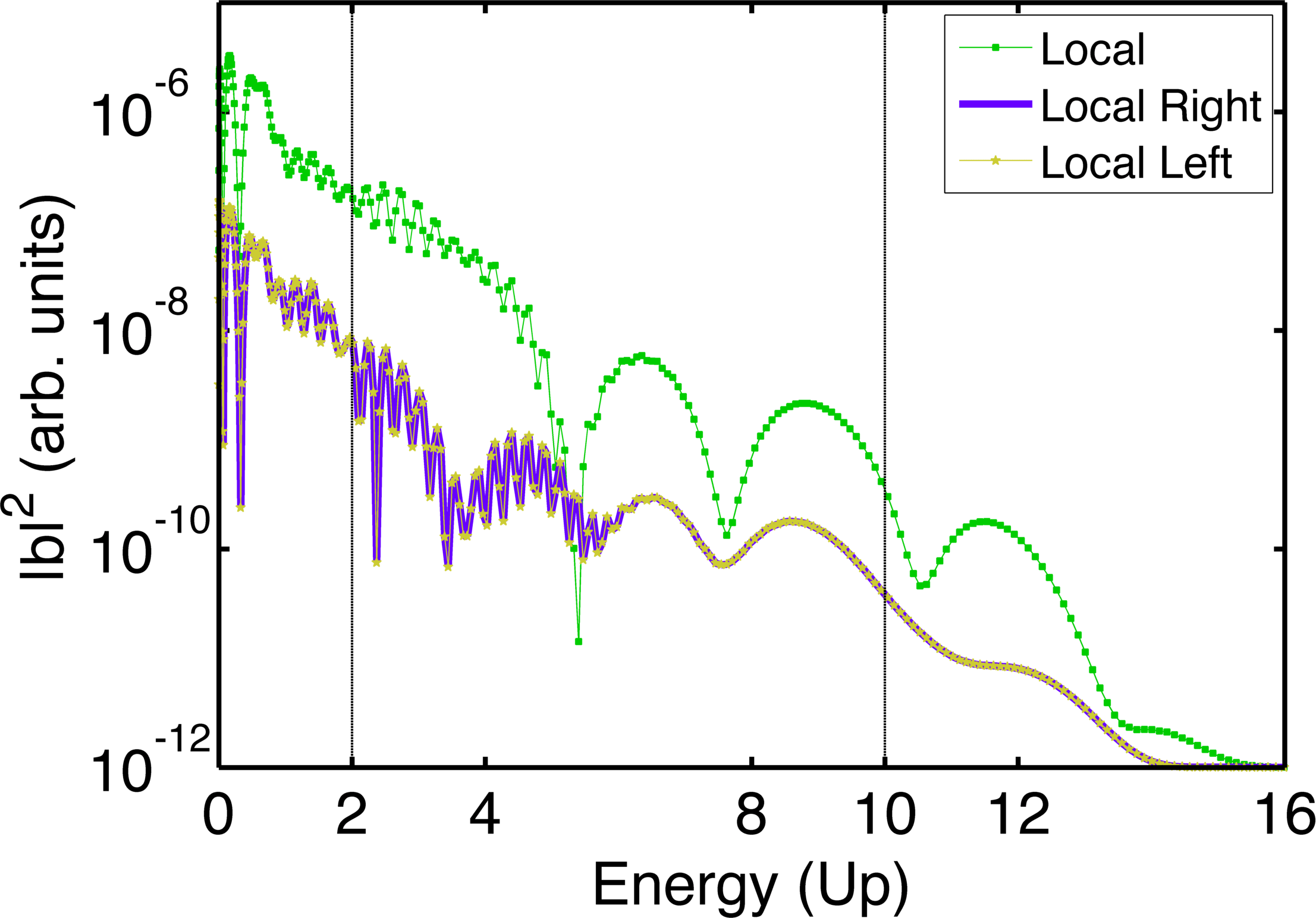}}
               \subfigure[]{ \includegraphics [width=0.4\textwidth] {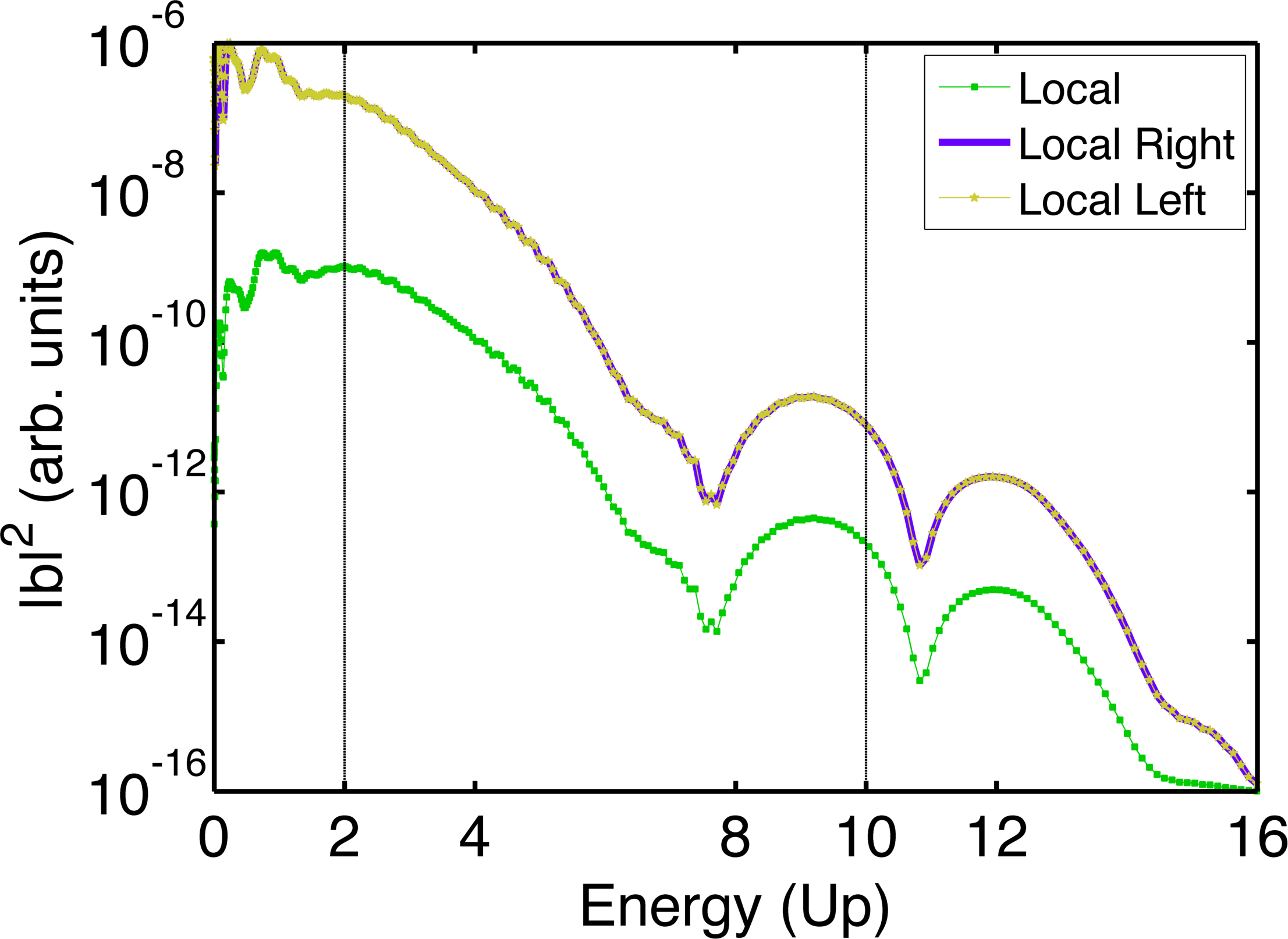}}

 \caption{(color online) Different rescattering processes contributions to the CO$_2$ molecular ATI spectra (in logarithmic scale) as a function of the electron energy in $U_p$ units. (a)-(c) spectra calculated using \textit{Model A}; (b)-(d) spectra computed using \textit{Model B}. In both calculations the CO$_2$ molecule is oriented parallel to the laser polarization (see text for more details).}
		  	\label{Fig:CO2_B1}
		\end{figure}

In Fig.~\ref{Fig:CO2_BT} we present the ATI spectra, computed by using both the \textit{Model A} [Fig.~\ref{Fig:CO2_BT}(a)] and \textit{Model B} [Fig.~\ref{Fig:CO2_BT}(b)]. Here, we show the different contributions: the total $|b(\textbf{p},t)|^2$  (solid red line), the \textit{direct} $|b_0(\textbf{p},t)|^2$ (blue solid line) and the \textit{rescattering} $|b_1(\textbf{p},t)|^2$ (dark brown line) ones for each model.  In both models we see that the direct processes contribute only in the low energy region of the spectra, $E_{p}\lesssim6 U_p$, being negligible at high energies, where the rescattering terms are dominant. In this case we also observe an overestimation of the direct terms and a difference of four orders of magnitude in the total yield between the \textit{Model A} [Fig.~\ref{Fig:CO2_BT}(a)] and \textit{Model B} [Fig.~\ref{Fig:CO2_BT}(b)]. Besides of this difference in amplitude, the shape of both spectra is quite similar: the change between direct and rescattering dominance is around the same energy ($\sim5 U_p$). On the other hand, we would like to attract the attention to the high energy part of the ATI spectra $E_{p}\gtrsim 4 U_p$. As can be seen, the two models show the same number of minima at around the same positions $\approx5 U_p$, $\approx8 U_p$ and $\approx11 U_p$.  In order to investigate if these minima are generated by the interference between the local and nonlocal+cross terms in Fig.~\ref{Fig:CO2_B1} we split the different \textit{rescattering} processes contributions. 

As can be seen in Figs.~\ref{Fig:CO2_B1}(a) and~\ref{Fig:CO2_B1}(b), both \textit{Local} and \textit{NL+Cross} contributions have almost the same yield over all the electron energy range and only minor differences are visible.  As a consequence the minima appear to be generated by the destructive interference between electrons tunnel-ionized and rescattered in the same ion core.  In Figs.~\ref{Fig:CO2_B1}(c) and~\ref{Fig:CO2_B1}(d), we present a split of the \textit{local} processes, namely, $|b_{1,11}(\textbf{p},t)|^2$ (solid yellow line with stars) and  $|b_{1,33}(\textbf{p},t)|^2$ (solid purple line). As we can see the contribution from the O atoms, placed at the end of the molecule, is equal in amplitude and shape in both models. On the contrary, the contribution of the C atom, placed at the origin, is almost negligible (not shown in the Figure).

Regarding the deep minima, if we take a look at the Figs.~\ref{Fig:CO2_B1}(c) and~\ref{Fig:CO2_B1}(d), we see that the minima are present in the independent contribution $|b_{1,11}(\textbf{p},t)|^2$ and $|b_{1,33}(\textbf{p},t)|^2$. This reinforce the hypothesis that internal interferences, inside of the atoms, are the responsible of those minima. We can also observe that, in the case of the \textit{Model A}, the local contributions (Right/Left) add up together to enhance the total local contribution. In the case of the \textit{Model B}, those two local contributions interfere each other leading up a total contribution with lower amplitude and exactly the same shape. This is a direct consequence of both the bound state wavefunction and the HOMO shape. 

\begin{figure}[htb]

		 \subfigure[]{\includegraphics[width=0.45\textwidth]{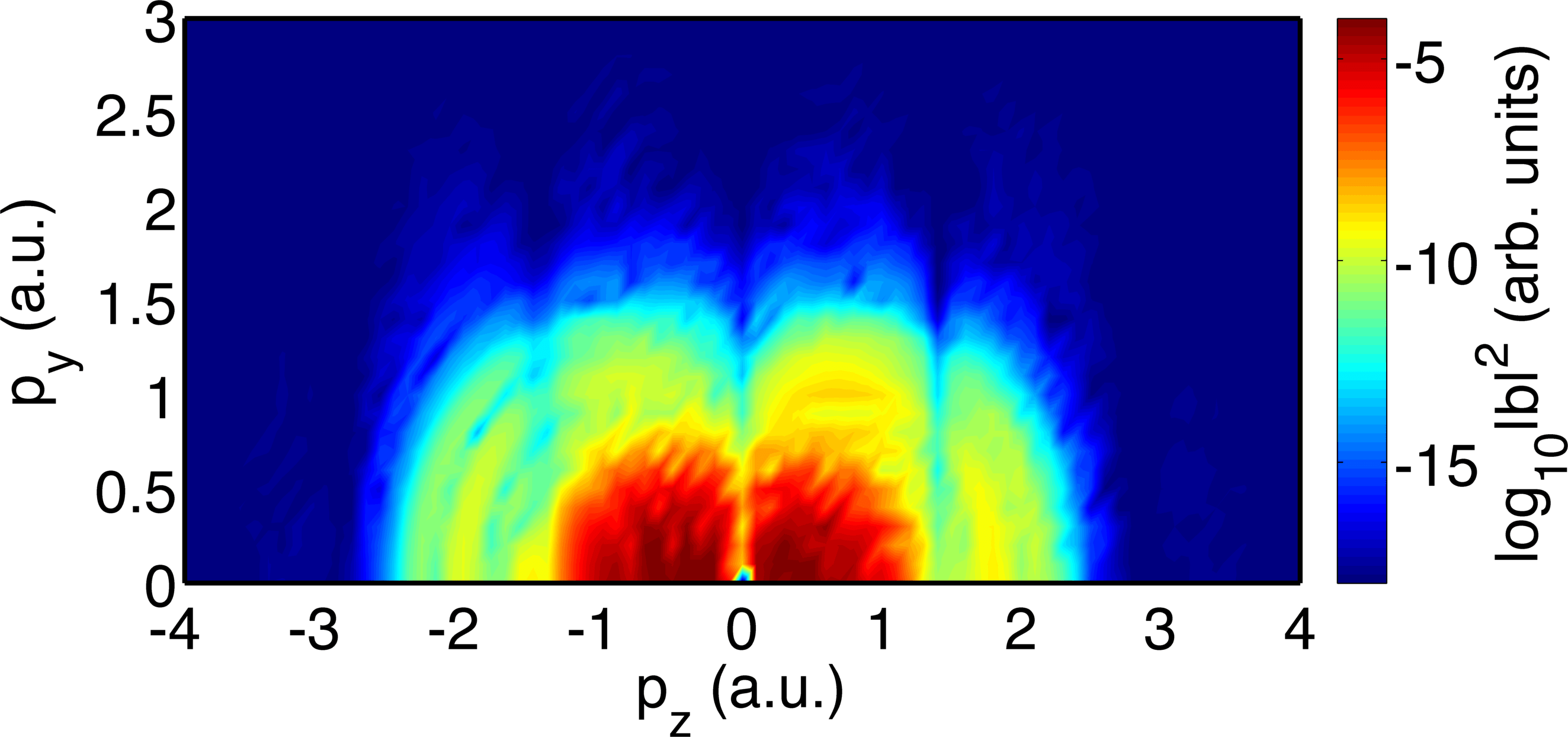}}
		\subfigure[]{\includegraphics[width=0.45\textwidth]{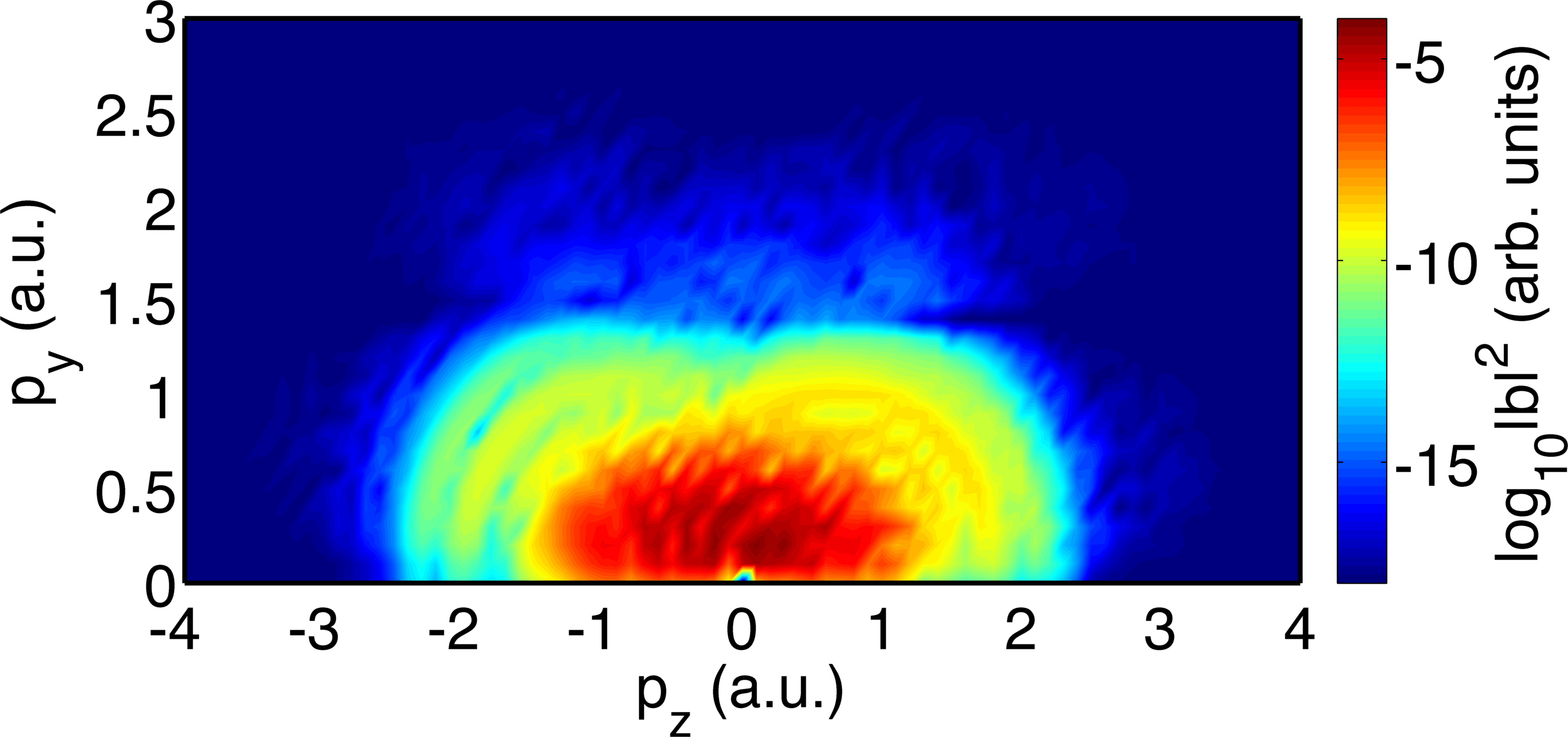} }
\caption{(color online) 2D-total ATI photoelectron spectra (in logarithmic scale) for the CO$_2$ molecule as a function of the $(p_z, p_y)$ electron momenta computed using the \textit{Model B}. (a) the molecule is oriented parallel to the laser field polarization, (b) the same as in (a), but the molecule is oriented perpendicular to the laser field polarization (see text for more details).
                                    }                  \label{Fig:CO2_2D}                
                    \end{figure}

Let us next analyze the effect of the molecular orientation on the ATI spectra. In order to do this we compute the final photoelectron spectra for the molecule oriented parallel and perpendicular with respect to the laser field polarization. In both cases we use the \textit{Model B} and in Fig.~\ref{Fig:CO2_2D} we show the results.

In the parallel configuration, $\theta=0^{\circ}$ [Fig.~\ref{Fig:CO2_2D}(a)], we can see the typical interference pattern with deep minima located at around p$_z=\pm1.4$ a.u. The position of these minima is in agreement with the second minimum predicted by the two slit interference formula~\cite{PRANoslen2016} for two radiant point separated a distance $R=2.2$ a.u, i.e.~just the separation between the oxygen and the carbon atoms. On the contrary, in the perpendicular configuration, $\theta=90^{\circ}$ [Fig.~\ref{Fig:CO2_2D}(b)], there is no trace of two-center interferences.
                  
\subsubsection{CS$_{2}$ molecule}

For the CS$_2$ molecule we focus our study in the dependency of the total photoelectron spectra with the molecular orientation. We perform calculations using both models for three different orientation angles. The parameters used in the nonlocal SR potential are $\Gamma=0.71$ and $\gamma=0.099$~a.u., respectively. With these values, we match the ionization potential $I_p=0.32$~a.u of the CS$_2$ molecule obtained with GAMESS. Additionally, the CS$_2$ HOMO is modelled in the 
\textit{Model B} as a combination of only $2p$ AOs.

 \begin{figure}[htb]
 
            \subfigure[]{ \includegraphics [width=0.45\textwidth] {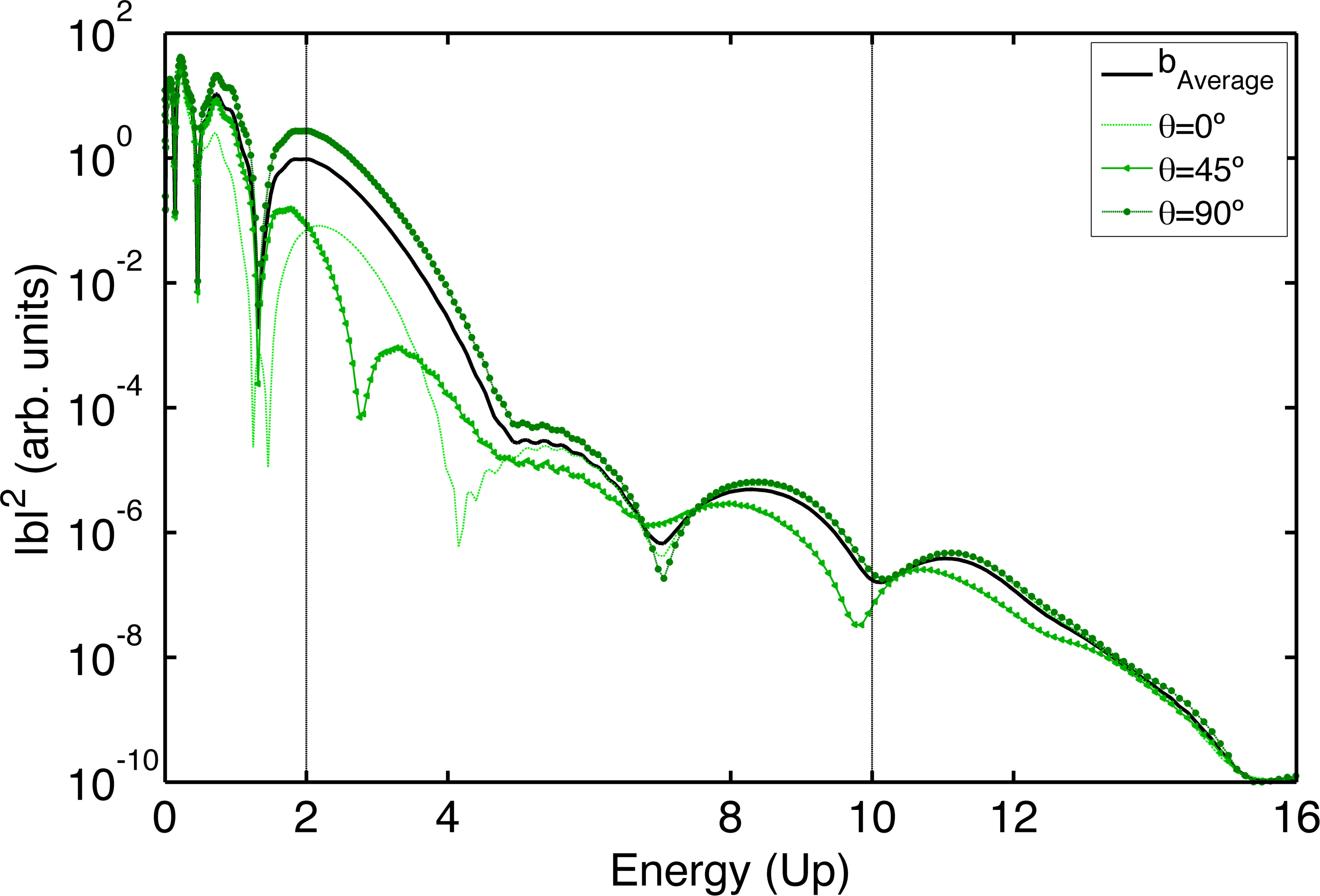}
            \label{fig:subfiga}}
		 \subfigure[]{\includegraphics[width=0.45\textwidth]{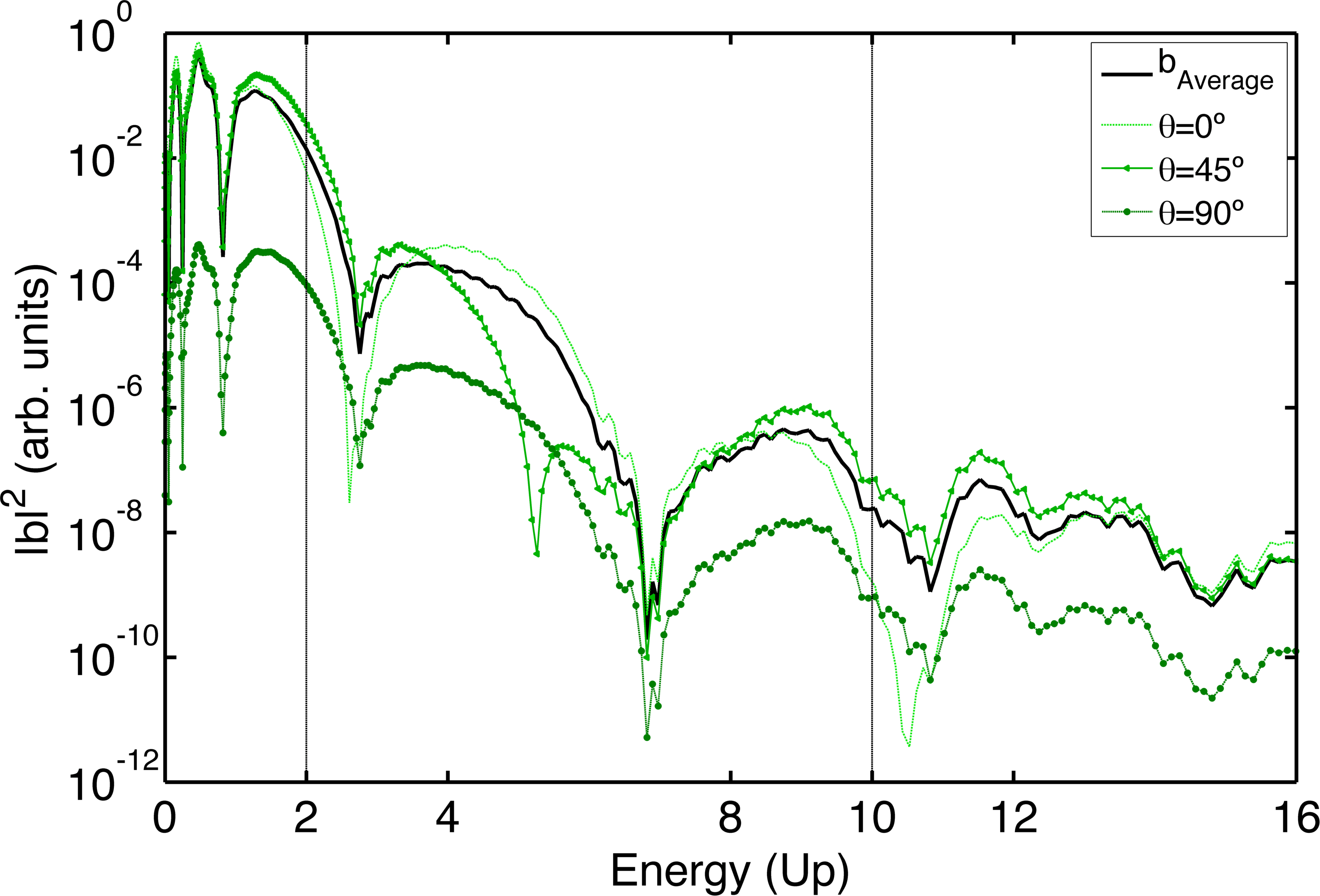}}

             \caption{(color online)  Total photoelectron spectra (in logarithmic scale) as a function the of the electron energy in $U_p$ units. (a) calculated by using \textit{Model A}; (b) calculated using \textit{Model B}. In both models the CS$_2$ molecule is at equilibrium $R=5.86$ a.u. (3.1 \AA). The peak laser intensity used in this calculation is set to $I_0= 1\times10^{14}$~W$\,\cdot$\,cm$^{-2}$ (see text for more details).
                                    }                  \label{Fig:CS2_Ang}                
                    \end{figure}

We consider the molecule oriented at $\theta=0^{\circ}$, $\theta=45^{\circ}$ and $\theta=90^{\circ}$ with respect to the laser field polarization and we also include an averaged-ATI spectra over these three orientations.  

The calculations using the \textit{Model A} [Fig.~\ref{Fig:CS2_Ang}(a)] show only minor dissimilarities in shape and amplitude for the three different orientations. The main differences appear in the low energy part, where the spectra depict different yield and the position of the interference minima change. In this case the most favorable orientation, i.e.~the one that gives the highest yield, is $\theta=90^{\circ}$, i.e.~the molecule is oriented perpendicular to the laser field polarization. This result is in agreement with our previous publication~\cite{PRANoslen2016}, where the HHG for a three-center molecule, CO$_2$, shows a similar behaviour.

For the ATI spectra obtained using the \textit{Model B} [Fig.~\ref{Fig:CS2_Ang}(b)], we observe that the behavior is completely the opposite: in the perpendicular case the total yield drops by more than three orders of magnitude and it is the parallel orientation the one that dominates. Additionally, the differences between the three orientations are now more visible. We could argue then that the \textit{Model B} is not only more accurate in the MO description but also more sensitive to the molecular orientation. 

\begin{figure}[htb]		
		 \subfigure[]{\includegraphics[width=0.45\textwidth]{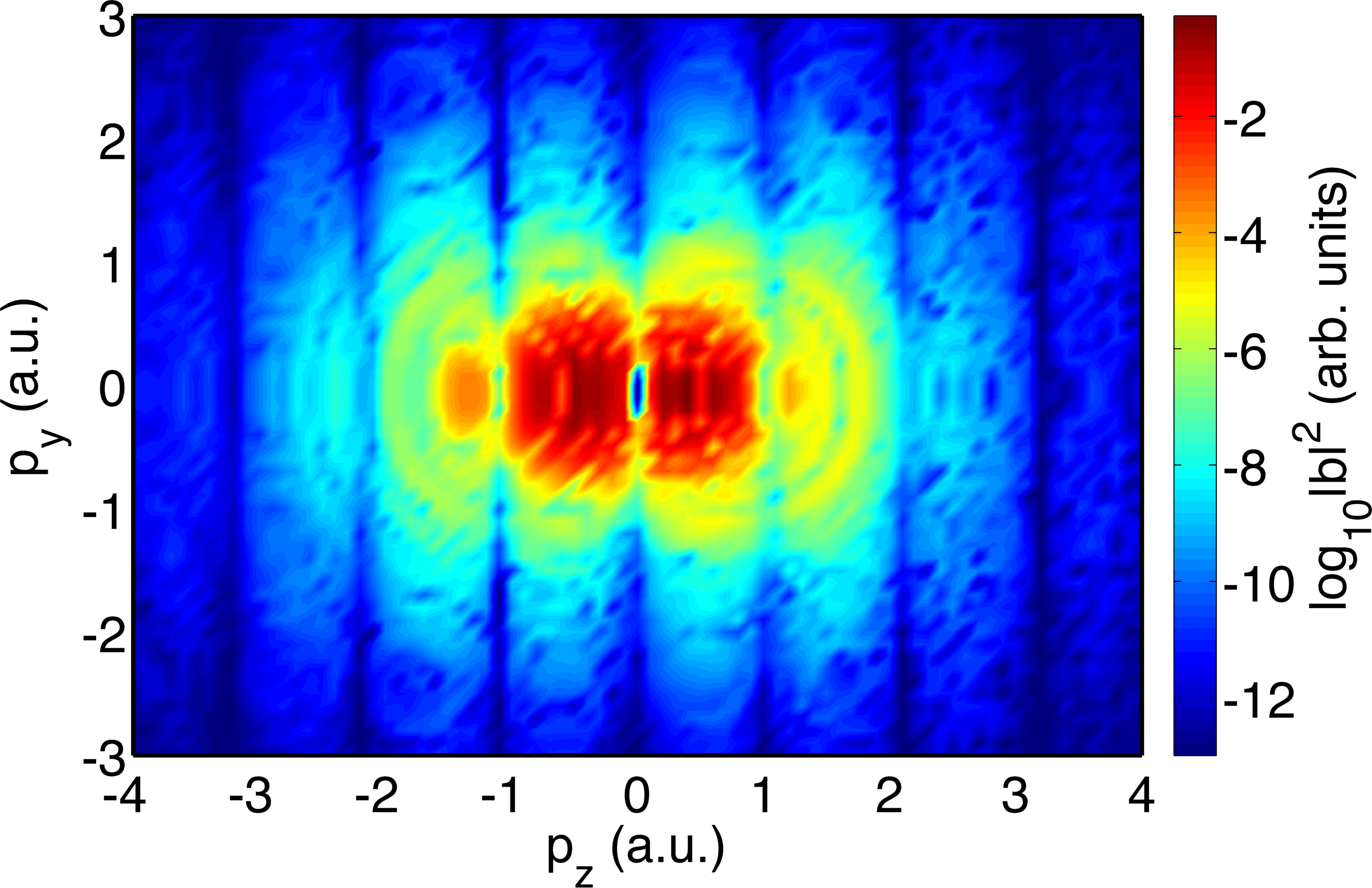}}
		 \subfigure[]{\includegraphics[width=0.45\textwidth]{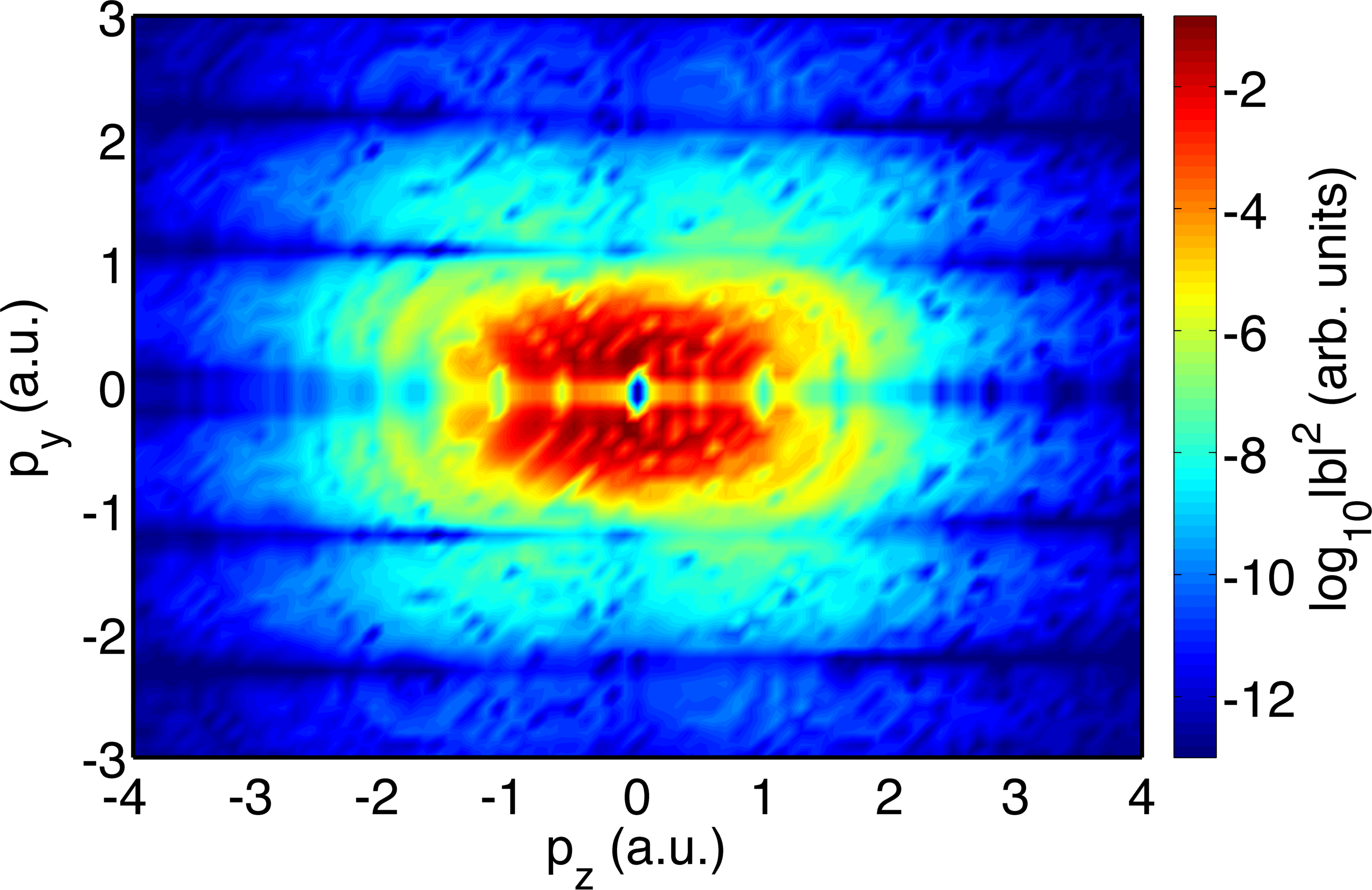} }
\caption{(color online) 2D-total ATI photoelectron spectra (in logarithmic scale) for the CS$_2$ molecule as a function of the $(p_z, p_y)$ electron momenta computed using our \textit{Model B}. (a) the molecule is oriented parallel to the laser field polarization, (b) the same as in (a), but the molecule is oriented perpendicular to the laser field polarization (see text for more details).
                                    }                  \label{Fig:CS2_2D}                
                    \end{figure}

In order to discuss differences and similarities with the CO$_2$ case, in Fig.~\ref{Fig:CS2_2D} we present 2D-total photoelectron spectra for a CS$_2$ molecule oriented at $\theta=0^{\circ}$ [Fig.~\ref{Fig:CS2_2D}(a)] and $\theta=90^{\circ}$ [Fig.~\ref{Fig:CS2_2D}(b)], with respect to the laser field polarization. 

The results obtained show the sensitivity of our model to the molecular orientation and the presence of interference minima now for the two orientations (this is in clear contrast to the CO$_2$ case, where for the perpendicular orientation, Fig.~\ref{Fig:CO2_2D}(b), there are not fingerprints of interferences). Furthermore, we observe that, for the parallel case, Fig.~\ref{Fig:CS2_2D}(a), the interference minima are placed for fixed $p_z$ values, i.e.~parallel to the $p_y$ axis, meanwhile that for the case of $\theta=90^{\circ}$ these minima are for fixed $p_y$  values, i.e. parallel to the $p_z$ axis. These features are related with the shape of the CS$_2$ HOMO, that it is inherited in the molecular bound-free matrix element.

\section{Conclusions and Outlook}

We present a quasiclassical approach that deals with molecular ATI within the SAE. Our model could be considered as a natural extension to the one introduced in Refs.~\cite{PRANoslen2015, PRANoslen2016}. The focus of the present study is on triatomic molecular systems, although the extension to more complex systems appears to be straightforward. 

First, we have shown our approach is able to capture the interference features, ubiquitously present in every molecular ATI process. As was already described, the core of our model are the saddle-point approximation and the linear combination of atomic orbitals (LCAO). One of the main advantages of our approach is the possibility to disentangle, in an easy and direct way, the different contributions to the total ATI. This is particularly important for complex systems, where there exists a large amount of direct and rescattering 'scenarios' that otherwise would be impossible to extricate. 
 
Second, we establish a comparison using two different ground states, one that uses a nonlocal SR potential, \textit{Model A}, and the other based on the LCAO, \textit{Model B}. Meanwhile both models allow us to formulate the ATI in a semi-analytical way, the latter (\textit{Model B}) gives a more accurate description of the MO. Nevertheless, we proved that, even when the former (\textit{Model A}) predicts an overestimation of the direct processes, the shape and the spectra features are well reproduced. Additionally, \textit{Model B} appears to be the adequate platform to investigate much more complex systems. For instance, the modeling of the DNA basis, formed by around a dozen of atoms, seems to be perfectly feasible. This will be object of future investigations.

\acknowledgments{This work was supported by the project ELI-Extreme Light Infrastructure-phase 2 (Grant No. CZ.02.1.01/0.0/0.0/15\_008/0000162) from European Regional Development Fund, Ministerio de Econom\'{\i}a y Competitividad through Plan Nacional (Grant No. FIS2016-79508-P, FISICATEAMO and Severo Ochoa Excellence Grant No. SEV-2015-0522), and funding from the European Unions Horizon 2020 research and innovation Programme under the Marie Sklodowska-Curie Grant Agreement No. 641272 and Laserlab-Europe (Grant No. EU-H2020 654148), Fundaci\'o Privada Cellex, and Generalitat de Catalunya (Grant No. SGR 874 and CERCA Programme). N.S. was supported by the Erasmus Mundus Doctorate Program Europhotonics (Grant No. 159224-1-2009-1-FR-ERAMUNDUS-EMJD). N.S., A.C., E.P., and M.L. acknowledge ERC AdG OSYRIS, EU FETPRO QUIC and National Science Centre Poland-Symfonia Grant No. 2016/20/W/ST4/00314. J.B. acknowledges Grant No. FIS2014-51478-ERC and the National Science Centre, Poland-Symfonia Grant No. 2016/20/W/ST4/00314. We thank Robert Moszynski, Kuba Zakrzewski and Francesca Calegari for fruitful discussions.}

\clearpage
\appendix

\section{Two-center systems. Bound-continuum and rescattering transition matrix elements}\label{App:01}

In this Appendix we present the equations to calculate the bound-continuum and the rescattering matrix elements for a diatomic molecule.
In this case we set $n=2$ -$n$ determines the number of atoms, in such a way to distinguish this formulation from the one presented for three-center systems.

Les us first recall the expression for the bound state of a diatomic system obtained in~\cite{PRANoslen2016}, i.e.~

\begin{equation}
\Psi_{2-0_{SR}}(\textbf{p}) =  \frac{ \mathcal{M}_2 \: e^{-i\textbf{R}_1 \cdot  \textbf{p}} }{ \sqrt{(p^2 + \Gamma^2})(\frac{p^2}{2} + I_p)} + \frac{ \mathcal{M}_2 \: e^{- i\textbf{R}_2 \cdot \textbf{p}}}{ \sqrt{(p^2 + \Gamma^2})(\frac{p^2}{2} + I_p)},
\label{Eq:BSMolec}
\end{equation} 
where the normalization constant $\mathcal{M}_2$ is,
\begin{equation}
\mathcal{M}_2= \frac{1}{2}{\Bigg[ \frac{ 2\pi^2}{( 2I_p-\Gamma^2 )^2} \Bigg\{  \frac{2\: e^{-R \Gamma}} {R}  - \frac{2 \: e^{-R \sqrt{2I_p}}} {R} -\frac{(2I_p -\Gamma^2) e^{-R \sqrt{2I_p}}}{\sqrt{2I_p}} + \frac{(\sqrt{2I_p} - \Gamma)^2}{\sqrt{2I_p}}   \Bigg\} \Bigg] }^{-1/2}.
\label{Eq:BNormConst}
 \end{equation}
Following the definition on Eq.~(\ref{Eq:DM1}), with now $n=2$, we have:
\begin{eqnarray}
\textbf{d}_{2-SR}( \textbf{p}_0)&=&\sum_{j=1}^2 {\bf d}_{2-SR_j}({\bf p}_0)=-2\textit{i}\: \mathcal{M}_{2}  \mathcal{ A} ( \textbf{p}_0) \:\Big\{  e^{-i\textbf{R}_1 \cdot \textbf{p}_0} + e^{-i\textbf{R}_2 \cdot \textbf{p}_0}\Big\},
\label{Eq:d2N-SR}
\end{eqnarray}
where $\mathcal{M}_2$ is a normalization constant defined in Eq.~(\ref{Eq:BNormConst}) and,
\begin{equation}
\mathcal{ A} ( \textbf{p}_0)=- \textbf{p}_0\: \frac{(3p_0^2 + 2I_p +2\Gamma^2)}{(p_0^2 + \Gamma^2)^{\frac{3}{2}}(p_0^2 + 2I_p)^2}.
\end{equation}
Using this last equation we have completely defined the direct transition amplitude for a two-center system by inserting Eq.~(\ref{Eq:d2N-SR}) in Eq.~(\ref{Eq:b_10}).

For the calculation of the rescattering and the total transition amplitudes we need to obtain the scattering states and the rescattering transition matrix elements. In this way, the scattering state reads as:
\begin{eqnarray}
\Psi_{2-{\bf p}_0}= \delta(\textbf{p}-\textbf{p}_0)+ \sum_{j=1}^{2}\: \delta\Psi_{\textbf{R}_j \textbf{p}_0}(\textbf{p}),
\end{eqnarray} 
where
\begin{eqnarray} 
\delta\Psi_{2-\textbf{R}_1 \textbf{p}_0}(\textbf{p})= \frac{ \mathcal{D}_{2-1} (\textbf{p}_0) \: e^{-i \textbf{R}_{1} \cdot (\textbf{p} - \textbf{p}_0 )} - \mathcal{D}_{2-2} (\textbf{p}_0) \: e^{-i \textbf{R}_{1}  \cdot (\textbf{p} + \textbf{p}_0 )}}{ \sqrt{p^2 + \Gamma^2} \:(p_0^2 -p^2 +\textit{i}\epsilon) },
\label{Eq:dphiML} \\
\delta\Psi_{2-\textbf{R}_2 \textbf{p}_0}(\textbf{p})= \frac{ \mathcal{D}_{2-1} (\textbf{p}_0) \: e^{-i \textbf{R}_{2} \cdot (\textbf{p} - \textbf{p}_0 )} - \mathcal{D}_{2-2} (\textbf{p}_0) \: e^{-i \textbf{R}_{2}  \cdot (\textbf{p} + \textbf{p}_0 )}}{ \sqrt{p^2 + \Gamma^2} \:(p_0^2 -p^2 +\textit{i}\epsilon) }.
 \label{Eq:dphiMR}
 \end{eqnarray}

Finally, let us obtain the explicit expressions for the rescattering transition matrix elements for diatomics, $\textbf{g}_{2-jj'}$.
\begin{eqnarray}
\textbf{g}_{2-11}(\textbf{p}_1, \textbf{p}_2) &=&  \mathcal{Q}_{2-1}(\textbf{p}_1, \textbf{p}_2) \: e^{- i \textbf{R}_{1} \cdot (\textbf{p}_1 - \textbf{p}_2) } ,\nonumber\\
\textbf{g}_{2-12}(\textbf{p}_1, \textbf{p}_2) &=& \mathcal{Q}_{2-2}(\textbf{p}_1, \textbf{p}_2) \: \:e^{- i \textbf{R}_2 \cdot\textbf{p}_1 +i \textbf{R}_1 \cdot\textbf{p}_2 },\nonumber\\
\textbf{g}_{2-22}(\textbf{p}_1, \textbf{p}_2) &=& \mathcal{Q}_{2-1}(\textbf{p}_1, \textbf{p}_2) \: e^{- i \textbf{R}_{2} \cdot (\textbf{p}_1 -  \textbf{p}_2 )},\nonumber\\
\textbf{g}_{2-21}(\textbf{p}_1, \textbf{p}_2) &=& \mathcal{Q}_{2-2}(\textbf{p}_1, \textbf{p}_2)\: e^{-i \textbf{R}_1 \cdot \textbf{p}_1 + i \textbf{R}_2 \cdot  \textbf{p}_2 } .
\label{Eq:g2N}
\end{eqnarray} 
 where
 \begin{eqnarray}
\mathcal{Q}_{2-1} (\textbf{p}_1, \textbf{p}_2) &=& i \Big[ \mathcal{D}_{2-1} (\textbf{p}_2) \mathcal{C}_1(\textbf{p}_1, \textbf{p}_2) - \mathcal{D}_{2_1}^*(\textbf{p}_1) \mathcal{C}_2(\textbf{p}_1, \textbf{p}_2) \Big]
\label{q2n1}
\end{eqnarray}
and
 \begin{eqnarray}
\mathcal{Q}_{2-2} (\textbf{p}_1, \textbf{p}_2) &=& -i \Big[ \mathcal{D}_{2-2} (\textbf{p}_2) \mathcal{C}_1(\textbf{p}_1, \textbf{p}_2) -\mathcal{D}_{2-2}^*(\textbf{p}_1) \mathcal{C}_2(\textbf{p}_1, \textbf{p}_2) \Big].
\label{q2n2}
\end{eqnarray}
The constants in Eqs.~(\ref{q2n1}) and~(\ref{q2n2}) are defined as:
\begin{equation}
\mathcal{D}_{2-1}(\textbf{p}_0)=\frac{\gamma}{\sqrt{p_0^2 + \Gamma^2}} \Bigg\{  \frac{1 + I'_1 }{ {I'_2}^2 - \big(1+I'_1  \big)^2  }  \Bigg\}; \mathcal{D}_{2-2}(\textbf{p}_0)=\frac{\gamma}{\sqrt{p_0^2 + \Gamma^2}} \Bigg\{  \frac{I'_2}{ {I'_2}^2 - \big(1+I'_1  \big)^2  }  \Bigg\},
\end{equation} 
\begin{eqnarray} 
I'_1 &=&  \frac{-  2\pi^2 \: \gamma }{\Gamma - i\sqrt{|p_0^2 + \textit{i}\:\epsilon|}},\label{Eq:I1}\\
I'_2 &=& \frac{- 2 \pi^2 \:\gamma} {R\:(p_0^2 + \Gamma^2  + i\epsilon)} \bigg[ e^{iR\:\sqrt{p_0^2 +i\epsilon}} - e^{-R\: \Gamma} \bigg ]\label{Eq:I2}
\end{eqnarray}
and   
\begin{equation}
\mathcal{C}_1 (\textbf{p}_1, \textbf{p}_2) =  \Bigg[  \frac{ \textbf{p}_1(3p^2_1 -p^2_2 + 2\Gamma^2 ) }{  (p_1^2 +\Gamma^2) ^{\frac{3}{2}}(p_2^2 - p_1^2 +i\epsilon)^2 }\Bigg],
\mathcal{C}_2(\textbf{p}_1, \textbf{p}_2) =  \Bigg[  \frac{\textbf{p}_2(3 p^2_2- p^2_1 + 2\Gamma^2 )}{  (p_2^2 +\Gamma^2)^{\frac{3}{2}} (p_1^2 - p_2^2 -\textit{i}\epsilon)^2 } \Bigg].
\end{equation}

\clearpage
\newpage


%

\end{document}